\documentclass[12pt]{article}

\usepackage{jheppub,hyperref,float,array,adjustbox,mathtools,physics, xcolor}
\usepackage{graphicx,latexsym} 
\usepackage{amsthm,lscape}
\usepackage{comment}
\usepackage{amsmath,amssymb,amsfonts,amsxtra,mathrsfs,makeidx,graphicx,amsthm,epsfig,bm,longtable,float,color,tikz,mathtools,xfrac,footnote,rotating,lscape,makecell,environ,mathtools,empheq,physics,cleveref,tensor,
slashed,subfiles,natbib,youngtab,multirow}
\usepackage[font=small]{caption}
\usepackage{subcaption}
\usetikzlibrary{calc}
\usetikzlibrary{backgrounds}
\usepackage{lipsum}

\newcommand{\USp}{\mathrm{USp}}
\newcommand{\SU}{\mathrm{SU}}
\newcommand{\U}{\mathrm{U}}

\newcommand{\ee}{\mathrm{e}}
\newcommand{\mi}{\mathrm{i}}
\newcommand{\Pf}[1]{\,\mathrm{Pf} #1}

\definecolor{USPcol}{rgb}{0.95, 0.75, 0.78} %chiarancione
\definecolor{SUcol}{rgb}{0.44, 0.68, 0.75} %pietrolio

\title{
\begin{center}
On the zoology of 2d $\mathcal{N}=(0,2)$ dualities of gauge theories with antisymmetric matter
\end{center}
}

\author[a]{Antonio Amariti,}	
\author[a,b]{Pietro Glorioso,}	
\author[a,b]{Chiara Mascherpa,}
\author[a,b]{and Andrea Zanetti}

\affiliation[a]{INFN, Sezione di Milano, Via Celoria 16, I-20133 Milano, Italy}
\affiliation[b]{Dipartimento di Fisica, Università degli studi di Milano, Via Celoria 16, I-20133}

\emailAdd{antonio.amariti@mi.infn.it}
\emailAdd{pietro.glorioso@mi.infn.it}
\emailAdd{chiara.mascherpa@mi.infn.it}
\emailAdd{andrea.zanetti@mi.infn.it}

\abstract{
In this paper we investigate and propose new dualities involving 2d gauge theories with $\mathcal{N}=(0,2)$ supersymmetry.
In the first part of the paper we focus on $\SU(n)$ gauge theories with two  antisymmetric chirals. The gauge theories are non-anomalous if we consider, in addition to such matter content, $n_f$ fundamental and $n_a$ antifundamental chirals, provided the constraint $n_f+n_a=4$. 
By exploring the five possibile scenarios arising from this constraint we provide in each case evidences of a dual LG description, by matching the 't Hooft anomalies and deriving the relation between the elliptic genera in terms of other \emph{more fundamental} dualities.
In the second part of the paper we provide a 4d origin for a gauge/LG duality already stated in the literature, that does not descend from any known 
s-confining duality.
In the last part of the paper we focus on dualities for $\SU(n)$ and $\USp(2n)$ models with antisymmetric Fermi multiplets, obtained from dimensional reduction of 4d parent dualities.
}

\begin{document}
\maketitle
\flushbottom
\allowdisplaybreaks 

\section{Introduction}

The existence of IR dualities is a fascinating aspect of quantum field theories, because it opens up the possibility of alternative field theoretical descriptions of the same physical setup.
Ubiquitous examples have been obtained in the supersymmetric setup, where dualities played a prominent role in different dimensions and degree of supersymmetry. An ultimate goal would be achieving a connection with the real world by uncovering dualities beyond the supersymmetric case. For this reason it is useful to take an intermediate step by finding dualities with a low amount of supercharges. While in 4d the lowest possible cases require four supercharges, in lower dimensional physics the number of supersymmetry generators can be lowered.
An interesting setup involves 2d dualities with $\mathcal{N}=(0,2)$ supersymmetry, as in this case a whole class of dualities can be achieved by dimensional reduction of 4d Seiberg-like dualities from a twisted compactification on a two sphere. 

Typically, such compactification of 4d dual pairs results in relations between collection of theories in lower dimension, obscuring the possible claims on pure 2d dualities. However, in \cite{Gadde:2015wta}, elaborating on the results of localization, the authors discuss a prescription that allows to infer the existence of 2d dualities starting from 4d ones. The prescription requires to fix the $R$ charges of the 4d fields to be non-negative integers. In this way the whole contribution to the 4d $S^2 \times T^2$ topologically twisted index \cite{Benini:2015noa} comes from the zero flux sector only, discarding the sum over the remaining ones.
By translating the same constraints imposed on the electric side to the magnetic one via the 4d duality map, one obtains a candidate 2d $\mathcal{N}=(0,2)$ duality. The 't Hooft anomaly matching are automatically satisfied and the equivalence of the elliptic genera is a consequence of the equivalence of the 4d topologically twisted indices.

In principle one cannot trust such dualities without any hesitation and further checks are necessary. 
For example the theories obtained in this way usually have non-compact direction in the target space. For this reason the c-extremization procedure of \cite{Benini:2012cz,Benini:2013cda} requires some care. Furthermore, the matching of the elliptic genera in this case is reliable only in the massive case.

Remarkable checks of such dualities have been performed in the literature when one restricts to the s-confining limit of Seiberg duality \cite{Seiberg:1994pq} and Intriligator-Pouliot duality \cite{Intriligator:1995ne}. In such cases the dual theories correspond to Landau-Ginzburg (LG) models.
Furthermore, as observed in \cite{Gadde:2015wta}, in the $\USp(2n)$ case the prescription described so far does not apply to the reduction of the Intriligator-Pouliot beyond the s-confining regime, whereas this is allowed for Seiberg duality.

Restricting to the reduction of s-confining gauge theories, it has been shown in \cite{Sacchi:2020pet,Jiang:2024ifv,Amariti:2024usp} that the other 4d s-confining dualities reduce to 2d gauge/LG dualities that can be derived in a pure 2d picture by generalizing the tensor deconfinement technique of \cite{Berkooz:1995km,Luty:1996cg}. This approach of deriving dualities from dualities was used recently in various dimensions 
\cite{Pasquetti:2019uop,Benvenuti:2020wpc,Etxebarria:2021lmq,Benvenuti:2021nwt,Bottini:2022vpy,Bajeot:2022lah,Bajeot:2022wmu,Amariti:2022wae,Amariti:2023wts,Amariti:2024sde,Amariti:2024gco,Benvenuti:2024glr,Hwang:2024hhy}
in order to show that several dualities in presence of tensor matter are consequence of more \emph{basic} ones with only fundamental matter.
A relevant result for our discussion was obtained in \cite{Bajeot:2022kwt}, where it was shown that the classification of \cite{Csaki:1996zb} for $\SU(N)$ and $\USp(2n)$ dualities can be obtained using only two \emph{basic} s-confining dualities, i.e. the s-confining limits of Seiberg and Intriligator-Pouliot dualities. The same phenomenon has been generalized in \cite{Amariti:2024usp} for 2d dualities obtained 
by applying the prescription of \cite{Gadde:2015wta} to such s-confining cases\footnote{The sporadic cases have not been studied in detail in \cite{Amariti:2024usp}, but the same result is expected to hold for them as well.}.

Furthermore, the idea of deriving 2d dualities with tensorial matter in terms of \emph{basic} gauge/LG dualities can be applied to other cases, where a 4d parent s-confining description is absent. Some examples have already been studied in \cite{Jiang:2024ifv,Amariti:2024usp}, but more cases are expected. A reason behind such expectations is that the constraints imposed by gauge anomalies are milder in 2d than in 4d. In addition, the similarities between the dualities found in \cite{Jiang:2024ifv,Amariti:2024usp} and the 3d ones with a similar field content obtained in \cite{Nii:2019ebv,Benvenuti:2021nwt,Amariti:2022wae,Amariti:2024gco} suggest that 2d gauge/LG dualities should exist also for $\SU(N)$ gauge theories with two antisymmetric chirals.

Motivated by this last observation, in this paper we continue our investigations of 2d $\mathcal{N}=(0,2)$ SUSY QFTs dual to LG models, providing a generalization of the previous results found in \cite{Amariti:2024usp}. We mostly focus on $\SU(N)$ gauge theories with antisymmetric and fundamental matter, considering both chiral and Fermi fields.

The first class of models under investigation consists of gauge/LG dualities without a known 4d origin, \textit{i.e.} no 4d confining dualities available in the literature give rise to the 2d dualities argued here through a topological twist.
We discuss $\SU(N)$ gauge theories with two antisymmetric, $n_f$  fundamental and $n_a$  antifundamental chirals, such that $n_f + n_a = 4$.
Notice that, even if such dualities do not descend from any 4d parent, they share a similarity with analogous 3d dualities proposed in \cite{Nii:2019ebv}, and further checked in \cite{Amariti:2024gco,Toappear}. Such relation agrees with the results of \cite{Amariti:2024usp}, where the various 2d dualities were shown to be in line, even in absence of a 4d parent, with analogous 3d dualities.
The main checks of the dualities proposed here consist of 't Hooft anomaly matching and of the study of the elliptic genus. While the first check is straightforward, the matching of the elliptic genera in this case cannot be inferred from the one of any 4d partition function, and its analysis is a crucial test of the duality.
Here we show that the matching of the elliptic genera reduces to the one of simpler dualities, that hold in absence of any tensorial matter field and that have a 4d origin.  
The analysis is very similar to the one performed in \cite{Sacchi:2020pet,Jiang:2024ifv,Amariti:2024usp} and it requires a duality that is not completely under control in 2d. The auxiliary duality involves an $\USp(2n)$ gauge theory with $2N+3$ fundamental chirals and $1$ fundamental Fermi. This model is dual to an LG, provided that a symmetry of the electric theory is obstructed\footnote{We will often refer, with a slight abuse of notation, to the electric and the magnetic phase for the 2d models under investigation, in analogy, when it exists, with the 4d counterpart.}. Such obstruction is due to the choice of $R$ charges used to perform the twist on the two-sphere. As observed in \cite{Gadde:2015wta,Sacchi:2020pet} this is reminiscent of the obstruction to the generation of axial (and topological) symmetries in 3d $\mathcal{N}=2$ gauge theories. Here in the 2d picture we do not have an analogous candidate playing the role of the monopole superpotential\footnote{The notion of boundary monopoles in \cite{Dimofte:2017tpi} seems promising for understanding a dynamical origin for such obstructions.}, nevertheless we claim that the identity derived from the reduction of the zero flux sector of the $S^2 \times T^2$ topologically twisted index to the elliptic genus is reliable, and can be used to provide a derivation of the dualities under investigation.
Assuming the matching of the elliptic genera for such duality, we will provide the relative identity for the duality involving $\SU(N)$ with two antisymmetric chirals and four fundamentals.
Such proof requires a recursive analysis that leads to the towers of operators, expected from the field theory duality, that one can conjecture from the 3d analogous found in \cite{Nii:2019ebv}.
Once this duality with $n_f=4$ is established, we will show that all the other cases with $n_a\neq 0$ can be derived from the one with $n_f=4$, \textit{i.e.} in the recursive derivation we always find a deformed version of the model with $n_f=4$.

In the second part of the paper we revisit the problem of 2d dualities without a known 4d parent, by considering the case of $\SU(2n)$ with an antisymmetric flavor and four fundamentals discussed in \cite{Amariti:2024usp}. Here we present a 4d parent duality, with a non-vanishing superpotential, that reduces to the 2d expected one by twisting along an assignation of non-negative integer $R$ charges.

The last part of the paper focuses on $\SU(N)$ and $\USp(2n)$ dualities in presence of a charged antisymmetric Fermi field. These models are obtained by reducing 4d dualities that have been guessed in the literature  from the analysis of the superconformal index \cite{Spiridonov:2009za}. A physical derivation of such 4d parent dualities was then provided in \cite{Amariti:2023wts,Amariti:2024sde}.
We study the twist of these dualities on the two sphere by choosing integer non-vanishing $R$ symmetries and in this way we argue the existence of new 2d dualities.
In the first case the duality involves an $\SU(N)$ theory with $N$ fundamentals and one antisymmetric Fermi whose dual model is identified with an LG model. The second case regards a duality between an $\USp(2n)$ theory and an $\USp(2N-2)$ theory, both with an antisymmetric Fermi, $4N$ fundamentals and non-trivial $J$-terms.

\section{Basic dualities}

\label{basic}

In this section we review some of the relevant basic dualities that we will use in the paper in order to derive the matching of the elliptic general for the dualities proposed below.
These dualities are already known, but they have been less discussed in the literature, for this reason we provide a detailed discussion and some further comments in order to employ them below. 

In general we consider the  4d/2d reduction of dualities  by compactifying on $S^2$ and by applying the prescription introduced in \cite{Gadde:2015wta}: \textit{i.e.}  given a $4$d chiral multiplet with $R$ charge $R$, this reduces to 
\begin{itemize}
    \item $R-1$ Fermi multiplets if $R>1$\,,
    \item $1-R$ chiral multiplets if $R<1$\,,
    \item no multiplets if $R=1$\,,
\end{itemize}
while the $4$d vector multiplet reduces to a $2$d vector multiplet.
As discussed in \cite{Gadde:2015wta} this set of rules, provided that we have integer and non-negative $R$ charges, allows to obtain a $2$d $\mathcal{N}=(0,2)$ theory from a $4$d $\mathcal{N}=1$ one.
Here we further restrict our attention to cases without any field with $R>2$ in order to avoid the generation of possible non-abelian symmetries in 2d.

\subsection{$\USp(2n)$ with $2n+3$ $\square$ chirals and $1$ $\square$ Fermi}
\label{USpallaSacchi}

We start our survey by discussing a
duality with a 4d origin that will play a relevant role in our search of new 2d gauge/LG dualities. It corresponds to $\USp(2n)$
with $2n+3$ fundamental chirals and one Fermi.
The duality can be derived from 4d using $\USp(2n)$ with $2n+4$ 
fundamentals by assigning an $R$ charge equal to 2 
to one fundamental and $R$ charges equal to zero to the remaining $2n+3$  fundamentals.
The 2d field content is anomaly free and the dual meson splits into a
chiral antisymmetric meson with $(2n+3)(n+2)$ components $\Phi_{M_{ab}}$ and $2n+3$ Fermi $\Psi_{a}$.

The superpotential of the 2d theory is
\begin{equation}
\label{spotUSpconfFermi}
W_{2d} = \epsilon_{a_1,\dots,a_{2n+3}}
\left(\Phi_{M_{a_1 a_2}} \dots \Phi_{M_{a_{2n+1} a_{2n+2}}} \Psi_{a_{2n+3}} \right)
\equiv \Phi_M^{n+1} \Psi
\end{equation}
and it descends from the 4d superpotential of the confining theory, corresponding to the limiting case of the Intriligator-Pouliot duality \cite{Intriligator:1995ne}.
In the following, we will adopt a 2d notation, where we will refer to the $J$-term for the Fermi $ \Psi_{a}$ instead of the 2d superpotential. Moreover, in all the examples discussed in this paper the $E$-terms are vanishing.

%%%%%%%%%%%%%%%%%%%%%%%%%%%%%%%%%%%%%%%%%%%%%%%%%%%%%%%%%%%%%%%%%%%%%%%
% FIGURE: Duality USp(2n+3)
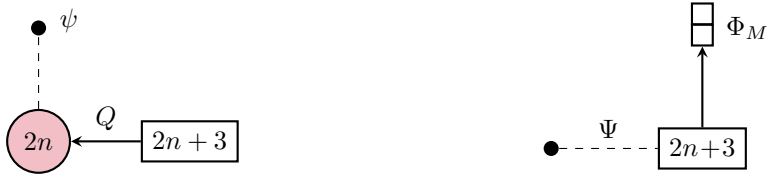
\begin{figure}[h!]
    \centering
        % STEP 1
        \begin{minipage}{0.45\textwidth}
            \centering
            \makebox[\textwidth][c]{
            \begin{tikzpicture}[
                every node/.style={font=\footnotesize},
                box/.style={rectangle, draw, thick},
                box2/.style={rectangle, draw, dashed}
            ]
            \pgfmathsetmacro{\x}{0.4}
            \pgfmathsetmacro{\y}{1.6}
            \pgfmathsetmacro{\a}{2}
            % Nodi
            \node[box] (r) at (\a,0) {$2n+3$};
            \node[fill=USPcol,circle,draw,thick] (center) at (0, 0)
             {$2n$};
            \node[circle, fill=black, inner sep=2pt] (dot) at (0, 1.5) {};
            % Fermi
            \node at (\x, \y) {$\psi$};
            \draw[-,dashed,>=stealth] (center.north) to[out=90, in=-90, looseness=0] (dot.south);
            % Fondamentali
            \draw[->,thick,>=stealth] (r) -- node[above] {$Q$} (center);
            \end{tikzpicture}
            }
        \end{minipage}
        % STEP 2
        \begin{minipage}{0.45\textwidth}
            \centering
            \makebox[\textwidth][c]{
            \begin{tikzpicture}[
                every node/.style={font=\footnotesize},
                box/.style={rectangle, draw, thick},
                box2/.style={rectangle, draw, dashed}
            ]
            \pgfmathsetmacro{\x}{0.6}
            \pgfmathsetmacro{\y}{1.6}
            \pgfmathsetmacro{\a}{2}
            % Nodi
            \node[box] (l) at (0,0) {$2n\!+\!3$};
            \node[circle, fill=black, inner sep=2pt] (dot) at (-\a, 0) {};
            % Antisimmetriche
            \node[box, minimum size=0.2cm] (square1r) at (0, 1.5) {};
            \node[box, minimum size=0.2cm] (square2r) at (0, 1.78) {};
            \node at (\x, \y) {$\Phi_M$};
            \draw[<-,thick,>=stealth] (square1r.south) to[out=-90, in=90, looseness=0] (l.north);
            % Fermi
            \draw[dashed] (dot) -- node[above] {$\Psi$} (l);
            \end{tikzpicture}
            }
        \end{minipage}
    \caption{In this figure we represents the quivers associated to the duality between $\USp(2n)$ with $2n+3$ fundamental chirals and one Fermi, and the LG model. Dashed lines are for Fermi fields while continuous lines are for chiral multiplets. The dots refer to the singlets.}
    \label{Figconfining}
\end{figure}
%%%%%%%%%%%%%%%%%%%%%%%%%%%%%%%%%%%%%%%%%%%%%%%%%%%%%%%%%%%%%%%%%%%%%%%

The global charges of the fields are
\begin{equation}
\begin{array}{c|cc}
&\U(1)_Q&  \SU(2n+3) \\
\hline
Q &1&\square\\
\psi&-2n-3&\cdot\\
\hline
\Phi_M= Q^2 &2&
\begin{array}{c}
\square \vspace{-2.85mm} \\
\square 
\end{array}  \\
\Psi=Q \psi &-2n-2&\square 
\end{array}
\end{equation}
As a first check one can see that the anomalies 
\begin{equation}
    \kappa_{QQ} = -4n(n+1)(2n+3)\,,
    \qquad
    \kappa_{\SU(2n+3)^2}=n\,,
\end{equation}
are equal on both sides of the duality.

Here is a crucial aspect of the duality proposed above:
on the gauge theory side of the duality we have not included an axial symmetry, which gives opposite charge to the $2n+3$ chirals and to the Fermi.
However, such symmetry is not forbidden by the field content and by the requirements from anomaly cancellation.
At the level of localization such a symmetry is not generated in the reduction of the $S^2 \times T^2$ index to the 2d elliptic genus along the lines of the prescription of \cite{Gadde:2015wta}, i.e. by assigning vanishing $R$ charge to $2n+2$ fields and $R=2$ to the $(2n+4)$-th  4d $\mathcal{N}=1$ chiral. As discussed originally in \cite{Gadde:2015wta} and elaborated in \cite{Sacchi:2020pet}, in this case the absence of such a symmetry is very similar to the obstruction to the generation of an axial symmetry in the 4d/3d reduction of dualities due to finite size effects \cite{Aharony:2013dha}. In the 4d/2d reduction of \cite{Gadde:2015wta}, restricting our attention to the cases with $R=0,1$ and $2$, such analogy is ubiquitous for all the anomaly free $R$ charge assignations where there are no fields with $R=1$, \textit{i.e.} each 4d field gives origin either to a 2d $\mathcal{N}=(0,2)$ chiral or to a 2d 
$\mathcal{N}=(0,2)$ Fermi.

We expect that analogous finite size effects have to be considered in such 2d cases as well. Even if we postpone a physical discussion about the origin of such effect to future investigation, it is worth here to comment more on such issue. 
At the level of localization the obstruction imposed by the anomalies translates into a constraint on the fugacities. Borrowing the mathematical terminology, such constraints are denoted as balancing conditions in the literature of the 4d superconformal index (see for example \cite{Spiridonov:2009za}). Similar constraints arise in the analysis of the 3d three sphere partition function when monopole superpotential (for example KK monopole superpotentials signalling the presence of finite size effects) are considered.
Similarly, in the 4d/2d reduction of the $S^2 \times T^2$ index to the elliptic genus, the 4d constraints on the fugacities survive when only fields with $R$ charges $R=0$ and $R=2$ are considered in 4d.
For this reason we expect that such balancing conditions signal the fact that finite size effects have to be considered in such cases.
We will return on such issue below in the paper when we will apply the duality discussed here in order to deconfine the antisymmetric tensors\footnote{Observe that this 2d confining duality has been recently used in \cite{Jiang:2024ifv} in order to prove 2d confining dualities with antisymmetric matter.}.

At the level of the elliptic genus the duality discussed in this subsection translates into the identity
\begin{equation}
\label{idsacchi}
I_{\USp(2n)}^{(2n+3;\;\cdot)}(x \vec u;x^{-2n-3};\cdot)
=
\frac{\prod_{a=1}^{2n+3} \theta\left(q u_a/x^{2n+2} \right)}{\prod_{1\leq a<b\leq 2n+3} \theta(x^2 u_a u_b)}\,,
\end{equation}
with $\prod_{a=1}^{2n+3} u_a=1$.
We refer the reader to Appendix \ref{convEG} for the conventions adopted here on the elliptic genus.

\subsection{$\USp(2n)$ with $2n+2$ fundamentals}
\label{sec:USP2Np2}

This duality has been obtained in \cite{Gadde:2015wta} and further studied in \cite{Dedushenko:2017osi} for $n=1$. It can be obtained from the reduction of the 4d s-confining limit of the $\USp(2n)$ Intriligator-Pouliot duality, \textit{i.e.} in presence of $2n+4$ fundamentals and vanishing superpotential. In this case the twist needs to be performed by fixing the $R$ charge of two fundamentals to $R=1$, and by assigning vanishing $R$ charge to the remaining ones.

The 2d duality maps an $\USp(2n)$ gauge theory with $2n+2$ fundamental chirals $Q$ to an LG theory with an antisymmetric chiral $A = Q^2$ in addition to a Fermi $\Psi$ with J-term $J_\Psi = \Pf A$. 
At the level of the elliptic genus the relative identity is
\begin{equation}
\label{ellipticdualUSP}
I_{\USp(2n)}^{(2n+2,\cdot)}(\vec u;\cdot;\cdot)
=
\frac{\theta\left(q ( u_1\dots u_{2n+2})^{-1} \right)}{\prod_{1\leq a<b\leq 2n+2} \theta(u_a u_b )}\, ,
\end{equation}
where we stress that the numerator on the RHS is written in this way in order to make apparent the structure of the $J$-term for the Fermi $\Psi$. Using the fact that $\theta(q/x) = \theta(x) $ we could also 
write the numerator as $\theta(\prod u_a)$, which would correspond to exchanging the $J$ term with an $E$ term.
We refer the reader to Appendix \ref{convEG} for the conventions adopted here on the elliptic genus.

\subsection{$\SU(N)$ with $N+x$ $\square$ chirals, $N-x+y$ $\overline \square$ chirals, $y$ $\overline \square$ Fermi}
\label{SUnGen}

\begin{figure}
    \centering
    \begin{minipage}[b]{0.45\linewidth}
        \centering
        \makebox[\textwidth][c]{
        \begin{tikzpicture}[
            every node/.style={font=\footnotesize},
            box/.style={rectangle, draw, thick},
        ]
        % Nodi
        \node[box] (n_plus_x) at (0, 0) {$N+x$};
        \node[fill=SUcol,circle,draw,thick] (n) at (2, 0) {$N$};
        \node[box] (n_minus_x_plus_y) at (4, 1) {$N-x+y$};
        \node[box] (y1) at (4, -1) {$y$};
        % Fondamentali
        \draw[->, thick, >=stealth] (n) -- node[above, xshift=2.5pt] {$Q$} (n_plus_x);
        \draw[->, thick, >=stealth] (n_minus_x_plus_y.west) -- node[above,xshift=-4pt] {$\tilde{Q}$} (n);
        % Fermi
        \draw[->, dashed, >=stealth] (y1.west) -- node[below, xshift=-2pt] {$\Psi$} (n);
        \end{tikzpicture}
        }
    \end{minipage}
    \begin{minipage}[b]{0.45\linewidth}
        \centering
        \makebox[\textwidth][c]{
        \begin{tikzpicture}[
            every node/.style={font=\footnotesize},
            box/.style={rectangle, draw, thick},
        ]
        % Nodi
        \node[box] (n_plus_x2) at (0, 0) {$N+x$};
        \node[fill=SUcol,circle,draw,thick] (x) at (2, 0) {$x$};
        \node[box] (n_minus_x_plus_y2) at (4, 1) {$N-x+y$};
        \node[box] (y2) at (4, -1) {$y$};
        % Fondamentali
        \draw[->, thick, >=stealth] (n_plus_x2) -- node[above] {$\tilde{q}$} (x);
        \draw[->, thick, >=stealth] (x) -- node[below,xshift=-2pt] {$q$} (y2.west);
        % Mesoni
        \draw[->, thick, >=stealth, bend left=-30] (n_minus_x_plus_y2.north west) to node[midway, above] {$\varphi$} (n_plus_x2.north east);   
        % Fermi
        \draw[->, dashed, >=stealth] (x) -- node[above,xshift=-4pt] {$\psi$} (n_minus_x_plus_y2.west);
        \draw[->, dashed, >=stealth, bend right=-30] (y2.south west) to node[midway, above] {$\lambda$} (n_plus_x2.south east);     
        \end{tikzpicture}
        }
        \end{minipage}
    \caption{Quiver diagrams of the electric theory (left) and the magnetic theory (right).}
    \label{quivSUnGen}
\end{figure}
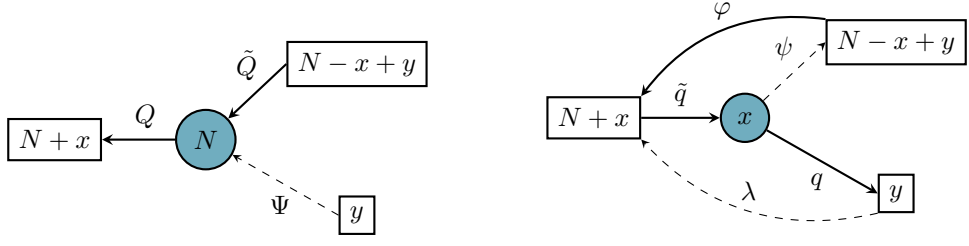

Here we present the duality introduced in \cite{Gadde:2015wta}.
We first start from a 4d $\SU(N)$ gauge theory with $N+x$ flavors. 
The Seiberg dual of this theory is a 4d $\SU(x)$ gauge theory with $N+x$ flavors, a chiral meson and a superpotential given by $W=\tilde{q}\varphi q$.
An anomaly-free $R$ symmetry for the electric theory requires that
\begin{equation}
    \label{Ranfree}
    \sum_{i=1}^{N+x} \frac{1}{2} \bigl(R(Q_i)-1\bigr) +
    \sum_{i=1}^{N+x} \frac{1}{2} \bigl(R(\tilde Q_i)-1\bigr) +
    N = 0\,.
\end{equation}
Any $R$ charge assignation must respect this constraint.
Using the 4d duality dictionary, in the magnetic theory we have
\begin{equation}
    \label{Rdual}
    \begin{aligned}
        &R(q_i) = \frac{1}{x} \sum_{j=1}^{N+x} R(\tilde Q_j) - R(\tilde Q_i)\,, \qquad 
        R(\tilde q_i) = \frac{1}{x} \sum_{j=1}^{N+x} R( Q_j) - R( Q_i)\,, \\
        &R(\varphi_{ij}) = R(Q_i) + R(\tilde Q_j)\,.
    \end{aligned}
\end{equation} 
To perform the reduction to 2d we assign $R=0$ to $N+x$ fundamental chirals and to $N-x+y$ antifundamental chirals, $R=1$ to some other $2(x-y)$ antifundamental chirals and $R=2$ to the last $y$ antifundamental chirals.
One can easily check that such assignation respects the constraint \eqref{Ranfree}.
Therefore we obtain a $2$d theory with $N+x$ fundamental chiral multiplets $Q$, $N-x+y$ antifundamental chiral multiplets $\tilde Q$ and $y$ antifundamental Fermi multiplets $\Psi$, as shown on the left side of the Figure \ref{quivSUnGen}.
There is a restriction on the allowed values of $x$ and $y$, indeed
Seiberg duality requires $x \ge 2$ and from the reduction rules we see that $x > y$\;\footnote{The case $x=y$ is similar to the one studied in subsection \ref{USpallaSacchi}, where a 4d anomalous axial symmetry is not generated in 2d. We will not need such duality in the present discussion and for this reason we omit to review this case. }.

On the other hand, from \eqref{Rdual} the field content of the 4d magnetic theory reduces as follows: the 4d antifundamental chirals become  $N+x$ antifundamental chirals $\tilde q$ in 2d, the 4d fundamental chirals split into $y$  fundamental chirals $q$ and $N-x+y$  fundamental Fermi fields\footnote{Or equivalently antifundamentals, trading the relative $J$ terms with opportune $E$-terms.}  $\psi$ in 2d.  The 4d chiral meson splits into a 2d chiral meson $\varphi$ and a 2d Fermi meson $\lambda$. 
The magnetic theory is represented on the right side of Figure \ref{quivSUnGen}.
The original 4d superpotential splits into two $J$-terms
\begin{equation}
    \label{2dsup}
J_{\lambda} =  \tilde q  q,\qquad  J_{\psi} =\tilde q \varphi \,.
\end{equation}

%%%%%%%%%%%%%%%%%%%%%%%%%%%%%%%%%%%%%%%%%%%%%%%%%%%%%%%%%%%%%%%%%%%%%%%
% FIGURE: SU(N) duality
%%%%%%%%%%%%%%%%%%%%%%%%%%%%%%%%%%%%%%%%%%%%%%%%%%%%%%%%%%%%%%%%%%%%%%%

In order to check this new $2$d duality we verify that the 't Hooft anomalies coincide.
The global symmetries of both electric and magnetic theories are listed below.

\begin{equation}
    \label{TableField}
    \begin{array}{c|c|c|c|c|c|c|c|}
     &\SU(N+x)&\SU(N-x+y)&\SU(y)&\U(1)_{Q}&\U(1)_{\tilde Q}&\U(1)_{\Psi}&\U(1)_{R}\\
     \hline
     Q&\overline{\square}&\cdot&\cdot&1&0&0&0\\
     \tilde Q&\cdot&\square&\cdot&0&1&0&0\\
     \Psi&\cdot&\cdot&\square&0&0&-1&1\\
     \hline
     q&\cdot&\cdot&\overline{\square}&-1-\frac{N}{x}&0&1&0\\
     \tilde q&\square&\cdot&\cdot&\frac{N}{x}&0&0&0\\
     \psi&\cdot&\overline{\square}&\cdot&-1-\frac{N}{x}&-1&0&1\\
     \varphi&\overline{\square}&\square&\cdot&1&1&0&0\\
     \lambda&\overline{\square}&\cdot&\square&1&0&-1&1\\
    \end{array}
\end{equation}\vspace{5pt}

Notice that we have an explicit duality dictionary to write $\tilde{q}$, $\varphi$ and $\lambda$ in terms of the electric variables, but not for $q$ and $\psi$. 
However, we can fix their $\U(1)$ charges from the superpotential \eqref{2dsup}.
From the table above one can read the 't Hooft anomalies and find out that they match:
\begin{align}
    &\kappa_{\SU(n)^2}=\tfrac{N}{2}\,,\quad 
    &&\kappa_{QQ}=-\kappa_{QR}=N(N+x)\,, \nonumber \\
    &\kappa_{RR}=N^2+1\,, \quad
    &&\kappa_{\tilde{Q}\tilde{Q}}=-\kappa_{\tilde Q R}=N(N-x+y)\,,\\
    &\kappa_{Q\tilde Q}=\kappa_{Q\Psi}=\kappa_{\tilde Q \Psi}=0\,,\quad
    &&\kappa_{\Psi \Psi}=-\kappa_{\Psi R}=-Ny\,.\nonumber 
\end{align}

At the level of the elliptic genus the identity that follows from the reduction of the index on $S^2 \times T^2$
is
\begin{equation}
\label{EGdualitySUgen}
I_{\SU(N)}^{(N+x;N-x+y;y;\cdot;\cdot)}
(\vec u;\vec v;  \vec h;\cdot;\cdot)
=
 \frac{
 \prod_{i=1}^{N+x} \prod_{a=1}^{y} \theta(q u_i  h_a)
 }
{
 \prod_{i=1}^{N+x} \prod_{j=1}^{N-x+y} \theta( u_i  v_j)
}
I_{\SU(x)}^{(y;N+x;N-x+y;\cdot;\cdot)}(
\vec {\tilde h}, \vec {\tilde u},   \vec {\tilde v};\cdot;\cdot )
\end{equation}
with $\tilde u_i = u_i^{-1}\prod_{\ell=1}^{N+x}  u_\ell^{\frac{1}{x}}$,
$\tilde v_j =  v_j^{-1} \prod_{\ell=1}^{N+x}  u_\ell^{-\frac{1}{x}}$ and
$\tilde h_a =  h_a^{-1} \prod_{\ell=1}^{N+x}  u_\ell^{-\frac{1}{x}}$.
Again, we refer the reader to Appendix \ref{convEG} for the conventions adopted here on the elliptic genus.

\section{$\SU(N)$ with two antisymmetric chirals}

In this section we discuss the existence of LG duals for 2d $\mathcal{N}=(0,2)$ $\SU(N)$ gauge theories with two antisymmetric tensors $A_{1,2}$, $n_f$ fundamentals and $n_a$ antifundamentals, with $n_f+n_a=4$. 
The cases with even $N=2n$ rank and with odd $N=2n+1$ rank are different for any allowed choice of $n_f$ and $n_a$, therefore we need to treat each case separately. 
In each one we propose the duality and check the 't Hooft anomaly matching. Furthermore, we provide a derivation of the matching of the elliptic genera by employing other identities obtained by the reduction of 4d Seiberg and Intriligator-Pouliot duality.

\subsection{$\SU(2n)$ with four fundamentals}
\label{subsec:3.1}

In this case the dual LG model has three gauge invariant operators corresponding to chiral fields interacting through a $J$-term with a Fermi multiplet.
The field content of the gauge theory and of the dual LG model is represented in the table below.
\begin{equation} 
\label{Table3.1}
\begin{array}{c|c|c|c|c|c|c|}
& \SU(2n) & \SU(2) & \SU(4) & \U(1)_A & \U(1)_Q & \U(1)_{R_0} \\
\hline
A & 
\begin{array}{c}
\square \vspace{-2.85mm} \\
\square 
\end{array} 
 & \square & \cdot & 1 & 0 & 0 \\
Q & \square & \cdot & \square & 0 & 1 & 0 \\       
\hline
T_{n} & \cdot & \otimes_{\mathrm{sym}}^{n} \square  & \cdot & n & 0 & 0 \\
T_{n-1} & \cdot &\otimes_{\mathrm{sym}}^{n-1} \square &\begin{array}{c}
\square \vspace{-2.85mm} \\
\square 
\end{array} & n-1 & 2 & 0 \\
T_{n-2} & \cdot &\otimes_{\mathrm{sym}}^{n-2} \square  & \cdot & n-2 & 4 & 0 \\
\Psi & \cdot &\otimes_{\mathrm{sym}}^{2n-2} \square& \cdot & -2n+2 & -4 & 1 \\
\end{array}
\end{equation}

The three chirals $T_j$ correspond to the following gauge invariant combinations of the charged fields
\begin{equation}
\label{chirals}
T_n = A^n,\quad
T_{n-1} = A^{n-1} Q^2,\quad
T_{n-2} = A^{n-2} Q^4\,,
\end{equation}
while the $J$-term is 
\begin{equation}
J_{\Psi} = T_n T_{n-2} + T_{n-1}^2\,,
\end{equation}

We have computed the 't Hooft anomalies in the two phases and we have observed that they match. Explicitly, we have
\begin{align}
    &\kappa_{\SU(2)^2} = \tfrac{n(2n-1)}{2}\,, \quad
    &&\kappa_{AA} = -\kappa_{AR_0} = 2n(2n-1)\,, \nonumber \\
    &\kappa_{\SU(4)^2} = n\,,\quad
    &&\kappa_{QQ} = -\kappa_{QR_0} = 8n\,,\\
    &\kappa_{R_0 R_0}  = 6n+1\,,\quad
    &&\kappa_{AQ} = 0\,. \nonumber
\end{align}

The further evidence that we provide to corroborate the existence of this duality consists of showing that the matching of the elliptic genera of the two phases is a consequence of other identities. These can be derived from the matching of the $S^2 \times T^2$ topologically twisted index of well stated 4d s-confining parent dualities for special unitary and symplectic SQCD.

%%%%%%%%%%%%%%%%%%%%%%%%%%%%%%%%%%%%%%%%%%%%%%%%%%%%%%%%%%%%%%%%%%%%%%%
% FIGURE: SU(2n) w 2A 4F deconfinement
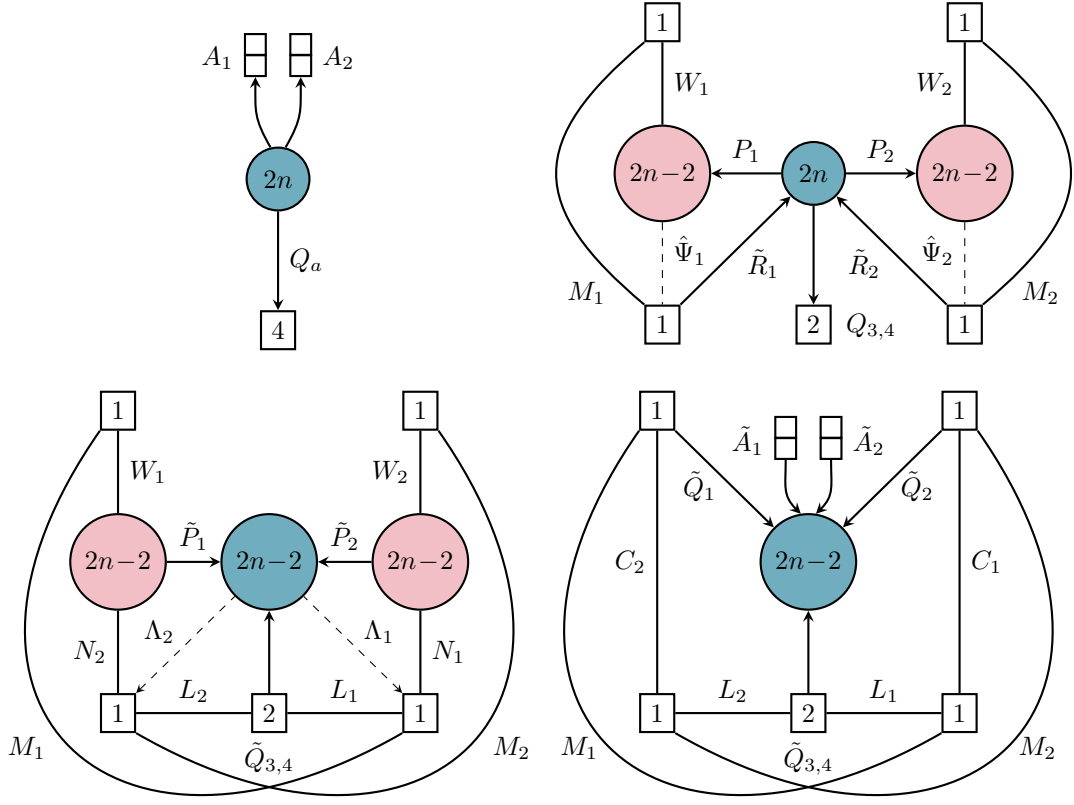
\begin{figure}[h!]
    \centering
        % STEP 1
        \begin{minipage}[b]{0.45\linewidth}
            \centering
            \makebox[\textwidth][c]{
            \hspace{-0.1em}
            \begin{tikzpicture}[
                every node/.style={font=\footnotesize},
                box/.style={rectangle, draw, thick}
            ]
            \pgfmathsetmacro{\x}{0.8}
            \pgfmathsetmacro{\y}{1.6}
            % Nodi
            \node[box] (bot) at (0, -2) {$4$};
            \node[fill=SUcol,circle,draw,thick] (center) at (0, 0) {$2n$};
            % Antisimmetriche
            \node[box, minimum size=0.2cm] (square1r) at (0.3, 1.5) {};
            \node[box, minimum size=0.2cm] (square2r) at (0.3, 1.78) {};
            \node[box, minimum size=0.2cm] (square1l) at (-0.3, 1.5) {};
            \node[box, minimum size=0.2cm] (square2l) at (-0.3, 1.78) {};
            \node at (\x, \y) {$A_2$};
            \node at (-\x, \y) {$A_1$};
            \draw[<-,thick,>=stealth] (square1r.south) to[out=-90, in=60, looseness=1.2] ($(center.north)+(0.1,-0.01)$);
            \draw[<-,thick,>=stealth] (square1l.south) to[out=-90, in=120, looseness=1.2] ($(center.north)+(-0.1,-0.01)$);
            % Fondamentali
            \draw[<-,thick,>=stealth] (bot.north) -- node[right] {$Q_{a}$} (center);
            \end{tikzpicture}
            }
        \end{minipage}
        % STEP 2
        \begin{minipage}[b]{0.45\linewidth}
            \centering
            \makebox[\textwidth][c]{
            \begin{tikzpicture}[
                every node/.style={font=\footnotesize},
                box/.style={rectangle, draw, thick}
            ]
            \pgfmathsetmacro{\a}{2}
            \pgfmathsetmacro{\b}{2}
            % Nodi
            \node[fill=USPcol,circle,draw,thick] (uspl) at (-\a,0) {$2n\!-\!2$};
            \node[fill=USPcol,circle,draw,thick] (uspr) at (\a, 0) {$2n\!-\!2$};
            \node[box] (botl) at (-\a, -\b) {$1$};
            \node[box] (botr) at (\a, -\b) {$1$};
            \node[box] (bot) at (0, -\a) {$2$};
            \node[fill=SUcol,circle,draw,thick] (center) at (0, 0) {$2n$};
            \node[box] (topl) at (-\a, \b) {$1$};
            \node[box] (topr) at (\a, \b) {$1$};
            % Fondamentali
            \draw[<-, thick, >=stealth] (uspl.east) -- node[above] {$P_1$} (center.west);
            \draw[<-, thick, >=stealth] (uspr.west) -- node[above] {$P_2$} (center.east);
            \draw[<-, thick, >=stealth] (bot.north) -- node[right, yshift=-27pt, xshift=8pt] {$Q_{3,4}$} (center);
            \draw[->,thick,>=stealth] (botl.north east) -- node[right,yshift=-5pt]{$\tilde R_1$}(center.south west);
            \draw[->,thick,>=stealth] (botr.north west) -- node[left,yshift=-5pt]{$\tilde R_2$}(center.south east);
            % Mesoni
            \draw[thick] (uspl) -- node[right]{$W_1$} (topl);
            \draw[thick] (uspr) -- node[left]{$W_2$} (topr);
            \draw[thick] (topl.south west) to[out=-140,in=140,looseness=1.5] node[right, yshift=-45pt]{$M_1$}(botl.north west);
            \draw[thick] (topr.south east) to[out=-40,in=40,looseness=1.5] node[left, yshift=-45pt]{$M_2$}(botr.north east);
            % Fermi
            \draw[dashed] (uspl) -- node[right, yshift=5pt] {$\hat \Psi_1$} (botl);
            \draw[dashed] (uspr) -- node[left, yshift=5pt] {$\hat \Psi_2$} (botr);
            \end{tikzpicture}
            }
        \end{minipage}
        \\[0.5cm]
        % STEP 3
        \begin{minipage}[b]{0.45\linewidth}
            \centering
            \makebox[\textwidth][c]{
            \begin{tikzpicture}[
                every node/.style={font=\footnotesize},
                box/.style={rectangle, draw, thick}
            ]
            \pgfmathsetmacro{\a}{2}
            \pgfmathsetmacro{\b}{2}
            % Nodi 
            \node[fill=USPcol,circle,draw,thick] (uspl) at (-\a,0) {$2n\!-\!2$};
            \node[fill=USPcol,circle,draw,thick] (uspr) at (\a, 0) {$2n\!-\!2$};
            \node[box] (botl) at (-\a, -\b) {$1$};
            \node[box] (botr) at (\a, -\b) {$1$};
            \node[box] (bot) at (0, -\b) {$2$};
            \node[fill=SUcol,circle,draw,thick] (center) at (0, 0) {$2n\!-\!2$};
            \node[box] (topl) at (-\a, \b) {$1$};
            \node[box] (topr) at (\a, \b) {$1$};
            % Fondamentali
            \draw[->, thick, >=stealth] (uspl.east) -- node[above] {$\tilde P_1$} (center.west);
            \draw[->, thick, >=stealth] (uspr.west) -- node[above] {$\tilde P_2$} (center.east);
            \draw[->, thick, >=stealth] (bot.north) -- (center);
            \node at (0,-\b-0.6) {$\tilde Q_{3,4}$};
            % Mesoni
            \draw[thick] (topl.south west) to[out=235, in=210, looseness=2] node[left, xshift=-5pt] {$M_1$} (botr.south west);
            \draw[thick] (topr.south east) to[out=-55, in=-30, looseness=2] node[right, xshift=5pt] {$M_2$}(botl.south east);
            \draw[thick] (botl) -- node[above] {$L_2$} (bot);
            \draw[thick] (botr) -- node[above] {$L_1$} (bot);
            \draw[thick] (uspl) -- node[right]{$W_1$} (topl);
            \draw[thick] (uspr) -- node[left]{$W_2$} (topr);
            \draw[thick] (uspl) -- node[left] {$N_2$} (botl);
            \draw[thick] (uspr) -- node[right] {$N_1$} (botr);
            % Fermi
            \draw[<-,dashed,>=stealth] (botl.north east) -- node[left,yshift=5pt]  {$\Lambda_2$}(center.south west);
            \draw[<-,dashed,>=stealth] (botr.north west) -- node[right,yshift=5pt]{$\Lambda_1$}(center.south east);
            \end{tikzpicture}
            }
        \end{minipage}
        % STEP 4
        \begin{minipage}[b]{0.45\linewidth}
            \centering
            \makebox[\textwidth][c]{
            \begin{tikzpicture}[
                every node/.style={font=\footnotesize},
                box/.style={rectangle, draw, thick},
            ]
            \pgfmathsetmacro{\a}{2}
            \pgfmathsetmacro{\x}{0.8}
            \pgfmathsetmacro{\y}{1.6}
            % Nodi
            \node[box] (topl) at (-\a,\a) {$1$};
            \node[box] (topr) at (\a, \a) {$1$};
            \node[box] (botl) at (-\a, -\a) {$1$};
            \node[box] (botr) at (\a, -\a) {$1$};
            \node[box] (bot) at (0, -\a) {$2$};
            \node[fill=SUcol,circle,draw,thick] (center) at (0, 0) {$2n\!-\!2$};
            % Antisimmetriche
            \node[box, minimum size=0.2cm] (square1r) at (0.3, 1.5) {};
            \node[box, minimum size=0.2cm] (square2r) at (0.3, 1.78) {};
            \node[box, minimum size=0.2cm] (square1l) at (-0.3, 1.5) {};
            \node[box, minimum size=0.2cm] (square2l) at (-0.3, 1.78) {};
            \node at (\x, \y) {$\tilde A_2$};
            \node at (-\x, \y) {$\tilde A_1$};
            \draw[->, thick, >=stealth] (square1r.south) to[out=-90, in=45, looseness=1.2] ($(center.north)+(0.1,-0.01)$);
            \draw[->, thick, >=stealth] (square1l.south) to[out=-90, in=135, looseness=1.2] ($(center.north)+(-0.1,-0.01)$);
            % Fondamentali
            \draw[->, thick, >=stealth] (topl.south east) -- node[pos=0.3, below, xshift=-2pt] {$\tilde Q_1$} (center.north west);
            \draw[->, thick, >=stealth] (topr.south west) -- node[pos=0.3, below, xshift=2pt] {$\tilde Q_2$} (center.north east);
            \draw[->, thick, >=stealth] (bot.north) -- (center);
            \node at (0,-\a-0.6) {$\tilde Q_{3,4}$};
            % Mesoni
            \draw[thick] (botl) -- node[above] {$L_2$} (bot);
            \draw[thick] (botr) -- node[above] {$L_1$} (bot);
            \draw[thick] (topl) -- node[left] {$C_2$} (botl);
            \draw[thick] (topr) -- node[right] {$C_1$} (botr);
            \draw[thick] (topl.south west) to[out=235, in=210, looseness=2] node[left, overlay] {$M_1$} (botr.south west);
            \draw[thick] (topr.south east) to[out=-55, in=-30, looseness=2] node[right, overlay] {$M_2$}(botl.south east);
            \end{tikzpicture}
        }
        \end{minipage}
    \vspace{-30pt}
    \caption{Deconfinement steps of $\SU(2n)$ with two antisymmetrics and four fundamentals.}
    \label{fig:SUpari}
\end{figure}
%%%%%%%%%%%%%%%%%%%%%%%%%%%%%%%%%%%%%%%%%%%%%%%%%%%%%%%%%%%%%%%%%%%%%%%

The identity that we want to prove is
\begin{eqnarray}
\label{toPRovesu2n4fond}
&&I^{(4,\cdot,\cdot,2,\cdot)}_{\SU(2n)}(\vec u;\cdot;\cdot;\vec t;\cdot) =\\ 
&&=\frac{
        \prod_{j=0}^{2n-2}
        \theta\left(
            q/\left(t_{1}^{2(2n-2-j)} t_2^{2j}\prod_{a=1}^{4} u_a\right)
        \right)
    }{
        \prod_{j=0}^{n} \theta\big(t_1^{2(n-j)}t_2^{2j}\big)
        \prod_{a<b}^{4}\prod_{j=0}^{n-1} \theta\big(t_1^{2(n-1-j)}t_2^{2j}u_a u_b\big)
        \prod_{j=0}^{n-2} \theta\big(t_1^{2(n-2-j)}t_2^{2j}\prod_{a=1}^{4}u_a\big)
    }\,,
    \nonumber
\end{eqnarray}
where we refer to the appendix \ref{convEG} for the details on the conventions on the elliptic genus used here.
The collection of $\theta$-functions on the RHS of the identity (\ref{toPRovesu2n4fond})
represents the Fermi $\Psi$ (numerator) and the chirals $T_j$ (denominator) in  the dual LG theory.

In the following, in order to prove the identity, we will break some of the global symmetries by moving all the theta functions from the denominator of the RHS to the numerator of LHS of  (\ref{toPRovesu2n4fond}). We will further employ the relation 
\begin{equation}
\label{inversion}
\theta(x) =\theta(q/x)\,.
\end{equation}
At the field theory level this operation corresponds to adding a $J$-term on the gauge theory side by setting a gauge invariant combination of chiral fields (the one that we have moved from the RHS to the LHS) to zero using the equations of motion.
The net effect of this operation on the elliptic genus consists of adding a Fermi on the gauge theory side and of removing a chiral on the LG side. 
We denote the new Fermi that appears on the LHS as Fermi \emph{flipper}, borrowing the  terminology used in higher dimensional theories with four supercharges for chiral multiplets.

In this case we also redefine the singlets $T_n$, $T_{n-1}$ and $T_{n-2}$ as
$\mathbf{T}_{n-j,j}$, for $j=0,\dots, n-1$,  $\mathbf{T}_{n-j-1,j}^{a b}$ for $j=0,\dots,n$ and $1\leq a<b\leq 4 $ and $\mathbf{T}_{n-j-2,j}$  for $j=0,\dots,n-2$ respectively.
This redefinition is quite useful in the following, because we will explicit break the $\SU(2)$ flavor symmetry and expressing the fields as $\mathbf{T}$ allows us to identify the $\U(1) \subset \SU(2)$ charges.
Furthermore, the Fermi $\Psi$ is redefined as $\mathbf{\Psi}_{2n-2-j,j}$ with $j=0,\dots,2n-2$ accordingly.
The electric theory is then modified by flippling all the combinations $\mathbf{T}$, which corresponds to adding the following $J$-terms
\begin{align}
\label{Jflippers}
&\Psi_{\mathbf{T}_{n-j,j}} = A_1^{n-j} A_2^j\,, 
&&j=0,\dots,n\,, \nonumber \\
&\Psi_{\mathbf{T}_{n-j-1,j}^{ab}} = A_1^{n-j-1} A_2^j Q_a Q_b\,, 
&&j=0,\dots,n-1,\, 1\leq a<b\leq 4\,,  \\
&\Psi_{\mathbf{T}_{n-j-2,j}} = A_1^{n-j-2} A_2^j Q_1 Q_2 Q_3 Q_4\,, 
&&j=0,\dots,n-2\,. \nonumber 
\end{align}
In the dual theory, we are then left with only the $2n-1$ free Fermi $\mathbf{\Psi}_{2n-2-j,j}$.

Then the crucial aspect of the analysis consists of finding an auxiliary dual quiver gauge theory with new gauge groups and new bifundamentals, where the original 
antisymmetric fields $A_{1,2}$ are absent.
In the case at hand, such an auxiliary quiver corresponds to the second one in Figure \ref{fig:SUpari}.

There are two symplectic gauge groups $\USp(2n-2)_{1,2}$ and each one is characterized by one $\USp/\SU$ bifundamental $P_{1,2}$, one fundamental $W_{1,2}$ and one Fermi $\hat \Psi_{1,2}$. There are also two new $\SU(2n)$ fundamentals $\hat R_{1,2}$ and two $J$-terms 
\begin{equation}
J_{\hat \Psi_{1,2}} = \tilde R_{1,2} P_{1,2}+W_{1,2} M_{1,2}\,.
\end{equation}
Observe that the procedure manifestly breaks the $\SU(4) \times \SU(2)$ global symmetry.
The original fields $A_{1,2}$ and $Q_{1,2}$ correspond here to the combinations $P_{1,2}^2$ and $P_{1,2} W_{1,2}$ respectively.
In addition, the $J$-terms (\ref{Jflippers}) are modified as
\begin{align}
\label{Jflippers2}
&\Psi_{\mathbf{T}_{n-j,j}} = P_1^{2n-2j} P_2^{2j}\,,
&&j=0,\dots,n\,, \nonumber \\
&\Psi_{\mathbf{T}_{n-j-1,j}^{1,2}} = P_1^{2(n-j-1)} P_2^{2j} P_1 W_1 P_2 W_2\,, 
&&j=0,\dots,n-1\,, \\
&\Psi_{\mathbf{T}_{n-j-1,j}^{\ell,a}} = P_1^{2(n-j-1)} P_2^{2j} P_\ell W_\ell Q_a\,,
&&j=0,\dots,n-1,\,\, \ell=1,2, \,\, a=3,4\,, \nonumber \\
&\Psi_{\mathbf{T}_{n-j-1,j}^{3,4}} = P_1^{2(n-j-1)} P_2^{2j} Q_3 Q_4\,,
&&j=0,\dots,n-1\,, \nonumber\\
&\Psi_{\mathbf{T}_{n-j-2,j}} = P_1^{2(n-j-2)} P_2^{2j} P_1 W_1 P_2 W_2 Q_3 Q_4\,,
&&j=0,\dots,n-2\,. \nonumber 
\end{align}

The original quiver in Figure \ref{fig:SUpari} can be recovered if we use the duality reviewed in subsection \ref{USpallaSacchi}, provided we remove an extra non-anomalous axial symmetry that is allowed by the field content.

As we stressed above, we do not have a dynamical mechanism that removes such a symmetry. A similar problem arises in 3d, where the extra symmetry is removed by the presence of a (linear) monopole superpotential. The procedure carried out here was indeed applied in \cite{Amariti:2024gco} to the analysis of a 3d duality with the same electric matter content\footnote{More precisely the duality that was proved in \cite{Amariti:2024gco} through tensor deconfinement had 4 fundamental and two antisymmetric chirals 3d $\mathcal{N}=2$.} found in \cite{Nii:2019ebv}, where in the auxiliary quiver two linear monopole superpotentials terms were added to the theory in order to remove the axial symmetries. This analogy is helpful in order to justify the existence of a dynamical constraint also in 2d, because the models discussed here can be obtained from the 3d ones as boundary dualities along the lines of the construction in \cite{Dimofte:2017tpi}.
If we apply such construction to the auxiliary quiver we then have to consider the contribution of the boundary monopoles, that forbids the generation of the axial symmetry in the half index.
Here a similar constraint exists also at the level of the elliptic genus. This last observation descends from the reduction of the $S^2 \times T^2$ partition function 
to the elliptic genus, where the choice of $R$ charges discussed in subsection \ref{USpallaSacchi}, prevents the extra symmetry to be generated.

Having this caveat in mind we can proceed by dualizing the $\SU(2n)$ gauge node.
This model has $4n-2$ fundamentals and $2$ antifundamentals and the necessary duality has been reviewed in subsection \ref{SUnGen}. Here we consider the case with $y=0$.

The model obtained in this case is represented in the third quiver of figure \ref{fig:SUpari}. The non-trivial aspect of this model is that the constraints that we imposed on the axial symmetries for the $\USp(2n-2)$ nodes above are not required here anymore. The reason is that there are linear $J$-term in the dual phase that remove some of the Fermi and chiral fields, and non-anomalous charges in the leftover field content are exactly the ones visible in the matching of the elliptic genera.

In this case the $J$-terms for the Fermi $\Lambda_{1,2}$ obtained in the duality are
\begin{eqnarray}
J_{\Lambda_{1}}&=& \sum_{a=3,4} \tilde Q_{a} L_{1,a} + \tilde P_1 M_1 W_1 + N_1 \tilde P_2\,, \nonumber \\
J_{\Lambda_{2}}&=& \sum_{a=3,4} \tilde Q_{a} L_{2,a} + \tilde P_2 M_2 W_2 + N_2 \tilde P_1\,.
\end{eqnarray}

The last step in Figure \ref{fig:SUpari} corresponds to confining the two $\USp(2n-2)$
gauge nodes. In this case, they both have $2n$ fundamentals and the dual models are LG theories, where the two combinations $\tilde P_{1,2}^2$ are conjugated antisymmetric chirals. There are also four $\SU(2n-2)$ antifundamentals, and the $J$-terms obtained after solving the equation of motion are
\begin{eqnarray}
J_{ \Psi^{(0)}_{1}}&=& \tilde Q_{1} \tilde A_1^{n-2} \Big( \sum_{a=3,4} \tilde Q_{a} L_{2,a} +M_2 \tilde Q_2 \Big)+C_1 \text{Pf} \tilde A_1\,, \\
J_{ \Psi^{(0)}_{2}}&=& \tilde Q_{2} \tilde A_2^{n-2} \Big( \sum_{a=3,4} \tilde Q_{a} L_{1,a} +M_1 \tilde Q_1 \Big)+C_2 \text{Pf} \tilde A_2\,.
\end{eqnarray}
Notice that we have denoted the Fermi flippers, arising from the gauge/LG duality of the symplectic gauge nodes discussed in subsection \ref{sec:USP2Np2}, as $ \Psi_1^{(0)}$ and $ \Psi_2^{(0)}$, as they will constitute the \emph{extremal} fields in the  Fermi \emph{tower} denoted as $\Psi$ in equation (\ref{Table3.1}). 

In this way we have obtained a duality between the original $\SU(2n)$ model with two antisymmetric and four fundamental chirals, and an $\SU(2n-2)$ model with two conjugate antisymmetric and four antifundamental chirals, in addition to a set of Fermi and chiral singlets.

We now iterate the procedure until we reach an $\SU(2n-4)$ gauge theory with the same charged field content of the original one, \textit{i.e.} two antisymmetric and four fundamental chirals.
In this way we have a simpler relation between the elliptic genera,
that can be recursively applied to obtain the final relation between the original model and an LG model. 
Such relation is 
\begin{eqnarray}
\label{recursive1}
I_{\SU(2n)}^{(4,\cdot,\cdot,2,\cdot)}(\vec u;\cdot;\cdot;\vec t;\cdot)
&&=
I_{\SU(2n-4)}^{(4,\cdot,\cdot,2,\cdot)}\left((t_1 t_2)^\frac{1}{n-2} \vec u;\cdot;\cdot;(t_1 t_2)^\frac{1}{n-2}\vec t ; \cdot\right) \nonumber
\\
&&
\times
\frac{\theta\left(q/(t_{2,1}^{4n-4} \prod_{a=1}^{4}u_a) \right)\,\theta\left(q/(t_{1,2}^2t_{2,1}^{4n-6} \prod_{a=1}^{4}u_a)\right)}{
\theta\left(t_{1,2}^{2n}\right)\,\theta\left(t_{1,2}^{2n-4}\prod_{a=1}^{4}u_a\right) \, \prod_{a<b} \theta\left(t_{1,2}^{2n-2} u_a u_b\right)
}\,.
\end{eqnarray}
The relation is derived by applying the various steps of deconfinement, duality and reconfinement discussed above and summarized graphically in Figure \ref{fig:SUpari} twice. The relations used to derive (\ref{recursive1}) have been summarized in Section \ref{basic} and they correspond to (\ref{idsacchi}), (\ref{EGdualitySUgen}) and (\ref{ellipticdualUSP}) respectively.
Observe that we kept the chiral fields unflipped  in the relation (\ref{recursive1}), while they appeared as Fermi fields in the field theory discussion in the $\SU(2n)$ side.

% {\color{red} Dovremmo scrivere la field theory a questo step in modo esplicito}
 
Each term in the second line of the (\ref{recursive1}) corresponds to one of the expected singlets of the duality, in which the procedure has apparently broken the global $\SU(2)$ non-abelian global symmetry\footnote{Also the $\SU(4)$ global symmetry is broken by our procedure, but it is actually recovered in the $\SU(2n-4)$ theory. }. 
More concretely:
\begin{itemize}
\item $\theta(t_{1,2}^{2n})$ corresponds to the combinations $A_{1,2}^{n}$
\item $\theta(t_{1,2}^{2n-4}\prod_{a=1}^{4}u_a)$ corresponds to $A_{1,2}^{n-2} Q_1 Q_2 Q_3 Q_4$
\item $\prod_{a<b} \theta(t_{1,2}^{2n-2} u_a u_b) $ corresponds to $A_{1,2}^{n-1} Q_a Q_b$, antisymmetric in the $\SU(4)$ indices 
\item $\theta\left(q/(t_{2,1}^{4n-4} \prod_{a=1}^{4}u_a)\right)$ correspond to the two Fermi $\Psi_{2n-2,0}^{(0)}$ and $\Psi_{0,2n-2}^{(0)}$
\item $\theta\left(q/(t_{1,2}^2t_{2,1}^{4n-6}\prod_{a=1}^{4}u_a)\right)$ correspond to the two Fermi $\Psi_{2n-1,1}^{(0)}$ and $\Psi_{1,2n-1}^{(0)}$.
\end{itemize}
The $(0)$ superscript in the Fermi fields indicates that they have been created at the \emph{zero}-th iteration of the procedure.

A recursive application of this relation leads to an $\SU(4)$ gauge theory  model for even $n$ or to an $\SU(2)$ gauge theory for odd $n$.
In general, at the $j$-th step we obtain an $\SU(2n-4j)$ gauge theory, and the leftover integral is 
\begin{equation}
 I_{\SU(2n-4j)}^{(4,\cdot,\cdot,2,\cdot)}\left((t_1 t_2)^\frac{j}{n-2j}\vec u;\cdot;\cdot;(t_1 t_2)^\frac{j}{n-2j}\vec t;\cdot \right)\,.
\end{equation}
The Fermi and chiral singlets generated at this step contribute as
\begin{itemize}
\item $\theta(t_{1,2}^{2(n-j+1)} t_{2,1}^{2(j-1)})$ corresponding  to the combinations $A_{1,2}^{n-j+1} A_{2,1}^{j-1}$
\item $\prod_{a<b} \theta(t_{1,2}^{2(n-j)}t_{2,1}^{2(j-1)} u_a u_b)$ corresponding to $A_{1,2}^{n-j} A_{2,1}^{j-1} Q_a Q_b$, antisymmetric in the $\SU(4)$ indices
\item $\theta(t_{1,2}^{2(n-j-1)} t_{2,1}^{2(j-1)}\prod_{a=1}^{4}u_a)$ corresponding to $A_{1,2}^{n-j-1} A_{2,1}^{j-1} Q_1 Q_2 Q_3 Q_4$
\item $\theta(q/(t_{1,2}^{4(n-j)} t_{2,1}^{4(j-1)} \!\prod_{a=1}^{4}\!u_a))$ associated to the two Fermi $\Psi_{2n-2j,2j-2}^{(j)}$ and $\Psi_{2j-2,2n-2j}^{(j)}$
\item $\theta(q/(t_{1,2}^{2(2j-1)} t_{2,1}^{2(2n-2j-1)} \prod_{a=1}^{4}u_a))$ corresponding to the two Fermi $\Psi_{2n-2j-1,2j-1}^{(j)}$ and $\Psi_{2j-1,2n-2j-1}^{(j)}$.
\end{itemize}

The Fermi generated at this stage constitute another bit of the tower of states $\Psi$.

In order to complete the proof of (\ref{toPRovesu2n4fond})  we must  distinguish the cases $n=2m$ and $n=2m+1$ as discussed above.

\subsubsection*{The case of $\SU(4m)$}
In this case we can use the recurrence relations until we get to an $\SU(4)$ gauge theory with two antisymmetric $\mathbf{A}_{1,2}$ and four fundamentals $\mathbf{Q}_{1,2,3,4}$. Their fugacities are $(t_1 t_2)^{\frac{m-1}{2}}t_{1,2}$ and $(t_1 t_2)^\frac{m-1}{2}u_a$ respectively.

This model is dual (see \cite{Amariti:2024usp}) to an LG theory with chirals $\varphi_i$ 
that contribute to the elliptic genus as
\begin{itemize}
\item $\varphi_1=\mathbf{A}_{2}\rightarrow  \prod_{a<b} \theta\left(t_1^{2m-2} t_2^{2m}u_a u_b\right)$
\item $\varphi_2=\text{Pf} \mathbf{A}_{1} \rightarrow  \theta\left(t_1^{2m+2} t_2^{2m-2}\right)$
\item $\varphi_3= \mathbf{A}_{1} \mathbf{Q}^2\rightarrow  \prod_{a<b} \theta\left(t_1^{2m} t_2^{2m-2}u_a u_b\right)$
\item $\varphi_4=\mathbf{Q}^4\rightarrow  \theta\left((t_1 t_2)^{2(m-1)}\prod_{a=1}^{4}u_a\right)$
\item $\varphi_5= \text{Pf} \mathbf{A}_{2}\rightarrow   \theta\left(t_1^{2m-2} t_2^{2m+2}\right)$
\item $\varphi_6= \mathbf{A}_{1}\mathbf{A}_{2}\rightarrow \theta\left((t_1 t_2)^{2m}\right)$.
\end{itemize}
There are in addition three Fermi singlets, whose charges (and contribution to the elliptic genus) can be read from the $J$-terms\footnote{Here we have flipped all the chiral combinations $T_{n,n-1,n-2}$. Then, the LG model will only have free Fermi multiplets and vanishing $J$-terms. Nevertheless we can read their charges from the un-flipped duality discussed in \cite{Amariti:2024usp}.}. We have
\begin{itemize}
\item $J_{\Psi_{11}} \supset \varphi_1^2 \rightarrow 
\theta(q/(t_1^{4m} t_2^{4(m-1)}  \prod_{a=1}^{4}u_a))$
\item $J_{\Psi_{33}} \supset \varphi_3^2\rightarrow 
\theta(q/(t_1^{4(m-1)} t_2^{4m}  \prod_{a=1}^{4}u_a))$
\item $J_{\Psi_{13}} \supset \varphi_1  \varphi_3\rightarrow 
\theta(q/(t_1^{2(2m-1)} t_2^{2(2m-1)}  \prod_{a=1}^{4}u_a))$\,.
\end{itemize}
These last Fermi singlets are the remaining ones needed for the full Fermi tower $\Psi$.

Once the contributions arising from the last duality are added to the ones obtained from the recursive relation above, we restore the original $\SU(2)$ symmetry and obtain the 
relation (\ref{toPRovesu2n4fond}) for $n=2m$.

\subsubsection*{The case of $\SU(4m+2)$}

In order to complete the proof of (\ref{toPRovesu2n4fond}) we need to discuss the case $\SU(4m+2)$ as well. In this case we arrive at an $\SU(2)$ gauge theory with four fundamentals $\mathbf{Q}_{1,2,3,4}$ with fugacities $(t_1 t_2)^{2m} u_a$ after $m$ iterations.
The two antisymmetric are singlets and they contribute to the elliptic genus as
$\theta\big(t_{1,2}^{2(m+1)}t_{2,1}^{2(m-1)}\big)$.
This gauge theory is dual to an LG model, where the singlets are given by the $\SU(4)$ two-index antisymmetric chiral $\mathbf{Q}_a \mathbf{Q}_b$ and a Fermi.
The contribution of the chiral to the elliptic genus is $\prod_{a<b} \theta\big((t_1 t_2)^{2m}u_a u_b \big)$, while the Fermi contributes as $\theta\big(q/((t_1 t_2)^{4m} \prod_{a=1}^{4}u_a)\big)$, compatibly with the $J$-terms that would be generated in the duality in absence of flippers in the original $\SU(2n)$ model.
Again, once the contributions of the last duality are added to the ones obtained from the recursive relation above, we restore the original $\SU(2)$ symmetry and obtain the 
relation (\ref{toPRovesu2n4fond}) for $n=2m+1$ as well.

This concludes the proof of the identity (\ref{toPRovesu2n4fond}).
Observe that at the level of the interaction we end up with a model with only Fermi multiplets, as we have flipped all the chirals $T_j$ in the original gauge theory. 
In the following, we will actually use the un-flipped (or partially flipped) duality in order to prove that the other cases of $\SU(N)$ with two antisymmetric chirals, $n_f < 4$  fundamental chirals  and $n_a > 1$ antifundamental chirals are dual to LG models.

\subsection{$\SU(2n+1)$ with four fundamentals}
\label{subsec:3.2}

In this case the dual LG model has two chiral gauge invariant operators corresponding to chiral fields interacting through a $J$-term with a Fermi multiplet.
The field content of the gauge theory and of the dual LG model is represented in the following table
\begin{equation} 
\label{Table3.2}
    \begin{array}{c|c|c|c|c|c|c|}
    & \SU(2n+1) & \SU(2) & \SU(4) & \U(1)_A & \U(1)_Q & \U(1)_{R_0} \\
    \hline
    A & \begin{array}{c}
\square \vspace{-2.85mm} \\
\square  %\otimes_{\mathrm{sym}}^{n-1} \square
\end{array}  & \square & \cdot & 1 & 0 & 0 \\
    Q & \square & \cdot & \square & 0 & 1 & 0 \\       
    \hline
    T_{n} & \cdot & \otimes_{\mathrm{sym}}^{n} \square& \square & n & 1 & 0 \\
    T_{n-1} & \cdot & \otimes_{\mathrm{sym}}^{n-1} \square& \overline{\square} & n-1 & 3 & 0 \\
    \Psi & \cdot & \otimes_{\mathrm{sym}}^{2n-1} \square& \cdot & -2n+1 & -4 & 2 \\
    \end{array}
\end{equation}
The chirals $T_j$ correspond to the following gauge invariant combinations of the charged fields
\begin{equation}
T_n = A^n Q\,,\qquad
T_{n-1} = A^{n-1} Q^3\,,
\end{equation}
while the $J$-term is 
\begin{equation}
\label{JPSI2}
J_{\Psi} = T_n T_{n-1}\,.
\end{equation}
We have computed the 't Hooft anomalies in the two phases and we have observed that they match. Explicitly we have
\begin{align}
    &\kappa_{\SU(2)^2} = \tfrac{n(2n+1)}{2}\,,\quad
    &&\kappa_{AA} = -\kappa_{AR_0} = 2n(2n+1)\,,\nonumber\\
    &\kappa_{\SU(4)^2} = \tfrac{2n+1}{2}\,,\quad
    &&\kappa_{QQ} = -\kappa_{QR_0} = 8n+4\,,\\
    &\kappa_{R_0 R_0}  = 6n+4\,,\quad
    &&\kappa_{AQ} = 0\,. \nonumber
\end{align}

In the following we will provide a further check of this duality by showing that the elliptic genera of the two models match, provided the validity of other identities that already appeared in the literature, and that descend from the reduction of 4d dualities to 2d.
The derivation is very similar to the one spelled out above, and for this reason we will be more sketchy and refer the reader to the derivation in subsection \ref{subsec:3.1} for further details.
The identity that we need to prove in order to check the validity of the duality is
\begin{equation}
\label{toprovesec3.2}
    I^{(4,\cdot,\cdot,2,\cdot)}_{\SU(2n+1)}=
    \frac{
        \prod_{j=0}^{2n-1}
        \theta\left(
            q/\left(t_{1}^{2(2n-1-j)} t_2^{2j}\prod_{a=1}^{4} u_a\right)
        \right)
    }{
        \prod_{a=1}^{4}\prod_{j=0}^{n} \theta\left(t_1^{2(n-j)}t_2^{2j}u_a\right)
        \prod_{a<b<c}^{4}\prod_{j=0}^{n-1} \theta\left(t_1^{2(n-1-j)}t_2^{2j}u_a u_b u_c\right)
    }\,.
\end{equation}

Again, we flip all the terms in the denominator of the RHS of (\ref{toprovesec3.2})
by moving the theta functions to the numerator of LHS and by using the relation (\ref{inversion}).
These theta functions are associated to Fermi flippers for the operators $T_{n}$ and $T_{n-1}$ on the gauge theory side. 
Then we re-define $T_n$ and $T_{n-1}$ as $\mathbf{T}_{n-j,j}^a$, for $j=0,\dots,n$ and $a=1,\dots,4$ and $\mathbf{T}_{n-j-1,j}^{a b c}$ for $j=0,\dots,n-1$ and $1\leq a<b<c\leq 4$ and
the Fermi $\Psi$ is redefined as  $\mathbf{\Psi}_{2n-1-j,j}$ with $j=0,\dots,2n-1$ accordingly.

%%%%%%%%%%%%%%%%%%%%%%%%%%%%%%%%%%%%%%%%%%%%%%%%%%%%%%%%%%%%%%%%%%%%%%%
% FIGURE: SU(2n+1) w 2A 4F deconfinement
\begin{figure}[h!]
    \centering
        % STEP 1
        \begin{minipage}{0.45\textwidth}
            \centering
            \makebox[\textwidth][c]{
            \begin{tikzpicture}[
                every node/.style={font=\footnotesize},
                box/.style={rectangle, draw, thick}
            ]
            \pgfmathsetmacro{\x}{0.8}
            \pgfmathsetmacro{\y}{1.6}
            % Nodi
            \node[box] (bot) at (0, -1.75) {$4$};
            \node[fill=SUcol,circle,draw,thick] (center) at (0, 0) {$2n\!+\!1$};
            % Antisimmetriche
            \node[box, minimum size=0.2cm] (square1r) at (0.3, 1.5) {};
            \node[box, minimum size=0.2cm] (square2r) at (0.3, 1.78) {};
            \node[box, minimum size=0.2cm] (square1l) at (-0.3, 1.5) {};
            \node[box, minimum size=0.2cm] (square2l) at (-0.3, 1.78) {};
            \node at (\x, \y) {$A_2$};
            \node at (-\x, \y) {$A_1$};
            \draw[<-,thick,>=stealth] (square1r.south) to[out=-90, in=60, looseness=1.2] ($(center.north)+(0.1,-0.01)$);
            \draw[<-,thick,>=stealth] (square1l.south) to[out=-90, in=120, looseness=1.2] ($(center.north)+(-0.1,-0.01)$);
            % Fondamentali
            \draw[<-,thick,>=stealth] (bot.north) -- node[right] {$Q$} (center);
            \end{tikzpicture}
            }
        \end{minipage}
        % STEP 2
        \begin{minipage}{0.45\textwidth}
            \centering
            \makebox[\textwidth][c]{
            \begin{tikzpicture}[
                every node/.style={font=\footnotesize},
                box/.style={rectangle, draw, thick}
            ]
            \pgfmathsetmacro{\a}{2}
            \pgfmathsetmacro{\b}{2}
            % Nodi
            \node[fill=USPcol,circle,draw,thick] (uspl) at (-\a,0) {$2n\!-\!2$};
            \node[fill=USPcol,circle,draw,thick] (uspr) at (\a, 0) {$2n\!-\!2$};
            \node[box] (botl) at (-\a, -\b) {$1$};
            \node[box] (botr) at (\a, -\b) {$1$};
            \node[box] (bot) at (0, -\a) {$4$};
            \node[fill=SUcol,circle,draw,thick] (center) at (0, 0) {$2n\!+\!1$};
            % Fondamentali
            \draw[<-, thick, >=stealth] (uspl.east) -- node[above] {$P_1$} (center.west);
            \draw[<-, thick, >=stealth] (uspr.west) -- node[above] {$P_2$} (center.east);
            \draw[<-, thick, >=stealth] (bot.north) -- (center);
            \node at (0.65,-\a) {$Q$};
            \draw[->,thick,>=stealth] (botl.north east) -- node[right,yshift=-5pt]{$\tilde R_1$}(center.south west);
            \draw[->,thick,>=stealth] (botr.north west) -- node[left,yshift=-5pt]{$\tilde R_2$}(center.south east);
            % Fermi
            \draw[dashed] (uspl) -- node[left] {$\hat \Psi_1$} (botl);
            \draw[dashed] (uspr) -- node[right] {$\hat \Psi_2$} (botr);
            \end{tikzpicture}
            }
        \end{minipage}
        \\[0.5cm]
        % STEP 3
        \begin{minipage}{0.45\textwidth}
            \centering
            \makebox[\textwidth][c]{
            \begin{tikzpicture}[
                every node/.style={font=\footnotesize},
                box/.style={rectangle, draw, thick}
            ]
            \pgfmathsetmacro{\a}{2}
            \pgfmathsetmacro{\b}{2}
            % Nodi 
            \node[fill=USPcol,circle,draw,thick] (uspl) at (-\a,0) {$2n\!-\!2$};
            \node[fill=USPcol,circle,draw,thick] (uspr) at (\a, 0) {$2n\!-\!2$};
            \node[box] (botl) at (-\a, -\b) {$1$};
            \node[box] (botr) at (\a, -\b) {$1$};
            \node[box] (bot) at (0, -\b) {$4$};
            \node[fill=SUcol,circle,draw,thick] (center) at (0, 0) {$2n\!-\!1$};
            % Fondamentali
            \draw[->, thick, >=stealth] (uspl.east) -- node[above] {$\tilde P_1$} (center.west);
            \draw[->, thick, >=stealth] (uspr.west) -- node[above] {$\tilde P_2$} (center.east);
            \draw[->, thick, >=stealth] (bot.north) -- (center);
            \node at (0,-\a-0.55) {$\tilde Q$};
            % Mesoni
            \draw[thick] (botl) -- node[below] {$L_2$} (bot);
            \draw[thick] (botr) -- node[below] {$L_1$} (bot);
            \draw[thick] (uspl) -- node[left] {$N_2$} (botl);
            \draw[thick] (uspr) -- node[right] {$N_1$} (botr);
            % Fermi
            \draw[<-,dashed,>=stealth] (botl.north east) -- node[right,yshift=-5pt]  {$\Lambda_2$}(center.south west);
            \draw[<-,dashed,>=stealth] (botr.north west) -- node[left,yshift=-5pt]{$\Lambda_1$}(center.south east);
            \end{tikzpicture}
            }
        \end{minipage}
        % STEP 4
        \begin{minipage}{0.45\textwidth}
            \centering
            \makebox[\textwidth][c]{
            \begin{tikzpicture}[
                every node/.style={font=\footnotesize},
                box/.style={rectangle, draw, thick},
            ]
            \pgfmathsetmacro{\a}{1.75}
            \pgfmathsetmacro{\x}{0.8}
            \pgfmathsetmacro{\y}{1.6}
            % Nodi        
            \node[box] (botl) at (-\a+0.5, -\a) {$1$};
            \node[box] (botr) at (\a-0.5, -\a) {$1$};
            \node[box] (bot) at (0, -\a) {$4$};
            \node[fill=SUcol,circle,draw,thick] (center) at (0, 0) {$2n\!-\!1$};
            % Antisimmetriche
            \node[box, minimum size=0.2cm] (square1r) at (0.3, 1.5) {};
            \node[box, minimum size=0.2cm] (square2r) at (0.3, 1.78) {};
            \node[box, minimum size=0.2cm] (square1l) at (-0.3, 1.5) {};
            \node[box, minimum size=0.2cm] (square2l) at (-0.3, 1.78) {};
            \node at (\x, \y) {$\tilde A_2$};
            \node at (-\x, \y) {$\tilde A_1$};
            \draw[->, thick, >=stealth] (square1r.south) to[out=-90, in=45, looseness=1.2] ($(center.north)+(0.1,-0.01)$);
            \draw[->, thick, >=stealth] (square1l.south) to[out=-90, in=135, looseness=1.2] ($(center.north)+(-0.1,-0.01)$);
            % Fondamentali
            \draw[->, thick, >=stealth] (bot.north) -- node[right] {$\tilde Q$} (center);
            % Mesoni
            \draw[thick] (botl) -- node[below] {$L_1$} (bot);
            \draw[thick] (botr) -- node[below] {$L_2$} (bot);
            \end{tikzpicture}
            }
        \end{minipage}
    \caption{Deconfinement steps of $\SU(2n+1)$ with two antisymmetrics and four fundamentals.}
    \label{fig:SUdisp}
\end{figure}
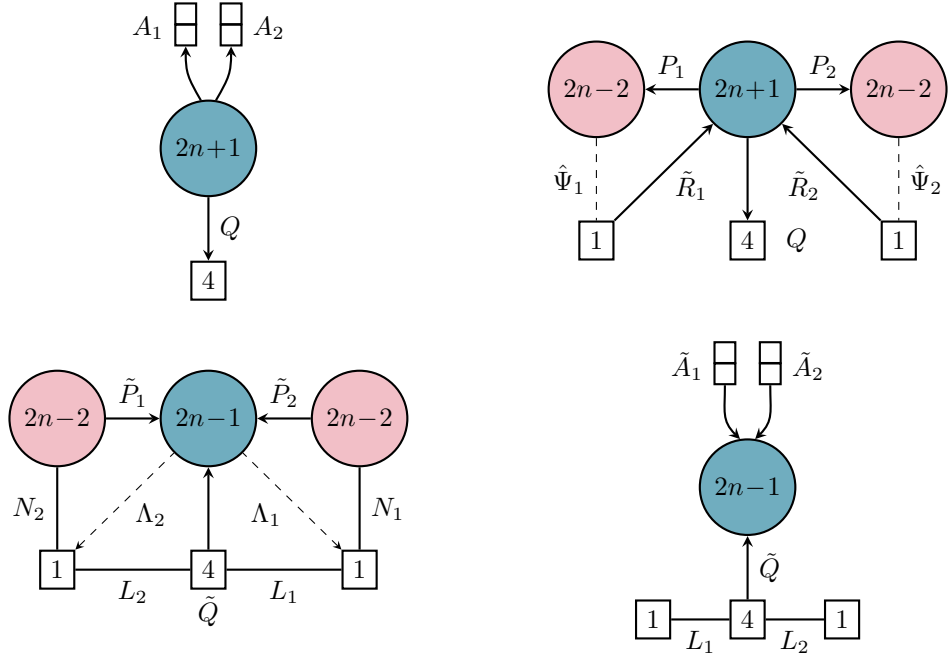
%%%%%%%%%%%%%%%%%%%%%%%%%%%%%%%%%%%%%%%%%%%%%%%%%%%%%%%%%%%%%%%%%%%%%%%

The electric theory is  modified by the addition of the following $J$-terms
\begin{align}
\label{Jflippers3}
&\Psi_{\mathbf{T}_{n-j,j}}^a = A_1^{n-j} A_2^j Q_a\,,
&&j=0,\dots,n, \quad a=1,\dots,4\,, \nonumber \\
&\Psi_{\mathbf{T}_{n-j-1,j}^{abc}} = A_1^{n-j-1} A_2^j Q_a Q_b Q_c\,,
&&j=0,\dots,n-1,\, 1\leq a<b<c\leq 4\,.
\end{align}
In the dual theory we are then left only with the  free Fermi  fields
$\mathbf{\Psi}_{2n-1-j,j}$.

We proceed by finding an auxiliary quiver, shown in Figure \ref{fig:SUdisp}, where the two antisymmetric chirals $A_{1,2}$ are traded with  two $\USp(2n-2)$ auxiliary gauge groups.
In this case the process does not break the $\SU(4)$ global symmetry, differently from the case studied in subsection \ref{subsec:3.1}.

There are two $\USp/\SU$ bifundamental  $P_{1,2}$, two $\SU(2n+1)$ fundamentals 
$\tilde R_{1,2}$ and two Fermi fields $\hat \Psi_{1,2}$. There are also two  $J$-terms 
$J_{\hat \Psi_{1,2}} = \tilde R_{1,2} P_{1,2}$.
The original fields $A_{1,2}$  correspond here to the combinations $P_{1,2}^2$.
In addition, the $J$-terms (\ref{Jflippers3}) are modified as
\begin{equation}
\label{Jflippers4}
\Psi_{\mathbf{T}_{n-j,j}^a} = P_1^{2n-2j} P_2^{2j} Q_a\,,\quad 
\Psi_{\mathbf{T}_{n-j-1,j}^{abc}} = P_1^{2(n-j-1)} P_2^{2j} Q_a Q_b Q_c\,.
\end{equation}
The original quiver in Figure \ref{fig:SUdisp} can be recovered if we exploit the duality reviewed in subsection \ref{USpallaSacchi} provided that we remove an extra, non-anomalous, axial symmetry that is allowed by the field content.

This requirement is translated into a constraint imposed on the elliptic genus, which prevents fugacities for the axial symmetries to be turned on both for the chirals and the Fermi charged under the two $\USp(2n-2)$ factors. Again, we interpret the constraint as in the discussion of subsection (\ref{subsec:3.1}).

We  proceed by dualizing the $\SU(2n+1)$ gauge node.
This model has  $4n$ fundamentals and $2$ antifundamentals and the duality we need has been reviewed in subsection \ref{SUnGen}. Here we consider the case with $y=0$.
The model obtained in this case is represented in the third quiver of Figure \ref{fig:SUdisp}. Again, the constraints imposed on the axial symmetries for the $\USp(2n-2)$ nodes above are not required here anymore. 
In this case the $J$-terms for the Fermi $\Lambda_{1,2}$ obtained in the duality are
\begin{equation}
J_{\Lambda_{1}}= L_1  \tilde Q + N_1 \tilde P_2\,, \qquad
J_{\Lambda_{2}}= L_2  \tilde Q + N_2 \tilde P_1\,. 
\end{equation}
The last step in Figure \ref{fig:SUdisp} corresponds to confine the two $\USp(2n-2)$
gauge nodes. In this case they are both characterized by $2n$ fundamentals, and the dual models are LG theories, where the two combinations $\tilde P_{1,2}^2$ define conjugated antisymmetric chirals. There are also four $\SU(2n-2)$ antifundamentals, and the $J$-terms obtained after solving the equation of motion are
\begin{equation}
J_{ \Psi^{(0)}_{1}}= \tilde A_1^{n-1} L_2 \tilde Q\,,
\qquad
J_{ \Psi^{(0)}_{2}}= \tilde A_2^{n-1} L_1 \tilde Q\,.
\end{equation}

Again we have denoted the Fermi flippers, arising from the gauge/LG duality of the symplectic gauge nodes discussed in subsection \ref{sec:USP2Np2}, as $ \Psi_1^{(0)}$ and $ \Psi_2^{(0)}$, as they will constitute the \emph{extremal} fields in the  Fermi \emph{tower} denoted as $\Psi$ in equation (\ref{Table3.2}).

In this way we have obtained a duality between the original $\SU(2n+1)$ model with two antisymmetric and four fundamental chirals and an $\SU(2n-1)$ model with two conjugate antisymmetric and four antifundamental chirals, in addition to a set of Fermi and chiral singlets.

At this point of the discussion we  iterate the procedure until we reach an $\SU(2n-3)$ gauge theory with the same charged field content of the original one, i.e. two  antisymmetric and four fundamental chirals.
In this way we get a simpler relation between the elliptic genera,
that can be recursively applied in order to obtain the final relation between the original model and an LG model.
Such relation is obtained by applying the relations 
(\ref{idsacchi}), (\ref{EGdualitySUgen}) and (\ref{ellipticdualUSP}), and it is
\begin{eqnarray}
\label{recursive2}
I_{\SU(2n+1)}^{(4,\cdot,\cdot,2,\cdot)}(\vec u;\cdot;\cdot;\vec t;\cdot)
&&=
I_{\SU(2n-3)}^{(4,\cdot,\cdot,2,\cdot)}\left((t_1 t_2)^\frac{2}{2n-3}\vec u;\cdot;\cdot;(t_1 t_2)^\frac{2}{2n-3}\vec t; \cdot\right) \nonumber
\\
&&
\times
\frac{\theta\left(q/(t_{1,2}^{2(2n-1)} \prod_{a=1}^{4}u_a)\right) \, \theta\left(q/(t_{1,2}^{2(2n-2)}t_{2,1}^2 \prod_{a=1}^{4}u_a)\right)}{
\prod_{a=1}^4 \theta\left(t_{1,2}^{2n}  u_a \right)
\prod_{a<b<c} \theta\left(t_{1,2}^{2n-2} u_a u_b u_c\right)
}\,.
\end{eqnarray}
Analogously to the case studied in subsection \ref{subsec:3.1} we keep the chiral fields unflipped at the level of the elliptic genus, while they appear as Fermi fields in the field theory discussion in the $\SU(2n+1)$ side.
%{\color{red} Dovremmo scrivere la field theory a questo step in modo esplicito.} 
Each term in the second line of the (\ref{recursive2}) corresponds to one of the expected singlets of the duality. These are consistent with the deconfinement procedure which apparently broke the global $\SU(2)$ non-abelian global symmetry.
We leave to the interested reader the details of the matching in this case.

A recursive application of this relation leads to an $\SU(5)$ gauge theory  model for even $n$ or to an $\SU(3)$ gauge theory for odd $n$.
In general, at the $j$-th step we obtain an $\SU(2n+1-4j)$ gauge theory, and the leftover integral is
\begin{equation}
\label{leftover2}
I_{\SU(2n-4j+1)}^{(4,\cdot,\cdot,2,\cdot)}\left((t_1 t_2)^\frac{2j}{2n+1-4j}\vec u;\cdot;\cdot; (t_1 t_2)^\frac{2j}{2n+1-4j}\vec t;\cdot \right)\,.
\end{equation}

In addition, the Fermi and chiral singlets generated at this step can be worked out
from the relation (\ref{recursive2}) in complete analogy to what was done in subsection
\ref{subsec:3.1}. 
The details of the derivation are straightforward, thus we omit them.

To conclude the proof of (\ref{toprovesec3.2}),  we must  distinguish the cases $n=2m$ and $n=2m+1$.

\subsubsection*{The case of $\SU(4m+1)$}
In this case we can use the recurrence relations until we obtain an $\SU(5)$ gauge theory with two antisymmetric $\mathbf{A}_{1,2}$ and four fundamentals $\mathbf{Q}_{1,2,3,4}$. Their fugacities are $(t_1 t_2)^{\frac{2(m-1)}{5}}t_{1,2}$ and $(t_1 t_2)^\frac{2(m-1)}{5}u_a$ respectively.

Now, we can further deconfine the two antisymmetric tensors $\mathbf{A}_{1,2}$, obtaining two $\USp$ gauge groups. 
By repeating the various steps in Figure \ref{fig:SUdisp} we get an $\SU(3)$ gauge theory, where the two conjugate antisymmetric correspond to two extra fundamentals. 
The $\SU(3)$ gauge theory has then two fundamentals and four antifundamentals and it is dual to an LG model (see \cite{Amariti:2024usp}).
Translating the discussion above to the corresponding identities among elliptic genera, we find the expected relation (\ref{toprovesec3.2}) in the case of $n=2m$.

\subsubsection*{The case of $\SU(4m+3)$}
In order to complete the proof of  (\ref{toprovesec3.2}) we need to discuss the case $\SU(4m+3)$ as well. In this case we get, after $m$ iterations, an $\SU(3)$ gauge theory with four fundamentals $\mathbf{Q}_{1,2,3,4}$ with fugacities $(t_1 t_2)^\frac{2m}{3}u_a$ and two antisymmetric chirals, corresponding to two extra antifundamentals, with fugacities $(t_1 t_2)^\frac{2m}{3}t_{1,2}$.
Again, the $\SU(3)$ gauge  is dual to an LG theory (see \cite{Amariti:2024usp}) and we find the expected identity (\ref{toprovesec3.2}) in the case of $n=2m+1$ as well.

This concludes the proof of the identity (\ref{toprovesec3.2}).
In the end we obtained a model with only Fermi multiplets, as all the chirals $T_j$ in the original gauge theory have been flipped. 
In the following we will actually use the un-flipped (or partially flipped) duality in order to prove the cases of $\SU(2n+1)$ with two antisymemtric chirals and $(4-j,j)$ pairs of fundamentals and antifundamental.

In the next subsections we study the other cases, in which we decrease the number of fundamentals and increase the number of antifundamentals.
As a general comment we will see that the derivation in such cases is simpler than above, because the proof does not require an iterative construction. Indeed after deconfining the two antisymmetric tensors and dualizing the unitary and the symplectic gauge nodes we will get $\SU(N)$ models with two conjugate antisymmetric and four antifundamental chirals, possibly in addition to a non-trivial set of $J$-terms. 
Such models are (modulo charge conjugation) exactly the ones studied so far.
It follows that we can use the results of subsection \ref{subsec:3.1} and  \ref{subsec:3.2} in order to prove that the other $\SU(N)$ models with two antisymmetric chirals,  
$n_f < 4$  fundamental chirals  and $n_a > 1$ antifundamental chirals are dual to LG theories.

\subsection{$\SU(2n)$ with three fundamentals and one antifundamental}
\label{subsec:3.3}

In this case the dual LG model has five chiral gauge invariant operators corresponding to chiral fields interacting through a $J$-term with a Fermi multiplet.
The field content of the gauge theory and of the dual LG model is represented in the table below.

\begin{equation} 
    \begin{array}{c|c|c|c|c|c|c|c|}
    & \SU(2n) & \SU(2) & \SU(3) & \U(1)_A & \U(1)_Q & \U(1)_{\tilde Q} & \U(1)_{R_0} \\
    \hline
    A & \begin{array}{c}
\square \vspace{-2.85mm} \\
\square 
\end{array}  & \square & \cdot & 1 & 0 & 0 & 0 \\
    Q & \square & \cdot & \square & 0 & 1 & 0 & 0 \\ 
    \tilde Q & \overline{\square} & \cdot & \cdot & 0 & 0 & 1 & 0 \\       
    \hline
    M & \cdot & \cdot & \square & 0 & 1 & 1 & 0 \\
    T_{n} & \cdot & \otimes_{\mathrm{sym}}^{n} \square& \cdot & n & 0 & 0 & 0 \\
    T_{n-1} & \cdot & \otimes_{\mathrm{sym}}^{n-1} \square& \overline{\square} & n-1 & 2 & 0 & 0 \\
    P_{n} & \cdot & \otimes_{\mathrm{sym}}^{n-2} \square& \square & n & 1 & 1 & 0 \\
    P_{n-1} & \cdot & \otimes_{\mathrm{sym}}^{n-3} \square& \cdot & n-1 & 3 & 1 & 0 \\
    \Psi & \cdot & \otimes_{\mathrm{sym}}^{2n-3} \square& \cdot & -2n+1 & -3 & -1 & 2 \\
    \end{array}
\end{equation}

The chirals $M$, $T_j$, $P_j$ correspond to the following gauge invariant combinations of the charged fields
\begin{equation}
\label{singlets3}
M = Q \tilde Q ,\quad
T_n = A^n \,,\quad
T_{n-1} = A^{n-1} Q^2, \quad
P_n = A^{n-1} (A \tilde Q) Q, \quad
P_{n-1} = A^{n-2} (A \tilde Q) Q^3,
\end{equation}
while the $J$-term is 
\begin{equation}
\label{JPSi3}
J_{\Psi} = M T_{n-1} T_n + T_n P_{n-1} + T_{n-1} P_n \,.
\end{equation}
We have computed the 't Hooft anomalies in the two phases and they match. Explicitly we have
\begin{equation}
    \begin{aligned}
        &\kappa_{\SU(2)^2} = \tfrac{n(2n-1)}{2}\,,\\
        &\kappa_{\SU(3)^2} = n\,,\\
        &\kappa_{R_0 R_0}  = 6n+1\,,\\
    \end{aligned}
    \qquad \qquad
    \begin{aligned}
        &\kappa_{AA} = -\kappa_{AR_0} = 2n(2n-1)\,,\\
        &\kappa_{QQ} = -\kappa_{QR_0} = 6n\,,\\
        &\kappa_{\tilde Q\tilde Q} = -\kappa_{\tilde QR_0} = 2n\,,\\
        &\kappa_{AQ} =\kappa_{A\tilde Q} = 0\,.
    \end{aligned}
\end{equation}

Again, we aim to corroborate the validity of the duality by matching the elliptic genera using other relations already derived in the literature.
In this case, differently from the cases above, we will not flip all the chirals in the dual LG description, but only the meson $M$. 
We will see that the expected $J$-term $J_{\Psi}$
can be reconstructed with our procedure starting from the $J$-term (\ref{JPSI2}) 
that we expect from the duality of subsection \ref{subsec:3.2}.

%%%%%%%%%%%%%%%%%%%%%%%%%%%%%%%%%%%%%%%%%%%%%%%%%%%%%%%%%%%%%%%%%%%%%%%
% FIGURE: SU(2n) w 2A 3F+1aF deconfinement
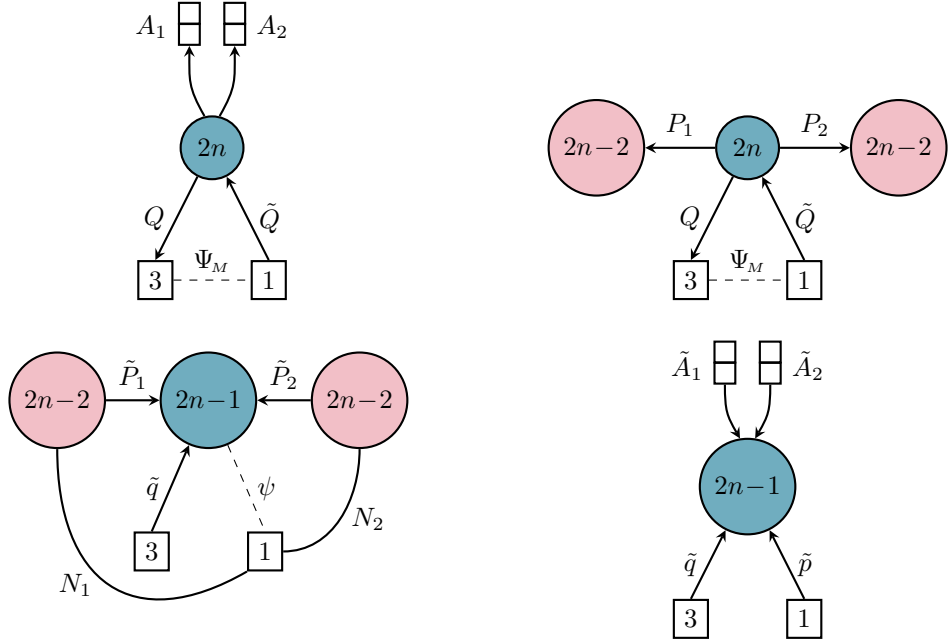
\begin{figure}[h!]
    \centering
        % STEP 1
        \begin{minipage}[b]{0.45\linewidth}
            \centering
            \makebox[\textwidth][c]{
            \hspace{-0.1em}
            \begin{tikzpicture}[
                every node/.style={font=\footnotesize},
                box/.style={rectangle, draw, thick}
            ]
            \pgfmathsetmacro{\c}{1.75}
            \pgfmathsetmacro{\b}{0.75}
            \pgfmathsetmacro{\x}{0.8}
            \pgfmathsetmacro{\y}{1.6}
            % Nodi
            \node[box] (bot) at (-\b, -\c) {$3$};
            \node[box] (bot2) at (\b, -\c) {$1$};
            \node[fill=SUcol,circle,draw,thick] (center) at (0, 0) {$2n$};
            % Antisimmetriche
            \node[box, minimum size=0.2cm] (square1r) at (0.3, 1.5) {};
            \node[box, minimum size=0.2cm] (square2r) at (0.3, 1.78) {};
            \node[box, minimum size=0.2cm] (square1l) at (-0.3, 1.5) {};
            \node[box, minimum size=0.2cm] (square2l) at (-0.3, 1.78) {};
            \node at (\x, \y) {$A_2$};
            \node at (-\x, \y) {$A_1$};
            \draw[<-,thick,>=stealth] (square1r.south) to[out=-90, in=60, looseness=1.2] ($(center.north)+(0.1,-0.01)$);
            \draw[<-,thick,>=stealth] (square1l.south) to[out=-90, in=120, looseness=1.2] ($(center.north)+(-0.1,-0.01)$);
            % Fondamentali
            \draw[<-,thick,>=stealth] (bot.north) -- node[left] {$Q$} (center);
            \draw[<-,thick,>=stealth] (center) -- node[right] {$\tilde Q$} (bot2.north);
            % Fermi
            \node at (0, -1.5) {$\Psi_{\!\scriptscriptstyle M}$};
            \draw[-,dashed,>=stealth] (bot) to[out=0, in=180, looseness=0] (bot2);
            \end{tikzpicture}
            }
        \end{minipage}
        % STEP 2
        \begin{minipage}[b]{0.45\linewidth}
            \centering
            \makebox[\textwidth][c]{
            \begin{tikzpicture}[
                every node/.style={font=\footnotesize},
                box/.style={rectangle, draw, thick}
            ]
            \pgfmathsetmacro{\d}{1.75}
            \pgfmathsetmacro{\c}{0.75}
            \pgfmathsetmacro{\a}{2}
            \pgfmathsetmacro{\b}{2}
            % Nodi
            \node[fill=USPcol,circle,draw,thick] (uspl) at (-\a,0) {$2n\!-\!2$};
            \node[fill=USPcol,circle,draw,thick] (uspr) at (\a, 0) {$2n\!-\!2$};
            \node[fill=SUcol,circle,draw,thick] (center) at (0, 0) {$2n$};
            \node[box] (bot) at (-\c, -\d) {$3$};
            \node[box] (bot2) at (\c, -\d) {$1$};
            % Fondamentali
            \draw[<-,thick,>=stealth] (bot.north) -- node[left] {$Q$} (center);
            \draw[<-, thick, >=stealth] (uspl.east) -- node[above] {$P_1$} (center.west);
            \draw[<-, thick, >=stealth] (uspr.west) -- node[above] {$P_2$} (center.east);
            \draw[<-,thick,>=stealth] (center) -- node[right] {$\tilde Q$} (bot2.north);
            % Fermi
            \node at (0, -1.5) {$\Psi_{\!\scriptscriptstyle M}$};
            \draw[-,dashed,>=stealth] (bot) to[out=0, in=180, looseness=0] (bot2);
            \end{tikzpicture}
            }
        \end{minipage}
        \\[0.5cm]
        % STEP 3
        \begin{minipage}[b]{0.45\linewidth}
            \centering
            \makebox[\textwidth][c]{
            \begin{tikzpicture}[
                every node/.style={font=\footnotesize},
                box/.style={rectangle, draw, thick}
            ]
            \pgfmathsetmacro{\a}{2}
            \pgfmathsetmacro{\b}{0.75}
            % Nodi
            \node[fill=USPcol,circle,draw,thick] (uspl) at (-\a,0) {$2n\!-\!2$};
            \node[fill=USPcol,circle,draw,thick] (uspr) at (\a, 0) {$2n\!-\!2$};
            \node[fill=SUcol,circle,draw,thick] (center) at (0, 0) {$2n\!-\!1$};
            \node[box] (bot) at (-\b, -2) {$3$};
            \node[box] (bot2) at (\b, -2) {$1$};
            % Fondamentali
            \draw[->,thick,>=stealth] (bot.north) -- node[left] {$\tilde q$} (center);
            \draw[->, thick, >=stealth] (uspl.east) -- node[above] {$\tilde P_1$} (center.west);
            \draw[->, thick, >=stealth] (uspr.west) -- node[above] {$\tilde P_2$} (center.east);
            % Mesoni
            \draw[thick] (bot2.east) to[out=0, in=-90, looseness=1] node[right, overlay] {$N_2$} (uspr.south);
            \draw[thick] (bot2.south west) to[out=-150, in=-90, looseness=1.5] node[left, overlay, xshift=-2pt] {$N_1$} (uspl.south);
            % Fermi
            \draw[-,dashed,>=stealth] (center) -- node[right] {$\psi$} (bot2.north);
            \end{tikzpicture}
            }
        \end{minipage}
        % STEP 4
        \begin{minipage}[b]{0.45\linewidth}
            \centering
            \makebox[\textwidth][c]{
            \begin{tikzpicture}[
                every node/.style={font=\footnotesize},
                box/.style={rectangle, draw, thick},
            ]
            \pgfmathsetmacro{\c}{1.75}
            \pgfmathsetmacro{\b}{0.75}
            \pgfmathsetmacro{\x}{0.8}
            \pgfmathsetmacro{\y}{1.6}
            % Nodi
            \node[box] (bot) at (-\b, -\c) {$3$};
            \node[box] (bot2) at (\b, -\c) {$1$};
            \node[fill=SUcol,circle,draw,thick] (center) at (0, 0) {$2n\!-\!1$};
            % Antisimmetriche
            \node[box, minimum size=0.2cm] (square1r) at (0.3, 1.5) {};
            \node[box, minimum size=0.2cm] (square2r) at (0.3, 1.78) {};
            \node[box, minimum size=0.2cm] (square1l) at (-0.3, 1.5) {};
            \node[box, minimum size=0.2cm] (square2l) at (-0.3, 1.78) {};
            \node at (\x, \y) {$\tilde A_2$};
            \node at (-\x, \y) {$\tilde A_1$};
            \draw[->,thick,>=stealth] (square1r.south) to[out=-90, in=60, looseness=1.2] ($(center.north)+(0.1,-0.01)$);
            \draw[->,thick,>=stealth] (square1l.south) to[out=-90, in=120, looseness=1.2] ($(center.north)+(-0.1,-0.01)$);
            % Fondamentali
            \draw[<-,thick,>=stealth] (center) -- node[right] {$\tilde p$} (bot2.north);
            \draw[->,thick,>=stealth] (bot.north) -- node[left] {$\tilde q$} (center);
            \end{tikzpicture}
        }
        \end{minipage}
    \caption{Deconfinement steps of $\SU(2n)$ with two antisymmetrics, three fundamentals and one antifundamental.}
    \label{fig:SUpari31}
\end{figure}
%%%%%%%%%%%%%%%%%%%%%%%%%%%%%%%%%%%%%%%%%%%%%%%%%%%%%%%%%%%%%%%%%%%%%%%
The identity that we want to prove is
\begin{eqnarray}
\label{toprovesec3.3}
&&I^{(3,1,\cdot,2,\cdot)}_{\SU(2n)}(\vec u;v;\cdot;\vec t;\cdot)=
    \frac{\prod_{j=0}^{2n-3}  \theta\left(q/(t_1^{j+1}t_2^{2n-j-2} u_1 u_2 u_3 v)\right)}
{       \prod_{a=1}^{3} \theta \left(u_a v\right)  \prod_{j=0}^{n-2} \theta\left(t_1^{n-1-j} t_2^{j+1} u_a v\right)   
} \\
&&\times
\frac{1}{  \prod_{j=0}^{n} \theta \left(t_1^{j} t_2^{n-j}\right) \cdot
     \prod_{j=0}^{n-1} \prod_{1\leq a<b\leq 3 } \theta \left(t_1^{j} t_2^{n-j-1} u_a u_b\right) \cdot
    \prod_{j=0}^{n-3}  \theta\left(t_1^{n-2-j}t_2^{j+1}u_1 u_2 u_3 v\right)}\,.
    \nonumber
\end{eqnarray}

In this case, in order to prove this identity through the techniques used above, we will flip the chirals combinations $Q \tilde Q$ and $\Pf A_{1,2}$. The first one corresponds to the singlet $M$ in the $J$-term (\ref{JPSi3}), while the other descends from the singlet $T_n$, and such flip breaks the $\SU(2)$ flavor symmetry.

In the following we will redefine the singlets $T_n $, $T_{n-1} $, $P_n $ and $P_{n-1}$ as
$ \mathbf{T}_{n-j,j} $ with $j=0,\dots,n$, 
$ \mathbf{T}_{n-j-1,j}$ with $j=0,\dots,n-1$, 
$\mathbf{P}_{n-j-1,\,j+1}$ with $j=0,\dots,n-2$ and 
$\mathbf{P}_{n-j-2,\,j+1}$ with $j=0,\dots,n-3$.
Accordingly the Fermi $\Psi$ becomes $\mathbf{\Psi}_{-j-1,-2n+j+2}$  with $j=0,\dots,2n-3$.

The Fermi flippers of the combinations $Q \tilde Q$ and $\Pf A_{1,2}$ give origin to the following $J$-terms in the original $\SU(2n)$ gauge theory 
\begin{equation}
J_{\Psi_M} = Q \tilde Q\,,\qquad 
J_{\Psi_{A_{1,2}}} = \Pf A_{1,2}\,.
\end{equation}

We now proceed by trading the two antisymmetric with two $\USp(2n-2)_{1,2}$ gauge groups, each one connected to the original $\SU(2n)$ gauge group through a bifundamental $P_{1,2}$. In this case the flippers $\Psi_{A_{1,2}}$
disappear and we are left with only $J_{\Psi_M} $. Observe that in this case the original theory is recovered by dualizing the two symplectic gauge theory using the duality between $\USp(2n)$ 
with $2n+2$ fundamentals $Q$ and the LG model characterized by the singlets $M=Q^2$ and $\psi_0$ and $J$-term $J_{\psi_0} = \text{Pf } M$. This duality was originally proposed in \cite{Gadde:2015wta} and further studied in \cite{Dedushenko:2017osi,Sacchi:2020pet}. We have reviewed the basic aspects of the duality in subsection \ref{sec:USP2Np2}.

The $\SU(2n)$ gauge theory has $4n-1$ fundamentals and one antifundamental. It can be dualized according to the rules explained in subsection \ref{SUnGen}, and it gives origin to the third quiver in Figure \ref{fig:SUpari31}.
In this case the $J$-term is given by $J_\psi = N_2 \tilde P_2 + N_1 \tilde P_1$.
The last step consists of dualizing the two $\USp(2n-2)$ groups, each one with $2n$ chiral fundamentals to an LG model.
The final quiver is the fourth one in figure \ref{fig:SUpari31}.
In this case there are two $J$-terms associated to two Fermi singlets $\Psi_{1,2}^{(0)}$. They read $J_{\Psi_{1,2}^{(0)}} = \tilde A_{1,2}^{n-1} \tilde p$.

We ended up with an $\SU(2n-1)$ gauge theory with four antifundamentals and two conjugate antisymmetric. This is, up to a conjugation, the same field content of the confining duality studied in subsection 
\ref{subsec:3.2}. In addition we have the non-vanshing $J$-terms $J_{\Psi_{1,2}^{(0)}}$.

The theory is then dual to an LG model where the $J$-term 
(\ref{JPSI2}) is modified by symmetry breaking pattern implied by the presence of $J_{\Psi_{1,2}^{(0)}}$.

In order to take such breaking into account let us consider 
an $\SU(2N+1)$ gauge theory with four antifundamentals $\tilde Q=\{\tilde q_1,\tilde q_2,\tilde q_3,\tilde p\}$ and two conjugate antisymmetric tensors $\tilde A= \{\tilde A_1,\tilde A_2\}$.
The singlets of the unbroken phase are $\tilde T_N$ and 
$\tilde T_{N-1}$ and the 2d superpotential\footnote{Here we use the notion of 2d superpotential instead of the one of $J$-terms because the structure of the former is more intuitive}
\begin{equation}
W_{2d} = \sum_{j=0}^{2N-1} \sum_{k=0}^N \sum_{\ell=0}^{N-1}
|\epsilon_{abcd}| \psi_{j-2N+1,-j} \tilde T_{k,N-k}^{(a)}  
\tilde T_{\ell,N-1-\ell}^{(bcd)} \delta_{j+k+\ell,2N-1}\,,
\end{equation}
where we used the notations of \ref{subsec:3.2} for the $\SU(2)$ and $\SU(4)$ indices.
Then we fix $N=n-1$ and consider the effect of 
$J_{\Psi_{1,2}^{(0)}}$.
The singlets $\tilde T_{n-1}$  and $\tilde T_{n-2}$ are split as 
\begin{equation}
    \begin{aligned}
        & \tilde T_{j,n-1-j}^{(a)} = \tilde A_1^j \tilde A_2^{n-1-j} \tilde q_{a}\,,\\
        & \tilde  T_{j,n-2-j}^{(0)} = \tilde A_1^j \tilde A_2^{n-2-j} \tilde q_{1} \tilde q_{2} \tilde q_{3}\,,
    \end{aligned}
    \qquad
    \begin{aligned}
        & \tilde   T_{j,n-1-j}^{(0)} = \tilde A_1^j \tilde A_2^{n-1-j}  \tilde R_-\,,\\
        & \tilde  T_{j,n-2-j}^{(ab)} = \tilde A_1^j \tilde A_2^{n-2-j} \tilde q_{a} \tilde q_{b} \tilde R_-\,.
    \end{aligned}
\end{equation}
In this way the superpotential of the LG for the deformed theory is
\begin{eqnarray}
\label{final3}
W_{2d} &=& \sum_{j=0}^{2n-3} \sum_{k=0}^{n-1} \sum_{\ell=0}^{n-2}
 \psi_{j-2n+3,-j} (|\epsilon_{abc}| 
 \tilde T_{k,n-1-k}^{(a)}  
\tilde T_{\ell,n-2-\ell}^{(bc)} 
+
 \tilde T_{k,n-1-k}^{(0)}  
\tilde T_{\ell,n-2-\ell}^{(0)} 
)
\delta_{j+k+\ell,2n-3} 
\nonumber  \\
&+&
\psi_1^{(0)} \tilde T_{n-1,0}^{(0)}
+
\psi_2^{(0)} \tilde T_{0,n-1}^{(0)},
\end{eqnarray}
where the deformation in the second line can be integrated out using the equations of motion.

The superpotential (\ref{final3}) is equivalent to the one that can be read from the $J$-term (\ref{JPSi3}) once the dictionary between the singlets $\tilde T$ and the singlets 
(\ref{singlets3}) is specified and after setting the combinations $Q \tilde Q$ and $\Pf A_{1,2}$ in (\ref{JPSi3}) to zero.

    \begin{align}
        & \mathbf{T}_{n-1-j,\,j+1} = \tilde T_{j,\,n-2-j}^{(0)} \,,\quad && \mathbf{T}_{n-1-j,\,j}^{(ab)} = |\epsilon^{ab}_{\phantom{ab}c}| \tilde T_{j,\,n-1-j}^{(c)} \,, \nonumber \\
        & \mathbf{P}_{n-1-j,\,j+1}^{(a)} = |\epsilon^{a}_{\phantom{a}bc}| \tilde T_{j,\,n-2-j}^{(bc)} \,, \quad && \mathbf{P}_{n-1-j,\,j} =  \tilde T_{j\,,n-1-j}^{(0)} \,, \\
        & \Psi_{-j-1,\,j-2n+2} = \psi_{j-2n+3,\,-j}  \nonumber
    \end{align}

The same strategy explained in the field theory discussion can be applied at the level of the elliptic genus. In this case
the four steps in Figure \ref{fig:SUpari31} can be reproduced at the level of the elliptic genus as in the previous cases of $\SU(N)$ with two antisymmetric and four fundamentals. 
The last step corresponds to plugging the identity (\ref{toprovesec3.2})
inside the elliptic genus of the $\SU(2n-1)$ gauge theory.
In this way we obtain the relation (\ref{toprovesec3.3}) as expected.

\subsection{$\SU(2n+1)$ with three fundamentals and one antifundamental}

In this case the dual LG model has five chiral gauge invariant operators corresponding to chiral fields interacting through a $J$-term with a Fermi multiplet.
The field content of the gauge theory and of the dual LG model is represented in the following table
\begin{equation} 
    \begin{array}{c|c|c|c|c|c|c|c|}
    & \SU(2n+1) & \SU(2) & \SU(3) & \U(1)_A & \U(1)_Q & \U(1)_{\tilde Q} & \U(1)_{R_0} \\
    \hline
    A & \begin{array}{c}
\square \vspace{-2.85mm} \\
\square 
\end{array}  & \square & \cdot & 1 & 0 & 0 & 0 \\
    Q & \square & \cdot & \square & 0 & 1 & 0 & 0 \\ 
    \tilde Q & \overline{\square} & \cdot & \cdot & 0 & 0 & 1 & 0 \\       
    \hline
    M & \cdot & \cdot & \square & 0 & 1 & 1 & 0 \\
    T_{n} & \cdot & \otimes_{\mathrm{sym}}^{n} \square& \square & n & 1 & 0 & 0 \\
    T_{n-1} & \cdot & \otimes_{\mathrm{sym}}^{n-1} \square& \cdot & n-1 & 3 & 0 & 0 \\
    P_{n+1} & \cdot & \otimes_{\mathrm{sym}}^{n-1} \square& \cdot & n+1 & 0 & 1 & 0 \\
    P_{n} & \cdot & \otimes_{\mathrm{sym}}^{n-2} \square& \overline{\square} & n & 2 & 1 & 0 \\
    \Psi & \cdot & \otimes_{\mathrm{sym}}^{2n-2} \square& \cdot & -2n & -3 & -1 & 2 \\
    \end{array}
\end{equation}
We have computed the 't Hooft anomalies in the two phases and they match. Explicitly:
\begin{equation}
    \begin{aligned}
        &\kappa_{\SU(2)^2} = \tfrac{n(2n+1)}{2}\,,\\
        &\kappa_{\SU(3)^2} = \tfrac{2n+1}{2}\,,\\
        &\kappa_{R_0 R_0}  = 6n+4\,,\\
    \end{aligned}
    \qquad
    \begin{aligned}
        &\kappa_{AA} = -\kappa_{AR_0} = 2n(2n+1)\,,\\
        &\kappa_{QQ} = -\kappa_{QR_0} = 6n+3\,,\\
        &\kappa_{\tilde Q\tilde Q}= \kappa_{\tilde Q R_0}=2n+1\,,\\
        &\kappa_{AQ} =\kappa_{A\tilde Q} = 0\,.
    \end{aligned}
\end{equation}
The chirals $M$, $T_j$ and $P_j$ correspond to the following gauge invariant combinations of the charged fields
\begin{equation}
M =  Q \tilde Q\,,\quad
T_n = A^n Q\,,\quad
T_{n-1} = A^{n-1} Q^3\,, \quad
P_{n+1} = A^n (A \tilde Q)\,, \quad
P_{n} = A^{n-1} (A \tilde Q) Q^2
\end{equation}
while the $J$-term is 
\begin{equation}
\label{JPSI3}
J_{\Psi} = M T_n^2 + T_{n-1} P_{n+1} + T_n P_n\,.
\end{equation}

Again we aim to corroborate the validity of the duality by matching the elliptic genera using other relations already derived in the literature.
In this case we will flip the meson $M$ and the combinations $A_1^{n} Q_1$ and $A_2^n Q_2$ in the dual LG and we show that the expected $J$-term in (\ref{JPSI3}) 
can be reconstructed by our procedure  using the $J$-term 
expected from the duality of subsection \ref{subsec:3.1}.
The derivation is very similar to the one discussed  in  subsection \ref{subsec:3.3}, and for this reason we will omit many technical details in this case.

%%%%%%%%%%%%%%%%%%%%%%%%%%%%%%%%%%%%%%%%%%%%%%%%%%%%%%%%%%%%%%%%%%%%%%%
% FIGURE: SU(2n+1) w 2A 3F+1aF deconfinement
\begin{figure}[h!]
    \centering
        % STEP 1
        \begin{minipage}[b]{0.45\linewidth}
            \centering
            \makebox[\textwidth][c]{
            \hspace{-0.1em}
            \begin{tikzpicture}[
                every node/.style={font=\footnotesize},
                box/.style={rectangle, draw, thick}
            ]
            \pgfmathsetmacro{\c}{1.75}
            \pgfmathsetmacro{\b}{1.2}
            \pgfmathsetmacro{\x}{0.8}
            \pgfmathsetmacro{\y}{1.6}
            % Nodi
            \node[box] (bot) at (-\b, -\c) {$2$};
            \node[box] (bot2) at (\b, -\c) {$1$};
            \node[box] (bot3) at (0, -\c) {$1$};
            \node[fill=SUcol,circle,draw,thick] (center) at (0, 0) {$2n\!+\!1$};
            % Antisimmetriche
            \node[box, minimum size=0.2cm] (square1r) at (0.3, 1.5) {};
            \node[box, minimum size=0.2cm] (square2r) at (0.3, 1.78) {};
            \node[box, minimum size=0.2cm] (square1l) at (-0.3, 1.5) {};
            \node[box, minimum size=0.2cm] (square2l) at (-0.3, 1.78) {};
            \node at (\x, \y) {$A_2$};
            \node at (-\x, \y) {$A_1$};
            \draw[<-,thick,>=stealth] (square1r.south) to[out=-90, in=60, looseness=1.2] ($(center.north)+(0.1,-0.01)$);
            \draw[<-,thick,>=stealth] (square1l.south) to[out=-90, in=120, looseness=1.2] ($(center.north)+(-0.1,-0.01)$);
            % Fondamentali
            \draw[<-,thick,>=stealth] (bot.north) -- node[left] {$Q_{1,2}$} (center);
            \draw[<-,thick,>=stealth] (bot3.north) -- node[left] {$Q_{3}$} (center);
            \draw[<-,thick,>=stealth] (center) -- node[right] {$\tilde Q$} (bot2.north);
            % Fermi
            \node at (0.6, -1.5) {$\Psi_{\!\scriptscriptstyle M_3}$};
            \draw[-,dashed,>=stealth] (bot3) to[out=0, in=180, looseness=0] (bot2);
            \end{tikzpicture}
            }
        \end{minipage}
        % STEP 2
        \begin{minipage}[b]{0.45\linewidth}
            \centering
            \makebox[\textwidth][c]{
            \begin{tikzpicture}[
                every node/.style={font=\footnotesize},
                box/.style={rectangle, draw, thick}
            ]
            \pgfmathsetmacro{\d}{1.75}
            \pgfmathsetmacro{\c}{1}
            \pgfmathsetmacro{\a}{2}
            \pgfmathsetmacro{\b}{2}
            % Nodi
            \node[fill=USPcol,circle,draw,thick] (uspl) at (-\a,0) {$2n$};
            \node[fill=USPcol,circle,draw,thick] (uspr) at (\a, 0) {$2n$};
            \node[fill=SUcol,circle,draw,thick] (center) at (0, 0) {$2n\!+\!1$};
            \node[box] (bot) at (-\c, -\d) {$1$};
            \node[box] (bot2) at (\c, -\d) {$1$};
            \node[box] (botl) at (-\a, -\d) {$1$};
            \node[box] (botr) at (\a, -\d) {$1$};
            % Fondamentali
            \draw[<-,thick,>=stealth] (bot.north) -- node[left] {$Q_{3}$} (center);
            \draw[<-, thick, >=stealth] (uspl.east) -- node[above] {$P_1$} (center.west);
            \draw[<-, thick, >=stealth] (uspr.west) -- node[above] {$P_2$} (center.east);
            \draw[<-,thick,>=stealth] (center) -- node[right] {$\tilde Q$} (bot2.north);
            \draw[-,thick, >=stealth] (uspl) -- node[left] {$R_1$} (botl.north);
            \draw[-,thick, >=stealth] (uspr) -- node[right] {$R_2$} (botr.north);
            % Fermi
            \node at (0, -1.5) {$\Psi_{\!\scriptscriptstyle M_3}$};
            \draw[-,dashed,>=stealth] (bot) to[out=0, in=180, looseness=0] (bot2);
            \end{tikzpicture}
            }
        \end{minipage}
        \\[0.5cm]
        % STEP 3
        \begin{minipage}[b]{0.45\linewidth}
            \centering
            \makebox[\textwidth][c]{
            \begin{tikzpicture}[
                every node/.style={font=\footnotesize},
                box/.style={rectangle, draw, thick}
            ]
            \pgfmathsetmacro{\a}{2}
            \pgfmathsetmacro{\b}{0.75}
            \pgfmathsetmacro{\c}{1.5}
            % Nodi
            \node[fill=USPcol,circle,draw,thick] (uspl) at (-\a,0) {$2n$};
            \node[fill=USPcol,circle,draw,thick] (uspr) at (\a, 0) {$2n$};
            \node[fill=SUcol,circle,draw,thick] (center) at (0, 0) {$2n$};
            \node[box] (bot) at (-\b, -\c) {$1$};
            \node[box] (bot2) at (\b, -\c) {$1$};
            \node[box] (top) at (-\a, \c) {$1$};
            \node[box] (top2) at (\a, \c) {$1$};
            % Fondamentali
            \draw[->,thick,>=stealth] (bot.north) -- node[left] {$\tilde q_{3}$} (center);
            \draw[->, thick, >=stealth] (uspl.east) -- node[above] {$\tilde P_1$} (center.west);
            \draw[->, thick, >=stealth] (uspr.west) -- node[above] {$\tilde P_2$} (center.east);
            \draw[-,thick, >=stealth] (uspl) -- node[left] {$R_1$} (top);
            \draw[-,thick, >=stealth] (uspr) -- node[right] {$R_2$} (top2);
            % Mesoni
            \draw[thick] (bot2.east) to[out=0, in=-90, looseness=1] node[right, overlay] {$N_2$} (uspr.south);
            \draw[thick] (bot2.south west) to[out=-150, in=-90, looseness=1.5] node[left, overlay, xshift=-2pt] {$N_1$} (uspl.south);
            % Fermi
            \draw[-,dashed,>=stealth] (center) -- node[right] {$\psi$} (bot2.north);
            \end{tikzpicture}
            }
        \end{minipage}
        % STEP 4
        \begin{minipage}[b]{0.45\linewidth}
            \centering
            \makebox[\textwidth][c]{
            \begin{tikzpicture}[
                every node/.style={font=\footnotesize},
                box/.style={rectangle, draw, thick},
            ]
            \pgfmathsetmacro{\a}{2}
            \pgfmathsetmacro{\c}{1.5}
            \pgfmathsetmacro{\b}{0.75}
            \pgfmathsetmacro{\x}{0.8}
            \pgfmathsetmacro{\y}{1.6}
            % Nodi
            \node[box] (uspl) at (-\a,0) {$1$};
            \node[box] (uspr) at (\a, 0) {$1$};
            \node[box] (bot) at (-\b, -\c) {$1$};
            \node[box] (bot2) at (\b, -\c) {$1$};
            \node[fill=SUcol,circle,draw,thick] (center) at (0, 0) {$2n$};
            % Antisimmetriche
            \node[box, minimum size=0.2cm] (square1r) at (0.3, 1.5) {};
            \node[box, minimum size=0.2cm] (square2r) at (0.3, 1.78) {};
            \node[box, minimum size=0.2cm] (square1l) at (-0.3, 1.5) {};
            \node[box, minimum size=0.2cm] (square2l) at (-0.3, 1.78) {};
            \node at (\x, \y) {$\tilde A_2$};
            \node at (-\x, \y) {$\tilde A_1$};
            \draw[->,thick,>=stealth] (square1r.south) to[out=-90, in=60, looseness=1.2] ($(center.north)+(0.1,-0.01)$);
            \draw[->,thick,>=stealth] (square1l.south) to[out=-90, in=120, looseness=1.2] ($(center.north)+(-0.1,-0.01)$);
            % Fondamentali
            \draw[->, thick, >=stealth] (uspl.east) -- node[above] {$\tilde q_1$} (center.west);
            \draw[->, thick, >=stealth] (uspr.west) -- node[above] {$\tilde q_2$} (center.east);
            \draw[<-,thick,>=stealth] (center) -- node[right] {$\tilde q$} (bot2.north);
            \draw[->,thick,>=stealth] (bot.north) -- node[left] {$\tilde q_{3}$} (center);
            % Mesoni
            \draw[thick] (bot2.east) to[out=0, in=-90, looseness=1] node[right, overlay] {$M_2$} (uspr.south);
            \draw[thick] (bot2.south west) to[out=-150, in=-90, looseness=1.5] node[left, overlay, xshift=-2pt] {$M_1$} (uspl.south);
            \end{tikzpicture}
        }
        \end{minipage}
    \caption{Deconfinement steps of $\SU(2n+1)$ with two antisymmetrics, three fundamentals and one antifundamental.}
    \label{fig:SUdispari31}
\end{figure}
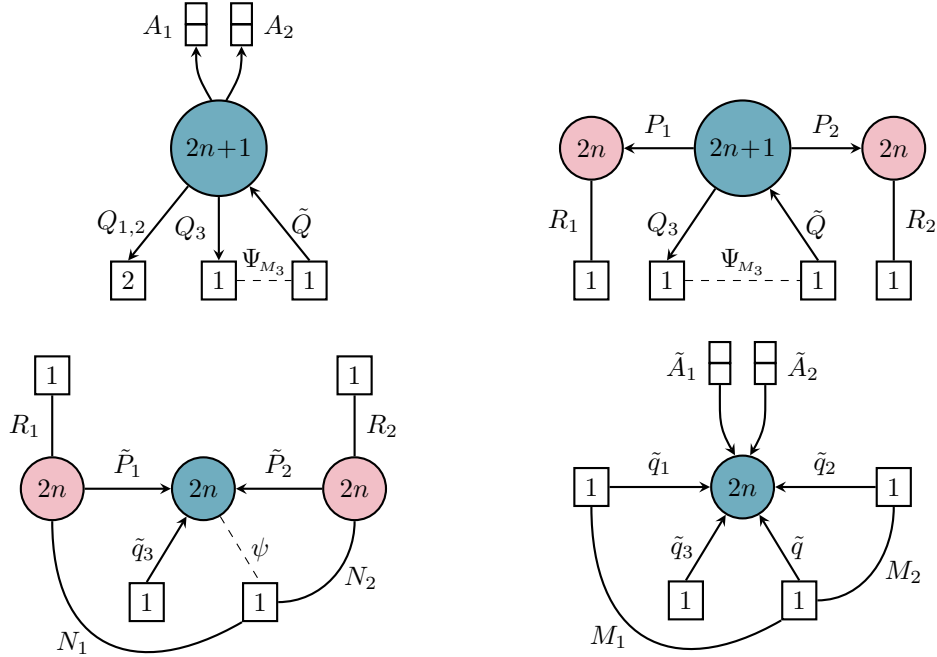
%%%%%%%%%%%%%%%%%%%%%%%%%%%%%%%%%%%%%%%%%%%%%%%%%%%%%%%%%%%%%%%%%%%%%%%

The identity we want is
\begin{eqnarray}
\label{toprovesec3.4}
&&    I^{(3,1,\cdot,2,\cdot)}_{\SU(2n+1)}(\vec u;v;\cdot;\vec t;\cdot)=    \frac{\prod_{j=0}^{2n-2}  \theta(q/(t_1^{j+1}t_2^{2n-j-1} u_1 u_2 u_3 v))}
{       
\prod_{a=1}^{3} \theta (u_a v)  
\prod_{j=0}^{n} \prod_a \theta (t_1^{j} t_2^{n-j} u_a) 
}     
\\
\times 
&&
\frac{1}{ 
\prod_{j=0}^{n-1} (\theta (t_1^{j} t_2^{n-j-1} u_1 u_2 u_3) 
\theta (t_1^{n-j} t_2^{j+1} v))
\prod_{j=0}^{n-2}\prod_{a<b}\theta(t_1^{n-j-1}t_2^{j+1}u_a u_b v)
}\,. \nonumber
\end{eqnarray}

In order to prove this relation we deconfine the two antisymmetric as in Figure \ref{fig:SUdispari31}, obtaining two $\USp(2n)$ gauge nodes, two bifundamentals $P_{1,2}$ and two new fundamentals $R_{1,2}$.
The original antisymmetric chirals $A_{1,2}$ correspond to the combinations $P_{1,2}^2$. Furthermore, two out of the three original fundamentals $Q_{1,2}$ correspond in this quiver to the composites $P_{1,2} R_{1,2}$.
The $J$-term flipping the operator $M$ splits here as
\begin{equation}
J_{\Psi_{M_1}} =\tilde Q P_1 R_1
\, , \quad
J_{\Psi_{M_2}} = \tilde Q P_2 R_2
\, , \quad
J_{\Psi_{M_3}} = \tilde Q Q_3\,,
\end{equation}
signalling the explicit breaking of the $\SU(3)$ flavor symmetry in the process.

The $\SU(2n+1)$ gauge theory has $4n+1$ fundamentals and one antifundamental. It can be dualized according to the rules explained in subsection \ref{SUnGen}, and it gives origin to the third quiver in Figure \ref{fig:SUdispari31}.
In this case the new $J$-term generated by the duality is given by $J_\psi = N_2 \tilde P_2 + N_1 \tilde P_1$.
Furthermore, the two $J$-terms for $\Psi_{M_{1,2}} $ become
$J_{\Psi_{M_{1,2}}} =\tilde Q N_{1,2}$.
The last step consists of dualizing the two $\USp(2n)$ groups, each one with $2n+2$ chiral fundamentals to an LG.
The final quiver is the fourth one in Figure \ref{fig:SUdispari31}.
In this case there are two $J$-terms associated to two Fermi singlets $\Psi_{1,2}^{(0)}$. They read $J_{\Psi_{1}^{(0)}} = \tilde A_{1}^{n-1} \tilde q_{1} \tilde q$ and $J_{\Psi_{2}^{(0)}} = \tilde A_{2}^{n-1} \tilde q_{2} \tilde q$.

Again we are in the situation described in subsection \ref{subsec:3.3}, where after deconfining the tensors, dualizing the unitary gauge group and confining the tensors back, we have, up to an overall charge conjugation, a model already described in \ref{subsec:3.1}. We can the use the duality of such model with an LG one in order to prove the  validity of (\ref{toprovesec3.4}) and to reconstruct (up to flippers) the expected $J$-term (\ref{JPSI3}).
The details of the derivation are straightforward and we leave them to the interested reader.

\subsection{$\SU(2n)$ with two fundamentals and two antifundamentals}

In this case the dual LG model has seven chiral gauge invariant operators corresponding to chiral fields interacting through a $J$-term with a Fermi multiplet.
The field content of the gauge theory and of the dual LG model is represented in the table below.
\begin{equation} 
    \begin{array}{c|c|c|c|c|c|c|c|c|}
    & \SU(2n) & \SU(2) & \SU(2) & \SU(2) & \U(1)_A & \U(1)_Q & \U(1)_{\tilde Q} & \U(1)_{R_0} \\
    \hline
    A & \begin{array}{c}
\square \vspace{-2.85mm} \\
\square 
\end{array}  & \square & \cdot & \cdot & 1 & 0 & 0 & 0 \\
    Q & \square & \cdot & \square & \cdot & 0 & 1 & 0 & 0 \\ 
    \tilde Q & \overline{\square} & \cdot & \cdot & \square & 0 & 0 & 1 & 0 \\       
    \hline
    M & \cdot & \cdot & \square & \square & 0 & 1 & 1 & 0 \\
    B & \cdot & \square & \cdot & \cdot & 1 & 0 & 2 & 0 \\
    T_{n} & \cdot & \otimes_{\mathrm{sym}}^{n} \square& \cdot & \cdot & n & 0 & 0 & 0 \\
    T_{n-1} & \cdot & \otimes_{\mathrm{sym}}^{n-1} \square& \cdot & \cdot & n-1 & 2 & 0 & 0 \\
    P_{n} & \cdot & \otimes_{\mathrm{sym}}^{n-2} \square& \square & \square & n & 1 & 1 & 0 \\
    P_{n+1} & \cdot & \otimes_{\mathrm{sym}}^{n-3} \square& \cdot & \cdot & n+1 & 0 & 2 & 0 \\
    R & \cdot & \otimes_{\mathrm{sym}}^{n-4} \square& \cdot & \cdot & n & 2 & 2 & 0 \\
    \Psi & \cdot & \otimes_{\mathrm{sym}}^{2n-4} \square& \cdot & \cdot & -2n & -2 & -2 & 2 \\
    \end{array}
\end{equation}

The chirals $B$, $M$, $T_j$, $P_j$ and $R$ correspond to the following gauge invariant combinations of the charged field
\begin{equation}
    \begin{aligned}
        &T_n = A^n\,,\\
        &T_{n-1} = A^{n-1} Q^2\,,
    \end{aligned}
    \quad
    \begin{aligned}
        &P_n = A^{n-1} (A \tilde Q) Q\,,\\
        &P_{n+1} = A^{n-1} (A \tilde Q)^2\,,
    \end{aligned}
    \quad
    \begin{aligned}
        &M = Q \tilde Q\,,\\
        &B=A \tilde Q^2\,,
    \end{aligned}
    \quad
    \begin{aligned}
        &R = A^{n-2} (A \tilde Q)^2 Q^2\,,\\
        &\phantom{x}
    \end{aligned}
\end{equation}
while the $J$-term is 
\begin{equation}
\label{JPSI4}
J_{\Psi} = M^2 T_n^2 + M P_{n}  T_n + T_n R + T_{n-1} P_{n+1}  + P_{n} ^2 + B T_{n-1} T_n \,.
\end{equation}
The 't Hooft anomalies match in the two phases and are explicitly
\begin{align}
    &\kappa_{\SU(2)_A^2} = \tfrac{n(2n-1)}{2}\,,\quad
    &&\kappa_{AA} = -\kappa_{AR_0} = 2n(2n-1)\,,\nonumber \\
    &\kappa_{\SU(2)_Q^2} =\kappa_{\SU(2)_{\tilde Q^2}} = n\,,\quad
    &&\kappa_{QQ} = -\kappa_{QR_0} =\kappa_{\tilde Q\tilde Q} = -\kappa_{\tilde QR_0} = 4n\,,\\  
    &\kappa_{R_0 R_0}  = 6n+1\,,\quad
    &&\kappa_{AQ} =\kappa_{A\tilde Q} = 0\,.\nonumber
\end{align}

In the following we show the matching of the elliptic genera along the same lines of the discussion performed in the other examples.
In this case we will flip the meson $M$ and the baryons $B$, $\Pf A_{1,2}$ in the dual LG, and we will show that the expected $J$-term in (\ref{JPSI4}) can be reconstructed by our procedure using the $J$-term expected from the duality of subsection \ref{subsec:3.1}.

\begin{figure}[h!]
    \centering
        % STEP 1
        \begin{minipage}[b]{0.45\linewidth}
            \centering
            \makebox[\textwidth][c]{
            \hspace{-0.1em}
            \begin{tikzpicture}[
                every node/.style={font=\footnotesize},
                box/.style={rectangle, draw, thick},
                box2/.style={rectangle, draw, dashed}
            ]
            \pgfmathsetmacro{\c}{1.75}
            \pgfmathsetmacro{\b}{0.75}
            \pgfmathsetmacro{\x}{0.8}
            \pgfmathsetmacro{\y}{1.6}
            % Nodi
            \node[box] (bot) at (-\b, -\c) {$2$};
            \node[box] (bot2) at (\b, -\c) {$2$};
            \node[fill=SUcol,circle,draw,thick] (center) at (0, 0) {$2n$};
            % Antisimmetriche
            \node[box, minimum size=0.2cm] (square1r) at (0.3, 1.5) {};
            \node[box, minimum size=0.2cm] (square2r) at (0.3, 1.78) {};
            \node[box, minimum size=0.2cm] (square1l) at (-0.3, 1.5) {};
            \node[box, minimum size=0.2cm] (square2l) at (-0.3, 1.78) {};
            \node at (\x, \y) {$A_2$};
            \node at (-\x, \y) {$A_1$};
            \draw[<-,thick,>=stealth] (square1r.south) to[out=-90, in=60, looseness=1.2] ($(center.north)+(0.1,-0.01)$);
            \draw[<-,thick,>=stealth] (square1l.south) to[out=-90, in=120, looseness=1.2] ($(center.north)+(-0.1,-0.01)$);
            % Fondamentali
            \draw[<-,thick,>=stealth] (bot.north) -- node[left] {$Q$} (center);
            \draw[<-,thick,>=stealth] (center) -- node[right] {$\tilde Q$} (bot2.north);
            % Fermi
            \draw[dashed] (bot) -- node[above] {$\Psi_{\!\scriptscriptstyle M}$} (bot2);
            % Fermi antisimmetrici
            %\pgfmathsetmacro{\u}{1.5}
            %\pgfmathsetmacro{\v}{1}
            %\pgfmathsetmacro{\w}{0.28}
            %\node[circle, fill=black, inner sep=2pt] (dot1) at (\u, -\v) {};
            %\node[circle, fill=black, inner sep=2pt] (dot2) at (\u+0.5, -\v) {};
            %\node at (\u+0.5, -\v+\w+0.1) {$\Psi_{\!\scriptscriptstyle B_2}$};
            %\node at (\u, -\v+\w+0.1) {$\Psi_{\!\scriptscriptstyle B_1}$};
            %\draw[dashed] (dot1.south) to[out=-90, in=0, looseness=1] ($(bot2.east)+(0,0.05)$);
            %\draw[dashed] (dot2.south) to[out=-90, in=0, looseness=1.1] ($(bot2.east)+(0,-0.05)$);
            \end{tikzpicture}
            }
        \end{minipage}
        % STEP 2
        \begin{minipage}[b]{0.45\linewidth}
            \centering
            \makebox[\textwidth][c]{
            \begin{tikzpicture}[
                every node/.style={font=\footnotesize},
                box/.style={rectangle, draw, thick}
            ]
            \pgfmathsetmacro{\d}{1.75}
            \pgfmathsetmacro{\c}{0.75}
            \pgfmathsetmacro{\a}{2}
            \pgfmathsetmacro{\b}{2}
            % Nodi
            \node[fill=USPcol,circle,draw,thick] (uspl) at (-\a,1) {$2n\!-\!2$};
            \node[fill=USPcol,circle,draw,thick] (uspr) at (\a,1) {$2n\!-\!2$};
            \node[fill=SUcol,circle,draw,thick] (center) at (0, 0) {$2n$};
            \node[box] (bot) at (-\c, -\d) {$2$};
            \node[box] (bot2) at (\c, -\d) {$2$};
            % Fondamentali
            \draw[<-,thick,>=stealth] (bot.north) -- node[left] {$Q$} (center);
            \draw[<-, thick, >=stealth] (uspl.south east) -- node[above, xshift=4pt] {$P_1$} (center.west);
            \draw[<-, thick, >=stealth] (uspr.south west) -- node[above, xshift=-4pt] {$P_2$} (center.east);
            \draw[<-,thick,>=stealth] (center) -- node[right] {$\tilde Q$} (bot2.north);
            % Fermi
            \draw[dashed] (bot) -- node[above] {$\Psi_{\!\scriptscriptstyle M}$} (bot2);
            % Fermi antisimmetrici
            %\pgfmathsetmacro{\u}{1.5}
            %\pgfmathsetmacro{\v}{1}
            %\pgfmathsetmacro{\w}{0.28}
            %\node[circle, fill=black, inner sep=2pt] (dot1) at (\u, -\v) {};
            %\node[circle, fill=black, inner sep=2pt] (dot2) at (\u+0.5, -\v) {};
            %\node at (\u+0.5, -\v+\w+0.1) {$\Psi_{\!\scriptscriptstyle B_2}$};
            %\node at (\u, -\v+\w+0.1) {$\Psi_{\!\scriptscriptstyle B_1}$};
            %\draw[dashed] (dot1.south) to[out=-90, in=0, looseness=1] ($(bot2.east)+(0,0.05)$);
            %\draw[dashed] (dot2.south) to[out=-90, in=0, looseness=1.1] ($(bot2.east)+(0,-0.05)$);
            \end{tikzpicture}
            }
        \end{minipage}
        \\[0.5cm]
        % STEP 3
        \begin{minipage}[b]{0.45\linewidth}
            \centering
            \vspace{4pt}
            \makebox[\textwidth][c]{
            \begin{tikzpicture}[
                every node/.style={font=\footnotesize},
                box/.style={rectangle, draw, thick}
            ]
            \pgfmathsetmacro{\a}{2}
            \pgfmathsetmacro{\b}{0.75}
            % Nodi
            \node[fill=USPcol,circle,draw,thick] (uspl) at (-\a,1) {$2n\!-\!2$};
            \node[fill=USPcol,circle,draw,thick] (uspr) at (\a, 1) {$2n\!-\!2$};
            \node[fill=SUcol,circle,draw,thick] (center) at (0, 0) {$2n\!-\!2$};
            \node[box] (bot) at (-\b, -2) {$2$};
            \node[box] (bot2) at (\b, -2) {$2$};
            % Fondamentali
            \draw[->,thick,>=stealth] (bot.north) -- node[left] {$\tilde q$} (center);
            \draw[->, thick, >=stealth] (uspl.south east) -- node[above, xshift=4pt] {$\tilde P_1$} (center.west);
            \draw[->, thick, >=stealth] (uspr.south west) -- node[above, xshift=-4pt] {$\tilde P_2$} (center.east);
            % Mesoni
            \draw[thick] (bot2.east) to[out=0, in=-90, looseness=1] node[right, overlay, xshift=0pt] {$N_2$} (uspr.south);
            \draw[thick] (bot2.south west) to[out=-150, in=-90, looseness=1.5] node[left, overlay, xshift=-7pt, yshift=10pt] {$N_1$} (uspl.south);
            % Fermi
            \draw[-,dashed,>=stealth] (center) -- node[right] {$\psi$} (bot2.north);
            % Fermi antisimmetrici
            %\pgfmathsetmacro{\u}{2.5}
            %\pgfmathsetmacro{\v}{1}
            %\pgfmathsetmacro{\w}{0.28}
            %\node[circle, fill=black, inner sep=2pt] (dot1) at (\u, -\v) {};
            %\node[circle, fill=black, inner sep=2pt] (dot2) at (\u+0.5, -\v) {};
            %\node at (\u+0.5, -\v+\w+0.1) {$\Psi_{\!\scriptscriptstyle B_2}$};
            %\node at (\u, -\v+\w+0.1) {$\Psi_{\!\scriptscriptstyle B_1}$};
            %\draw[dashed] (dot1.south) to[out=-90, in=0, looseness=1] ($(bot2.east)+(0,-0.05)$);
            %\draw[dashed] (dot2.south) to[out=-90, in=0, looseness=1.1] ($(bot2.east)+(0,-0.1)$);
            \end{tikzpicture}
            }
        \end{minipage}
        % STEP 4
        \begin{minipage}[b]{0.45\linewidth}
            \centering
            \makebox[\textwidth][c]{
            \begin{tikzpicture}[
                every node/.style={font=\footnotesize},
                box/.style={rectangle, draw, thick},
            ]
            \pgfmathsetmacro{\c}{2}
            \pgfmathsetmacro{\b}{0.75}
            \pgfmathsetmacro{\x}{0.8}
            \pgfmathsetmacro{\y}{1.6}
            \pgfmathsetmacro{\u}{2}
            \pgfmathsetmacro{\v}{1}
            \pgfmathsetmacro{\w}{0.28}
            % Nodi
            \node[box] (bot) at (-\b, -\c) {$2$};
            \node[box] (bot2) at (\b, -\c) {$2$};
            \node[fill=SUcol,circle,draw,thick] (center) at (0, 0) {$2n\!-\!2$};
            % Antisimmetriche
            \node[box, minimum size=0.2cm] (square1r) at (0.3, 1.5) {};
            \node[box, minimum size=0.2cm] (square2r) at (0.3, 1.78) {};
            \node[box, minimum size=0.2cm] (square1l) at (-0.3, 1.5) {};
            \node[box, minimum size=0.2cm] (square2l) at (-0.3, 1.78) {};
            \node at (\x, \y) {$\tilde A_2$};
            \node at (-\x, \y) {$\tilde A_1$};
            \draw[->,thick,>=stealth] (square1r.south) to[out=-90, in=60, looseness=1.2] ($(center.north)+(0.1,-0.01)$);
            \draw[->,thick,>=stealth] (square1l.south) to[out=-90, in=120, looseness=1.2] ($(center.north)+(-0.1,-0.01)$);
            % Fondamentali
            \draw[<-,thick,>=stealth] (center) -- node[right] {$\tilde p$} (bot2.north);
            \draw[->,thick,>=stealth] (bot.north) -- node[left] {$\tilde q$} (center);
            \end{tikzpicture}
        }
        \end{minipage}
    \caption{Deconfinement steps of $\SU(2n)$ with two antisymmetrics, two fundamentals and two antifundamentals.}
    \label{fig:SUpari22}
\end{figure}
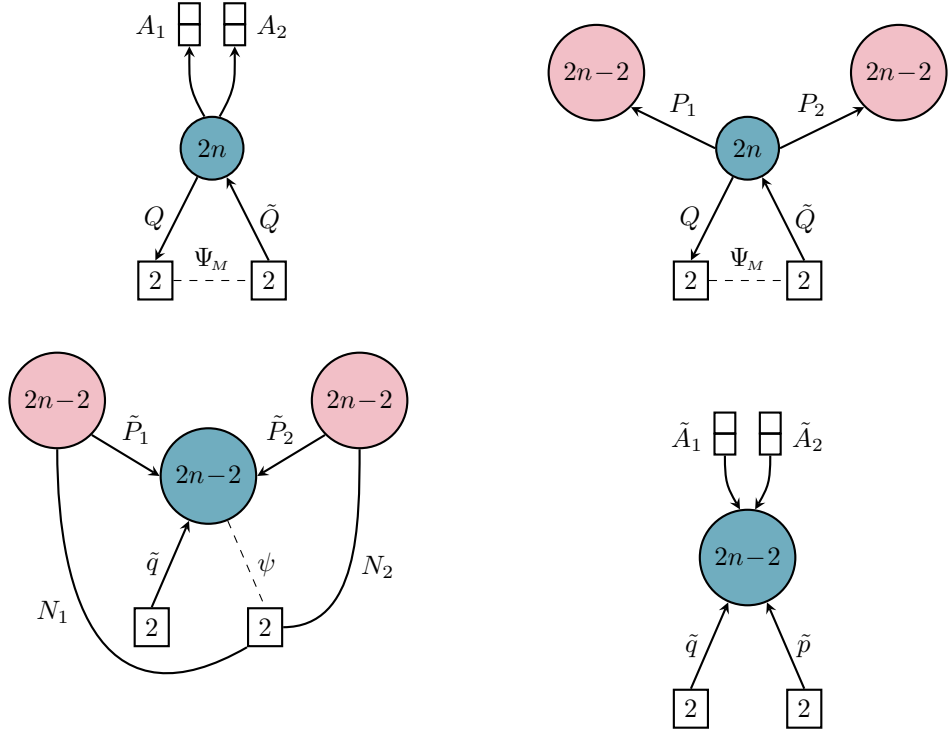

At the level of the elliptic genus the identity
associated to this duality is
\begin{eqnarray}
\label{toprovesec3.5}
&&
I^{(2,2,\cdot,2,\cdot)}_{\SU(2n)}(\vec u;\vec v;\cdot;\vec t;\cdot)=    \frac{\prod_{j=0}^{2n-4}  \theta(q/(t_1^{j+2}t_2^{2n-j-2} u_1 u_2 v_1 v_2))}
{       
\prod_{a,b=1}^{2} \theta (u_a v_b) 
\theta (t_{1,2} v_1 v_2) 
\prod_{j=0}^{n}  \theta (t_1^{j} t_2^{n-j}) 
\prod_{j=0}^{n-1} \theta(t_1^{n-1-j} t_2^j u_1 u_2)
}     
\nonumber \\
&& 
\times
\frac{1}{
\prod_{j=0}^{n-4} \theta(t_1^{n-j-2}t_2^{j+2} u_1 u_2 v_1 v_2)
\prod_{j=0}^{n-3} \theta(t_1^{n-j-1}t_2^{j+2} v_1 v_2)
\prod_{j=0}^{n-2} \prod_{a,b=1}^2 \theta(t_1^{n-j-1}t_2^{j+1} u_a v_b)
}\,. \nonumber 
\\
\end{eqnarray}
In order to prove this relation we deconfine the two antisymmetric as in Figure \ref{fig:SUpari22}, obtaining two $\USp(2n-2)$ gauge nodes, with two bifundamentals $P_{1,2}$.
The original antisymmetric chirals $A_{1,2}$ correspond to the combinations $P_{1,2}^2$.
The $J$-terms at this stage are the ones that flip the operators $M$ and $B$
\begin{equation}
J_{\Psi_{M}} =\tilde Q  Q
\, , \quad
J_{\Psi_{B_1}} = \tilde Q^2 P_1^2
\, , \quad
J_{\Psi_{B_2}} = \tilde Q^2 P_2^2\,.
\end{equation}
The $\SU(2n)$ gauge theory has $4n-2$ fundamentals and two antifundamentals. It can be dualized according to the rules explained in subsection \ref{SUnGen}, giving origin to the third quiver in Figure \ref{fig:SUpari22}.
In this case the new $J$-term generated by the duality is  
$J_{\psi} = N_2 \tilde P_2 + N_1\tilde P_1$.
In addition, the two $J$-terms for $\Psi_{B_{1}}$ and $\Psi_{B_{2}}$
become $J_{\Psi_{B_{1}}} = N_{1}^2$.
and  $J_{\Psi_{B_{2}}} = N_{2}^2$ respectively.

The last step consists in dualizing the two $\USp(2n-2)$ groups, each one with $2n$ chiral fundamentals, to an LG model.
The final quiver is the fourth one in Figure \ref{fig:SUpari22}.
In this case there are two $J$-terms associated to two Fermi singlets $\Psi_{1,2}^{(0)}$. They read $J_{\Psi_{1,2}^{(0)}} = \tilde A_{1,2}^{n-2} \tilde p^2$. 

Similarly to the cases discussed above we have, up to an overall charge conjugation, a model already described in \ref{subsec:3.1}. We can the use the duality of such model with an LG one in order to prove the validity of (\ref{toprovesec3.5}) and to reconstruct (up to flippers) the expected $J$-term (\ref{JPSI4}).
The details of the derivation are straightforward and we leave them to the interested reader.

%%%%%%%%%%%%%%%%%%%%%%%%%%%%%%%%%%%%%%%%%%%%%%%%%%%%%%%%%%%%%%%%%%%%%%%
% FIGURE: SU(2n) w 2A 2F+2aF deconfinement
%%%%%%%%%%%%%%%%%%%%%%%%%%%%%%%%%%%%%%%%%%%%%%%%%%%%%%%%%%%%%%%%%%%%%%%

\subsection{$\SU(2n+1)$ with two fundamentals and two antifundamentals}

In this case the dual LG model has six chiral gauge invariant operators corresponding to chiral fields interacting through a $J$-term with a Fermi multiplet.
The field content of the gauge theory and of the dual LG model is represented in the table below.
\begin{equation} 
    \begin{array}{c|c|c|c|c|c|c|c|c|}
    & \SU(2n+1) & \SU(2) & \SU(2) & \SU(2) & \U(1)_A & \U(1)_Q & \U(1)_{\tilde Q} & \U(1)_{R_0} \\
    \hline
    A & \begin{array}{c}
\square \vspace{-2.85mm} \\
\square 
\end{array}  & \square & \cdot & \cdot & 1 & 0 & 0 & 0 \\
    Q & \square & \cdot & \square & \cdot & 0 & 1 & 0 & 0 \\ 
    \tilde Q & \overline{\square} & \cdot & \cdot & \square & 0 & 0 & 1 & 0 \\       
    \hline
    M & \cdot & \cdot & \square & \square & 0 & 1 & 1 & 0 \\
    B & \cdot & \square & \cdot & \cdot & 1 & 0 & 2 & 0 \\
    T_{n} & \cdot & \otimes_{\mathrm{sym}}^{n} \square& \square & \cdot & n & 1 & 0 & 0 \\
    T_{n-1} & \cdot & \otimes_{\mathrm{sym}}^{n-3} \square& \square & \cdot & n+1 & 1 & 2 & 0 \\
    P_{n} & \cdot & \otimes_{\mathrm{sym}}^{n-1} \square& \cdot & \square & n+1 & 0 & 1 & 0 \\
    P_{n-1} & \cdot & \otimes_{\mathrm{sym}}^{n-2} \square& \cdot & \square & n & 2 & 1 & 0 \\
    \Psi & \cdot & \otimes_{\mathrm{sym}}^{2n-3} \square& \cdot & \cdot & -2n-1 & -2 & -2 & 2 \\
    \end{array}
\end{equation}

The chirals $T_j$, $P_j$ correspond to the following gauge invariant combinations of the charged fields
\begin{equation}
T_n = A^n Q\,,\quad
T_{n-1} = A^{n-1} (A \tilde Q)^2 Q\,, \quad
P_n = A^n (A \tilde Q)\,, \quad
P_{n-1} = A^{n-1} (A \tilde Q) Q^2\,.
\end{equation}
In addition there are a meson $M= Q \tilde Q$ and the baryons 
$B_{1,2} = A_{1,2} \tilde Q^2$,
while the expected $J$-term compatible with the global symmetry is
\begin{equation}
\label{JPSI5}
J_{\Psi} = M T_n P_n + B T_n^2 + T_n T_{n-1} + P_n P_{n-1}\,.
\end{equation}
We have matched the 't Hooft anomalies in the two phases. Explicitly we have
\begin{align}
    &\kappa_{\SU(2)_A^2} = \tfrac{n(2n+1)}{2}\,,\quad
    &&\kappa_{AA} = -\kappa_{AR_0} = 2n(2n+1)\,,\nonumber \\
    &\kappa_{\SU(2)_Q^2} =\kappa_{\SU(2)_{\tilde Q^2}} = \tfrac{2n+1}{2}\,,\quad
    &&\kappa_{QQ} = -\kappa_{QR_0} =\kappa_{\tilde Q\tilde Q} = -\kappa_{\tilde QR_0} = 4n+2\,,\\  
    &\kappa_{R_0 R_0}  = 6n+4\,,\quad
    &&\kappa_{AQ} =\kappa_{A\tilde Q} = 0\,.\nonumber
\end{align}

In the following we show the matching of the elliptic genera 
along the same lines of the discussion performed in the other examples.
In this case we will flip the combinations $Q \tilde Q$, $A_{1,2} \tilde Q^2$ and
 $A_{1,2}^n Q$ and we will show that the expected $J$-term in (\ref{JPSI5}) can be reconstructed by our procedure starting from the $J$-term expected from the duality of subsection \ref{subsec:3.2}.

%%%%%%%%%%%%%%%%%%%%%%%%%%%%%%%%%%%%%%%%%%%%%%%%%%%%%%%%%%%%%%%%%%%%%%%
% FIGURE: SU(2n+1) w 2A 2F+2aF deconfinement
\begin{figure}[h!]
    \centering
        % STEP 1
        \begin{minipage}[b]{0.45\linewidth}
            \centering
            \makebox[\textwidth][c]{
            \hspace{-0.1em}
            \begin{tikzpicture}[
                every node/.style={font=\footnotesize},
                box/.style={rectangle, draw, thick}
            ]
            \pgfmathsetmacro{\c}{1.75}
            \pgfmathsetmacro{\b}{0.75}
            \pgfmathsetmacro{\x}{0.8}
            \pgfmathsetmacro{\y}{1.6}
            % Nodi
            \node[box] (bot) at (-\b, -\c) {$2$};
            \node[box] (bot2) at (\b, -\c) {$2$};
            \node[fill=SUcol,circle,draw,thick] (center) at (0, 0) {$2n\!+\!1$};
            % Antisimmetriche
            \node[box, minimum size=0.2cm] (square1r) at (0.3, 1.5) {};
            \node[box, minimum size=0.2cm] (square2r) at (0.3, 1.78) {};
            \node[box, minimum size=0.2cm] (square1l) at (-0.3, 1.5) {};
            \node[box, minimum size=0.2cm] (square2l) at (-0.3, 1.78) {};
            \node at (\x, \y) {$A_2$};
            \node at (-\x, \y) {$A_1$};
            \draw[<-,thick,>=stealth] (square1r.south) to[out=-90, in=60, looseness=1.2] ($(center.north)+(0.1,-0.01)$);
            \draw[<-,thick,>=stealth] (square1l.south) to[out=-90, in=120, looseness=1.2] ($(center.north)+(-0.1,-0.01)$);
            % Fondamentali
            \draw[<-,thick,>=stealth] (bot.north) -- node[left] {$Q$} (center);
            \draw[<-,thick,>=stealth] (center) -- node[right] {$\tilde Q$} (bot2.north);
            % Fermi
            \draw[dashed] (bot) -- node[above] {$\Psi_{\!\scriptscriptstyle M}$} (bot2);
            % Fermi antisimmetrici
            %\pgfmathsetmacro{\u}{1.5}
            %\pgfmathsetmacro{\v}{1}
            %\pgfmathsetmacro{\w}{0.28}
            %\node[circle, fill=black, inner sep=2pt] (dot1) at (\u, -\v) {};
            %\node[circle, fill=black, inner sep=2pt] (dot2) at (\u+0.5, -\v) {};
            %\node at (\u+0.5, -\v+\w+0.1) {$\Psi_{\!\scriptscriptstyle B_2}$};
            %\node at (\u, -\v+\w+0.1) {$\Psi_{\!\scriptscriptstyle B_1}$};
            %\draw[dashed] (dot1.south) to[out=-90, in=0, looseness=1] ($(bot2.east)+(0,0.05)$);
            %\draw[dashed] (dot2.south) to[out=-90, in=0, looseness=1.1] ($(bot2.east)+(0,-0.05)$);
            \end{tikzpicture}
            }
        \end{minipage}
        % STEP 2
        \begin{minipage}[b]{0.45\linewidth}
            \centering
            \makebox[\textwidth][c]{
            \begin{tikzpicture}[
                every node/.style={font=\footnotesize},
                box/.style={rectangle, draw, thick}
            ]
            \pgfmathsetmacro{\d}{1.75}
            \pgfmathsetmacro{\c}{0.75}
            \pgfmathsetmacro{\a}{2}
            \pgfmathsetmacro{\b}{2}
            \pgfmathsetmacro{\x}{0.8}
            \pgfmathsetmacro{\y}{1.6}
            % Nodi
            \node[fill=USPcol,circle,draw,thick] (uspl) at (-\a,0) {$2n$};
            \node[fill=USPcol,circle,draw,thick] (uspr) at (\a, 0) {$2n$};
            \node[fill=SUcol,circle,draw,thick] (center) at (0, 0) {$2n$+1};
            \node[box] (botl) at (-\a, -\d) {$1$};
            \node[box] (botr) at (\a, -\d) {$1$};
            \node[box] (bot) at (0, -\d) {$2$};
            % Fondamentali
            \draw[-,thick,>=stealth] (botl.north) -- node[left] {$R_{1}$} (uspl);
            \draw[-,thick,>=stealth] (botr.north) -- node[right] {$R_{2}$} (uspr);
            \draw[<-, thick, >=stealth] (uspl.east) -- node[above] {$P_1$} (center.west);
            \draw[<-, thick, >=stealth] (uspr.west) -- node[above] {$P_2$} (center.east);
            \draw[<-,thick,>=stealth] (center) -- node[right] {$\tilde Q$} (bot.north);
            % Fermi
            \draw[dashed] (bot) -- node[above] {$\Psi_{\!\scriptscriptstyle M_1}$} (botl);
            \draw[dashed] (bot) -- node[above] {$\Psi_{\!\scriptscriptstyle M_2}$} (botr);
            % Fermi antisimmetrici
            %\pgfmathsetmacro{\u}{-0.3}
            %\pgfmathsetmacro{\v}{3.2}
            %\pgfmathsetmacro{\w}{0.28}
            %\pgfmathsetmacro{\cacca}{0.5}
            %\node[circle, fill=black, inner sep=2pt] (dot1) at (\u, -\v+\w) {};
            %\node[circle, fill=black, inner sep=2pt] (dot2) at (\u+0.6, -\v+\w) {};
            %\node at (\u-\cacca, -\v+\w) {$\Psi_{\!\scriptscriptstyle B_1}$};
            %\node at (\u+0.6+\cacca, -\v+\w) {$\Psi_{\!\scriptscriptstyle B_2}$};
            %\draw[dashed] (dot1.north) to[out=90, in=-120, looseness=1] ($(bot.south)+(-0.1,0)$);
            %\draw[dashed] (dot2.north) to[out=90, in=-60, looseness=1] ($(bot.south)+(0.1,0)$);
            \end{tikzpicture}
            }
        \end{minipage}
        \\[0.5cm]
        % STEP 3
        \begin{minipage}[b]{0.45\linewidth}
            \centering
            \makebox[\textwidth][c]{
            \begin{tikzpicture}[
                every node/.style={font=\footnotesize},
                box/.style={rectangle, draw, thick}
            ]
            \pgfmathsetmacro{\d}{1.75}
            \pgfmathsetmacro{\c}{0.75}
            \pgfmathsetmacro{\a}{2}
            \pgfmathsetmacro{\b}{2}
            \pgfmathsetmacro{\f}{1}
            \pgfmathsetmacro{\h}{0}
            % Nodi
            \node[fill=USPcol,circle,draw,thick] (uspl) at (-\a+\f,\f) {$2n$};
            \node[fill=USPcol,circle,draw,thick] (uspr) at (\a+\f, \f) {$2n$};
            \node[fill=SUcol,circle,draw,thick] (center) at (\f, \f) {$2n\!-\!1$};
            \node[box] (botl) at (-\a+\f-\h, -\d) {$1$};
            \node[box] (botr) at (\a+\f+\h, -\d) {$1$};
            \node[box] (bot) at (\f, -\d+\f) {$2$};
            % Fondamentali
            \draw[-,thick,>=stealth] (botl) -- node[left] {$R_{1}$} (uspl);
            \draw[-,thick,>=stealth] (botr) -- node[right] {$R_{2}$} (uspr);
            \draw[->, thick, >=stealth] (uspl.east) -- node[above] {$\tilde P_1$} (center.west);
            \draw[->, thick, >=stealth] (uspr.west) -- node[above] {$\tilde P_2$} (center.east);
            % Mesoni
            \draw[->,thick,>=stealth] (uspr.south west) to[out=-120, in=30, looseness=1] node[right, xshift=2pt] {$N_2$} (bot.east);
            \draw[->,thick,>=stealth] (uspl.south east) to[out=-60, in=150, looseness=1] node[left, xshift=-2pt] {$N_1$} (bot.west);
            % Fermi
            \draw[dashed] (center) -- node[right] {$\psi$} (bot.north);
            \draw[dashed] (bot.south east) to[out=-60, in=180, looseness=1] node[above, xshift=2pt] {$\Psi_{\!\scriptscriptstyle M_2}$} (botr.west);
            \draw[dashed] (bot.south west) to[out=-120, in=0, looseness=1] node[above, xshift=-2pt] {$\Psi_{\!\scriptscriptstyle M_1}$} (botl.east);
            % Fermi antisimmetrici
            %\pgfmathsetmacro{\x}{0.8}
            %\pgfmathsetmacro{\y}{1.6}
            %\pgfmathsetmacro{\u}{-0.3}
            %\pgfmathsetmacro{\v}{3.2}
            %\pgfmathsetmacro{\w}{0.28}
            %\pgfmathsetmacro{\giuda}{0.45}
            %\node[circle, fill=black, inner sep=2pt] (dot1) at (\u+\f, -\v+\w+\f) {};
            %\node[circle, fill=black, inner sep=2pt] (dot2) at (\u+0.6+\f, -\v+\w+\f) {};
            %\node at (\u+\f-\giuda, -\v+\w+\f) {$\Psi_{\!\scriptscriptstyle B_1}$};
            %\node at (\u+0.6+\f+\giuda, -\v+\w+\f) {$\Psi_{\!\scriptscriptstyle B_2}$};
            %\draw[dashed] (dot1.north) to[out=90, in=-120, looseness=1] ($(bot.south)+(-0.1,0)$);
            %\draw[dashed] (dot2.north) to[out=90, in=-60, looseness=1] ($(bot.south)+(0.1,0)$);
            \end{tikzpicture}
            }
        \end{minipage}
        % STEP 4
        \begin{minipage}[b]{0.45\linewidth}
            \centering
            \makebox[\textwidth][c]{
            \begin{tikzpicture}[
                every node/.style={font=\footnotesize},
                box/.style={rectangle, draw, thick},
            ]
            \pgfmathsetmacro{\c}{1.75}
            \pgfmathsetmacro{\b}{1.5}
            \pgfmathsetmacro{\x}{0.8}
            \pgfmathsetmacro{\y}{1.6}
            \pgfmathsetmacro{\u}{-0.3}
            \pgfmathsetmacro{\v}{3.2}
            \pgfmathsetmacro{\w}{0.28}
            % Nodi
            \node[box] (bot) at (0, -\c) {$2$};
            \node[box] (botl) at (-\b, -\c) {$1$};
            \node[box] (botr) at (\b, -\c) {$1$};
            \node[fill=SUcol,circle,draw,thick] (center) at (0, 0) {$2n\!-\!1$};
            % Antisimmetriche
            \node[box, minimum size=0.2cm] (square1r) at (0.3, 1.5) {};
            \node[box, minimum size=0.2cm] (square2r) at (0.3, 1.78) {};
            \node[box, minimum size=0.2cm] (square1l) at (-0.3, 1.5) {};
            \node[box, minimum size=0.2cm] (square2l) at (-0.3, 1.78) {};
            \node at (\x, \y) {$\tilde A_2$};
            \node at (-\x, \y) {$\tilde A_1$};
            \draw[->,thick,>=stealth] (square1r.south) to[out=-90, in=60, looseness=1.2] ($(center.north)+(0.1,-0.01)$);
            \draw[->,thick,>=stealth] (square1l.south) to[out=-90, in=120, looseness=1.2] ($(center.north)+(-0.1,-0.01)$);
            % Fondamentali
            \draw[<-,thick,>=stealth] (center) -- node[right] {$\tilde p$} (bot.north);
            \draw[->,thick,>=stealth] (botl.north east) -- node[left, xshift=-2pt] {$\tilde q_1$} (center);
            \draw[->,thick,>=stealth] (botr.north west) -- node[right,xshift=2pt] {$\tilde q_2$} (center);
            \end{tikzpicture}
        }
        \end{minipage}
    \caption{Deconfinement steps of $\SU(2n+1)$ with two antisymmetrics, two fundamentals and two antifundamentals.}
    \label{fig:SUdispari22}
\end{figure}
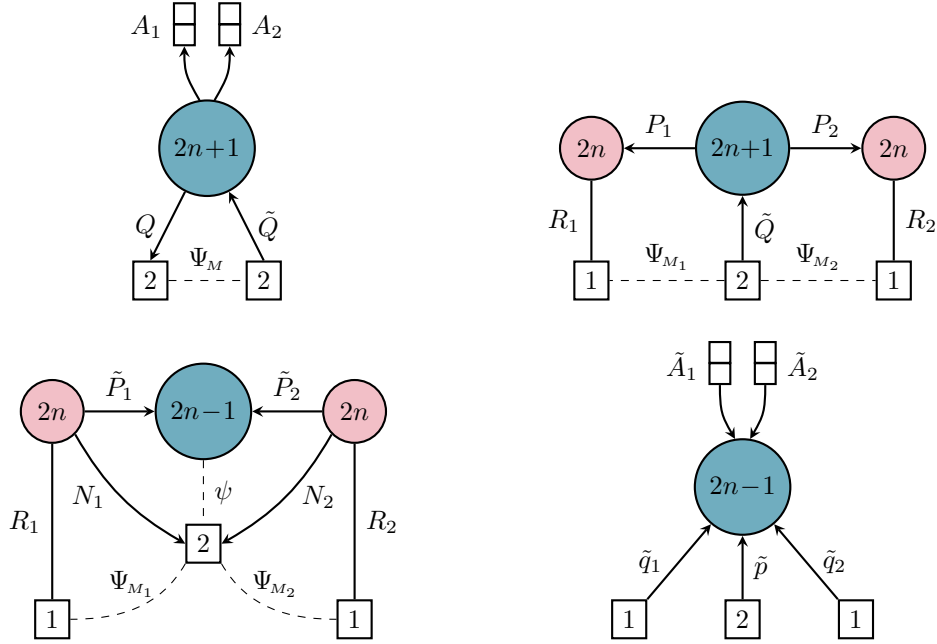
%%%%%%%%%%%%%%%%%%%%%%%%%%%%%%%%%%%%%%%%%%%%%%%%%%%%%%%%%%%%%%%%%%%%%%%

At the level of the elliptic genus the identity
associated to this duality is
\begin{eqnarray}
\label{toprovesec3.6}
&&I^{(2,2,\cdot,2,\cdot)}_{\SU(2n+1)}(\vec u;\vec v;\cdot;\vec t;\cdot)=    \frac{\prod_{j=0}^{2n-3}  \theta(q/(t_1^{j+2}t_2^{2n-j-1} u_1 u_2 v_1 v_2))}
{       
\theta (t_{1,2} v_1 v_2) 
\prod_{a=1}^{2} 
(\prod_{j=0}^{n}   \theta (t_1^{j} t_2^{n-j} u_a)  \cdot
\prod_{j=0}^{n-1} \theta(t_1^{n-j} t_2^{j+1} v_a))
}     
\nonumber \\
&& 
\times
\frac{1}{
\prod_{a,b=1}^{2} \theta (u_a v_b) 
\prod_{a=1}^2 
(\prod_{j=0}^{n-3} \theta(t_1^{n-1-j} t_2^{j+2}u_a v_1 v_2)
\cdot
\prod_{j=0}^{n-2} \prod_{a=1}^2 \theta(t_1^{n-1-j} t_2^{j+1}u_1 u_2 v_a)
)}\,. \nonumber \\
\end{eqnarray}

In order to prove this relation we deconfine the two antisymmetric as in Figure \ref{fig:SUdispari22}, obtaining two $\USp(2n)$ gauge nodes, with two bifundamentals $P_{1,2}$ and two fundamentals $R_{1,2}$.
The original antisymmetric chirals  $A_{1,2}$ correspond to the combinations $P_{1,2}^2$ while the original fundamentals $Q_{1,2}$ correspond to the combinations $P_{1,2} R_{1,2}$.
The $J$-terms at this stage are the ones flipping the operators $M$ and $B$ and they are
\begin{equation}
J_{\Psi_{M_a}} =\tilde Q  P_a R_a
\, , \quad
J_{\Psi_{B_a}} = \tilde Q^2 P_a^2\,,
\end{equation}
for $a=1,2$.
The $\SU(2n+1)$ gauge theory has $4n$ fundamentals and two antifundamentals. It can be dualized according to the rules explained in subsection \ref{SUnGen}, and it gives origin to the third quiver in Figure \ref{fig:SUdispari22}.
In this case the new $J$-term generated by the duality is  
$J_{\psi} = N_2 \tilde P_2 + N_1 \tilde P_1$.
In addition the  $J$-terms for $\Psi_{M_a}$ and $\Psi_{B_a}$
become 
$J_{\Psi_{M_a}} = R_a N_a$ and 
$J_{\Psi_{B_a}} = N_{a}^2$.

The last step consists of dualizing the two $\USp(2n)$ groups, each one with $2n+2$ chiral fundamentals, to an LG.
The final quiver is the fourth one  in  Figure \ref{fig:SUdispari22}.
In this case there are two $J$-terms associated to two Fermi singlets $\Psi_{1,2}^{(0)}$. They read $J_{\Psi_{1,2}^{(0)}} = \tilde A_{1,2}^{n-2} \tilde q_{1,2} \tilde p^2$.

Similarly to the cases discussed above we have, up to an overall charge conjugation, a model already described in \ref{subsec:3.2}. We can the use the duality of such model with an LG in order to prove the  validity of (\ref{toprovesec3.6}) and to reconstruct (up to flippers) the expected $J$-term (\ref{JPSI5}).
The details of the derivation are straightforward and we leave them to the interested reader.

\subsection{$\SU(2n)$ with one fundamentals and three antifundamentals}

In this case the dual LG model has six chiral gauge invariant operators corresponding to chiral fields interacting through a $J$-term with a Fermi multiplet.
The field content of the gauge theory and of the dual LG model is represented in the table below.

\begin{equation} 
    \begin{array}{c|c|c|c|c|c|c|c|}
    & \SU(2n) & \SU(2) & \SU(3) & \U(1)_A & \U(1)_Q & \U(1)_{\tilde Q} & \U(1)_{R_0} \\
    \hline
    A & \begin{array}{c}
\square \vspace{-2.85mm} \\
\square 
\end{array}  & \square & \cdot & 1 & 0 & 0 & 0 \\
    Q & \square & \cdot & \cdot & 0 & 1 & 0 & 0 \\ 
    \tilde Q & \overline{\square} & \cdot & \square & 0 & 0 & 1 & 0 \\       
    \hline
    M & \cdot & \cdot & \square & 0 & 1 & 1 & 0 \\
    B & \cdot & \square & \overline{\square} & 1 & 0 & 2 & 0 \\
    T_{n} & \cdot & \otimes_{\mathrm{sym}}^{n} \square& \cdot & n & 0 & 0 & 0 \\
    P_{n} & \cdot & \otimes_{\mathrm{sym}}^{n-2} \square& \square & n & 1 & 1 & 0 \\
    P_{n+1} & \cdot & \otimes_{\mathrm{sym}}^{n-3} \square& \overline{\square} & n+1 & 0 & 2 & 0 \\
    \Pi_{n+1} & \cdot & \otimes_{\mathrm{sym}}^{n-5} \square& \cdot & n+1 & 1 & 3 & 0 \\
    \Psi & \cdot & \otimes_{\mathrm{sym}}^{2n-5} \square& \cdot & -2n-1 & -1 & -3 & 2 \\
    \end{array}
\end{equation}

The chirals $T_j$, $P_j$ correspond to the following gauge invariant combinations of the charged fields
\begin{equation}
T_n = A^n\,,\quad
P_n = A^{n-1} (A \tilde Q) Q \,, \quad
P_{n+1}= A^{n-1} (A \tilde Q)^2\,, \quad
\Pi_{n+1} = A^{n-2} (A \tilde Q)^3 Q\,,
\end{equation}
In addition there are a meson $M= \tilde Q Q$ and the baryons 
$B_{1,2} = A_{1,2} \tilde Q^2$,
while the expected $J$-term compatible with the global symmetry is
\begin{equation}
\label{JPSI6}
J_{\Psi} = B T_n P_n + P_n P_{n+1} + T_n \Pi_{n+1} + M T_n P_{n+1} \,.
\end{equation}
We checked the matching of the 't Hooft anomalies in the two theories. Explicitly we have
\begin{equation}
    \begin{aligned}
        &\kappa_{\SU(2)^2} = \tfrac{n(2n-1)}{2}\,,\\
        &\kappa_{\SU(3)} = n\,,\\     
        &\kappa_{R_0 R_0}  = 6n+1\,,\\
    \end{aligned}
    \qquad \qquad
    \begin{aligned}
        &\kappa_{AA} = -\kappa_{AR_0} = 2n(2n-1)\,,\\
        &\kappa_{QQ} = -\kappa_{QR_0} =2n\,,\\
        &\kappa_{\tilde Q\tilde Q} = -\kappa_{\tilde QR_0} = 6n\,,\\
        &\kappa_{AQ} =\kappa_{A\tilde Q} = 0\,.\\
    \end{aligned}
\end{equation}
In the following we show the matching of the eblliptic genera 
along the same lines of the discussion performed in the other examples.
In this case we will flip the meson $M$ and the baryons $B_{1,2}$ and $\Pf A_{1,2}$ in the dual LG and we will show that the expected $J$-term in (\ref{JPSI6}) can be reconstructed by our procedure  using the $J$-term 
expected from the duality of subsection \ref{subsec:3.2}.
%%%%%%%%%%%%%%%%%%%%%%%%%%%%%%%%%%%%%%%%%%%%%%%%%%%%%%%%%%%%%%%%%%%%%%%
% FIGURE: SU(2n) w 2A 1F+3aF deconfinement
\begin{figure}[h!]
    \centering
        % STEP 1
        \begin{minipage}[b]{0.45\linewidth}
            \centering
            \makebox[\textwidth][c]{
            \hspace{-0.1em}
            \begin{tikzpicture}[
                every node/.style={font=\footnotesize},
                box/.style={rectangle, draw, thick},
                box2/.style={rectangle, draw, dashed}
            ]
            \pgfmathsetmacro{\c}{1.75}
            \pgfmathsetmacro{\b}{0.75}
            \pgfmathsetmacro{\x}{0.8}
            \pgfmathsetmacro{\y}{1.6}
            % Nodi
            \node[box] (bot) at (-\b, -\c) {$1$};
            \node[box] (bot2) at (\b, -\c) {$3$};
            \node[fill=SUcol,circle,draw,thick] (center) at (0, 0) {$2n$};
            % Antisimmetriche
            \node[box, minimum size=0.2cm] (square1r) at (0.3, 1.5) {};
            \node[box, minimum size=0.2cm] (square2r) at (0.3, 1.78) {};
            \node[box, minimum size=0.2cm] (square1l) at (-0.3, 1.5) {};
            \node[box, minimum size=0.2cm] (square2l) at (-0.3, 1.78) {};
            \node at (\x, \y) {$A_2$};
            \node at (-\x, \y) {$A_1$};
            \draw[<-,thick,>=stealth] (square1r.south) to[out=-90, in=60, looseness=1.2] ($(center.north)+(0.1,-0.01)$);
            \draw[<-,thick,>=stealth] (square1l.south) to[out=-90, in=120, looseness=1.2] ($(center.north)+(-0.1,-0.01)$);
            % Fondamentali
            \draw[<-,thick,>=stealth] (bot.north) -- node[left] {$Q$} (center);
            \draw[<-,thick,>=stealth] (center) -- node[right] {$\tilde Q$} (bot2.north);
            % Fermi
            \draw[dashed] (bot) -- node[above] {$\Psi_{\!\scriptscriptstyle M}$} (bot2);
            % Fermi antisimmetrici
            \pgfmathsetmacro{\u}{1.5}
            \pgfmathsetmacro{\v}{1}
            \pgfmathsetmacro{\w}{0.28}
            \node[box2, minimum size=0.2cm] (ssquare1r) at (\u, -\v) {};
            \node[box2, minimum size=0.2cm] (ssquare2r) at (\u, -\v+\w) {};
            \node[box2, minimum size=0.2cm] (ssquare1l) at (\u+0.5, -\v) {};
            \node[box2, minimum size=0.2cm] (ssquare2l) at (\u+0.5, -\v+\w) {};
            \node at (\u+0.6, -\v+\w+0.5) {$\Psi_{\!\scriptscriptstyle B_2}$};
            \node at (\u-0.1, -\v+\w+0.5) {$\Psi_{\!\scriptscriptstyle B_1}$};
            \draw[dashed] (ssquare1r.south) to[out=-90, in=0, looseness=1] ($(bot2.east)+(0,0.05)$);
            \draw[dashed] (ssquare1l.south) to[out=-90, in=0, looseness=1.1] ($(bot2.east)+(0,-0.05)$);
            \end{tikzpicture}
            }
        \end{minipage}
        % STEP 2
        \begin{minipage}[b]{0.45\linewidth}
            \centering
            \makebox[\textwidth][c]{
            \begin{tikzpicture}[
                every node/.style={font=\footnotesize},
                box/.style={rectangle, draw, thick},
                box2/.style={rectangle, draw, dashed}
            ]
            \pgfmathsetmacro{\d}{1.75}
            \pgfmathsetmacro{\c}{0.75}
            \pgfmathsetmacro{\a}{2}
            \pgfmathsetmacro{\b}{2}
            \pgfmathsetmacro{\merda}{1}
            % Nodi
            \node[fill=USPcol,circle,draw,thick] (uspl) at (-\a,\merda) {$2n\!-\!2$};
            \node[fill=USPcol,circle,draw,thick] (uspr) at (\a, \merda) {$2n\!-\!2$};
            \node[fill=SUcol,circle,draw,thick] (center) at (0, 0) {$2n$};
            \node[box] (bot) at (-\c, -\d) {$1$};
            \node[box] (bot2) at (\c, -\d) {$3$};
            % Fondamentali
            \draw[<-,thick,>=stealth] (bot.north) -- node[left] {$Q$} (center);
            \draw[<-, thick, >=stealth] (uspl.south east) -- node[above] {$P_1$} (center.west);
            \draw[<-, thick, >=stealth] (uspr.south west) -- node[above] {$P_2$} (center.east);
            \draw[<-,thick,>=stealth] (center) -- node[right] {$\tilde Q$} (bot2.north);
            % Fermi
            \draw[dashed] (bot) -- node[above] {$\Psi_{\!\scriptscriptstyle M}$} (bot2);
            % Fermi antisimmetrici
            \pgfmathsetmacro{\u}{1.5}
            \pgfmathsetmacro{\v}{1}
            \pgfmathsetmacro{\w}{0.28}
            \node[box2, minimum size=0.2cm] (ssquare1r) at (\u, -\v) {};
            \node[box2, minimum size=0.2cm] (ssquare2r) at (\u, -\v+\w) {};
            \node[box2, minimum size=0.2cm] (ssquare1l) at (\u+0.5, -\v) {};
            \node[box2, minimum size=0.2cm] (ssquare2l) at (\u+0.5, -\v+\w) {};
            \node at (\u+0.6, -\v+\w+0.5) {$\Psi_{\!\scriptscriptstyle B_2}$};
            \node at (\u-0.1, -\v+\w+0.5) {$\Psi_{\!\scriptscriptstyle B_1}$};
            \draw[dashed] (ssquare1r.south) to[out=-90, in=0, looseness=1] ($(bot2.east)+(0,0.05)$);
            \draw[dashed] (ssquare1l.south) to[out=-90, in=0, looseness=1.1] ($(bot2.east)+(0,-0.05)$);
            \end{tikzpicture}
            }
        \end{minipage}
        \\[0.5cm]
        % STEP 3
        \begin{minipage}[b]{0.45\linewidth}
            \centering
            \makebox[\textwidth][c]{
            \begin{tikzpicture}[
                every node/.style={font=\footnotesize},
                box/.style={rectangle, draw, thick},
                box2/.style={rectangle, draw, dashed}
            ]
            \pgfmathsetmacro{\a}{2}
            \pgfmathsetmacro{\b}{0.75}
            \pgfmathsetmacro{\merda}{1}
            % Nodi
            \node[fill=USPcol,circle,draw,thick] (uspl) at (-\a,\merda) {$2n\!-\!2$};
            \node[fill=USPcol,circle,draw,thick] (uspr) at (\a, \merda) {$2n\!-\!2$};
            \node[fill=SUcol,circle,draw,thick] (center) at (0, 0) {$2n\!-\!3$};
            \node[box] (bot) at (-\b, -2) {$1$};
            \node[box] (bot2) at (\b, -2) {$3$};
            % Fondamentali
            \draw[->,thick,>=stealth] (bot.north) -- node[left] {$\tilde q$} (center);
            \draw[->, thick, >=stealth] (uspl.south east) -- node[above] {$\tilde P_1$} (center.west);
            \draw[->, thick, >=stealth] (uspr.south west) -- node[above] {$\tilde P_2$} (center.east);
            % Mesoni
            \draw[thick] (bot2.east) to[out=0, in=-90, looseness=1] node[left, overlay] {$N_2$} (uspr.south);
            \draw[thick] (bot2.south west) to[out=-150, in=-90, looseness=1.5] node[left, overlay, xshift=-2pt] {$N_1$} (uspl.south);
            % Fermi
            \draw[-,dashed,>=stealth] (center) -- node[right] {$\psi$} (bot2.north);
            % Fermi antisimmetrici
            \pgfmathsetmacro{\u}{2.5}
            \pgfmathsetmacro{\v}{1}
            \pgfmathsetmacro{\w}{0.28}
            \node[box2, minimum size=0.2cm] (ssquare1r) at (\u, -\v) {};
            \node[box2, minimum size=0.2cm] (ssquare2r) at (\u, -\v+\w) {};
            \node[box2, minimum size=0.2cm] (ssquare1l) at (\u+0.5, -\v) {};
            \node[box2, minimum size=0.2cm] (ssquare2l) at (\u+0.5, -\v+\w) {};
            \node at (\u+0.6, -\v+\w+0.5) {$\Psi_{\!\scriptscriptstyle B_2}$};
            \node at (\u-0.1, -\v+\w+0.5) {$\Psi_{\!\scriptscriptstyle B_1}$};
            \draw[dashed] (ssquare1r.south) to[out=-90, in=0, looseness=1] ($(bot2.east)+(0,-0.05)$);
            \draw[dashed] (ssquare1l.south) to[out=-90, in=0, looseness=1.1] ($(bot2.east)+(0,-0.1)$);
            \end{tikzpicture}
            }
            %\caption{}
        \end{minipage}
        % STEP 4
        \begin{minipage}[b]{0.45\linewidth}
            \centering
            \makebox[\textwidth][c]{
            \begin{tikzpicture}[
                every node/.style={font=\footnotesize},
                box/.style={rectangle, draw, thick},
            ]
            \pgfmathsetmacro{\c}{1.75}
            \pgfmathsetmacro{\b}{0.75}
            \pgfmathsetmacro{\x}{0.8}
            \pgfmathsetmacro{\y}{1.6}
            \pgfmathsetmacro{\u}{2}
            \pgfmathsetmacro{\v}{1}
            \pgfmathsetmacro{\w}{0.28}
            % Nodi
            \node[box] (bot) at (-\b, -\c) {$1$};
            \node[box] (bot2) at (\b, -\c) {$3$};
            \node[fill=SUcol,circle,draw,thick] (center) at (0, 0) {$2n\!-\!3$};
            % Antisimmetriche
            \node[box, minimum size=0.2cm] (square1r) at (0.3, 1.5) {};
            \node[box, minimum size=0.2cm] (square2r) at (0.3, 1.78) {};
            \node[box, minimum size=0.2cm] (square1l) at (-0.3, 1.5) {};
            \node[box, minimum size=0.2cm] (square2l) at (-0.3, 1.78) {};
            \node at (\x, \y) {$\tilde A_2$};
            \node at (-\x, \y) {$\tilde A_1$};
            \draw[->,thick,>=stealth] (square1r.south) to[out=-90, in=60, looseness=1.2] ($(center.north)+(0.1,-0.01)$);
            \draw[->,thick,>=stealth] (square1l.south) to[out=-90, in=120, looseness=1.2] ($(center.north)+(-0.1,-0.01)$);
            % Fondamentali
            \draw[<-,thick,>=stealth] (center) -- node[right] {$\tilde p$} (bot2.north);
            \draw[->,thick,>=stealth] (bot.north) -- node[left] {$\tilde q$} (center);
            \end{tikzpicture}
        }
        \end{minipage}
    \caption{Deconfinement steps of $\SU(2n)$ with two antisymmetrics, one fundamental and three antifundamentals.}
    \label{fig:SUpari13}
\end{figure}
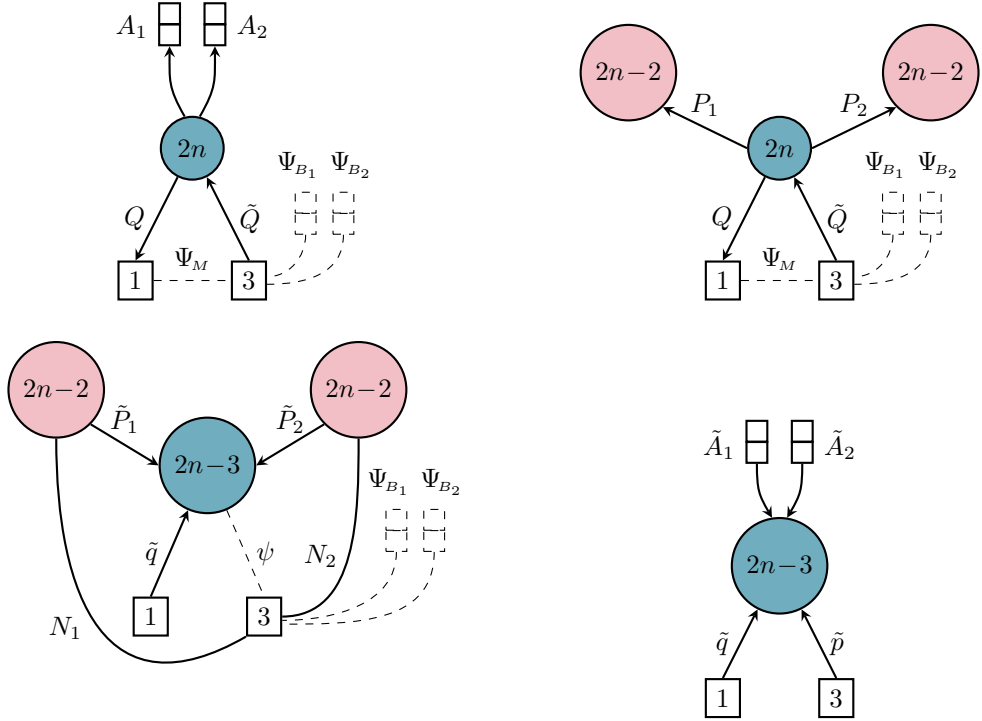
%%%%%%%%%%%%%%%%%%%%%%%%%%%%%%%%%%%%%%%%%%%%%%%%%%%%%%%%%%%%%%%%%%%%%%%
At the level of the elliptic genus the identity
associated to this duality is
\begin{eqnarray}
\label{toprovesec3.7}
&&    I^{(1,3,\cdot,2,\cdot)}_{\SU(2n)}( u;\vec v;\cdot;\vec t;\cdot)=    \frac{\prod_{j=0}^{2n-5}  
\theta(q/(t_1^{j+3}t_2^{2n-j-1} u v_1 v_2 v_3))}
{       
\prod_a  \theta (u v_a) 
\prod_{a<b} (\theta (t_{1,2} v_a v_b) )
\prod_{j=0}^{n-3}   \theta (t_1^{n-1-j} t_2^{j+2} v_a v_b)
}
 \nonumber \\
 && 
\times  \frac{1}{
 \prod_{j=0}^{n-5}   \theta (t_1^{n-j-2} t_2^{j+3} u v_1 v_2 v_3)
 \prod_{j=0}^n \theta(t_1^{n-j} t_2^j)
 \prod_{j=0}^{n-2} \prod_a \theta(t_1^{n-1-j} t_2^{j+1} u v_a)
} \,.
\end{eqnarray}

In order to prove this relation we deconfine the two antisymmetric as in Figure \ref{fig:SUpari13}, obtaining two $\USp(2n-2)$ gauge nodes, with two bifundamentals $P_{1,2}$.
The original antisymmetric chirals  $A_{1,2}$ correspond to the combinations $P_{1,2}^2$.
The $J$-term at this stage are the ones flipping the operators $M$ and $B$ and they read
\begin{equation}
J_{\Psi_{M}} =\tilde Q  Q
\, , \quad
J_{\Psi_{B_{1,2}}} = Q^2 P_{1,2}^2\,.
\end{equation}
The $\SU(2n)$ gauge theory has $4n-1$ fundamentals and one antifundamental. It can be dualized according to the rules explained in subsection \ref{SUnGen}, and it gives origin to the third quiver in Figure \ref{fig:SUpari13}.

In this case the new $J$-term generated by the duality is  
$J_{\psi} = N_2 \tilde P_2 + N_1 \tilde P_1$.
In addition the  $J$-terms for $\Psi_{B_1}$ and $\Psi_{B_2}$
become 
$J_{\Psi_{B_1}} = N_{1}^2$ and  $J_{\Psi_{B_2}} = N_{2}^2$ respectively.

The last step consists of dualizing the two $\USp(2n-2)$ groups, each one with $2n$ chiral fundamentals, to an LG.
The final quiver is the fourth one  in  Figure \ref{fig:SUpari13}.
In this case there are two $J$-terms associated to two Fermi singlets $\Psi_{1,2}^{(0)}$. They read $J_{\Psi_{1,2}^{(0)}} = \tilde A_{1,2}^{n-3} \tilde p^3$.

Again we have, up to an overall charge conjugation, a model already described in \ref{subsec:3.2}. We can the use the duality of such model with an LG in order to prove the  validity of (\ref{toprovesec3.7}) and to reconstruct (up to flippers) the expected $J$-term (\ref{JPSI6}).
The details of the derivation are straightforward and we leave them to the interested reader.

\subsection{$\SU(2n+1)$ with one fundamentals and three antifundamentals}
In this case the dual LG model has six chiral gauge invariant operators corresponding to chiral fields interacting through a $J$-term with a Fermi multiplet.
The field content of the gauge theory and of the dual LG model is represented in the table below.
\begin{equation} 
    \begin{array}{c|c|c|c|c|c|c|c|}
    & \SU(2n+1) & \SU(2) & \SU(3) & \U(1)_A & \U(1)_Q & \U(1)_{\tilde Q} & \U(1)_{R_0} \\
    \hline
    A & \begin{array}{c}
\square \vspace{-2.85mm} \\
\square 
\end{array}  & \square & \cdot & 1 & 0 & 0 & 0 \\
    Q & \square & \cdot & \cdot & 0 & 1 & 0 & 0 \\ 
    \tilde Q & \overline{\square} & \cdot & \square & 0 & 0 & 1 & 0 \\       
    \hline
    M & \cdot & \cdot & \square & 0 & 1 & 1 & 0 \\
    B & \cdot & \square & \overline{\square} & 1 & 0 & 2 & 0 \\
    T_{n} & \cdot & \otimes_{\mathrm{sym}}^{n} \square& \cdot & n & 1 & 0 & 0 \\
    P_{n+1} & \cdot & \otimes_{\mathrm{sym}}^{n-1} \square& \square & n+1 & 0 & 1 & 0 \\
    P_{n+2} & \cdot & \otimes_{\mathrm{sym}}^{n-4} \square& \cdot & n+2 & 0 & 3 & 0 \\
    R & \cdot & \otimes_{\mathrm{sym}}^{n-3} \square& \overline{\square} & n+1 & 1 & 2 & 0 \\
    \Psi & \cdot & \otimes_{\mathrm{sym}}^{2n-4} \square& \cdot & -2n-2 & -1 & -3 & 2 \\
    \end{array}
\end{equation}

The chirals $T_j$, $P_j$ and $R$ correspond to the following gauge invariant combinations of the charged fields
\begin{equation}
T_n = A^n Q\,,\quad
P_{n+1} = A^n (A \tilde Q)\,, \quad
P_{n+2} = A^{n-1} (A \tilde Q)^3\,, \quad
R = A^{n-1} (A \tilde Q)^2 Q\,,
\end{equation}
In additions there are other singlets identified with the meson 
$M= Q \tilde Q$ and the baryons $B_{1,2}=A_{1,2} \tilde Q^2$.
The $J$-term compatible with the global symmetry is 
\begin{equation}
\label{JPSI7}
J_{\Psi} = M P_{n+1}^2 + B P_{n+1} T_n + T_{n} P_{n+2} + P_{n+1} R\,.
\end{equation}
We checked the matching of the 't Hooft anomalies in the two theories. Explicitly we have
\begin{equation}
    \begin{aligned}
        &\kappa_{\SU(2)^2} = \tfrac{n(2n+1)}{2}\,,\\
        &\kappa_{\SU(3)^2} = \tfrac{2n+1}{2}\,\\
        &\kappa_{R_0 R_0}  = 6n+4\,,\\
    \end{aligned}
    \qquad \qquad
    \begin{aligned}
        &\kappa_{AA} = -\kappa_{AR_0} = 2n(2n+1)\,,\\
        &\kappa_{QQ} = -\kappa_{QR_0} = 2n+1\,,\\
        &\kappa_{\tilde Q\tilde Q} = -\kappa_{\tilde QR_0} = 6n+3\,,\\
        &\kappa_{AQ} =\kappa_{A\tilde Q} = 0\,.
    \end{aligned}
\end{equation}
%%%%%%%%%%%%%%%%%%%%%%%%%%%%%%%%%%%%%%%%%%%%%%%%%%%%%%%%%%%%%%%%%%%%%%%
% FIGURE: SU(2n+1) w 2A 1F+3aF deconfinement
\begin{figure}[h!]
    \centering
        % STEP 1
        \begin{minipage}[b]{0.45\linewidth}
            \centering
            \makebox[\textwidth][c]{
            \hspace{-0.1em}
            \begin{tikzpicture}[
                every node/.style={font=\footnotesize},
                box/.style={rectangle, draw, thick},
                box2/.style={rectangle, draw, dashed}
            ]
            \pgfmathsetmacro{\c}{1.75}
            \pgfmathsetmacro{\b}{0.75}
            \pgfmathsetmacro{\x}{0.8}
            \pgfmathsetmacro{\y}{1.6}
            % Nodi
            \node[box] (bot) at (-\b, -\c) {$1$};
            \node[box] (bot2) at (\b, -\c) {$3$};
            \node[fill=SUcol,circle,draw,thick] (center) at (0, 0) {$2n\!+\!1$};
            % Antisimmetriche
            \node[box, minimum size=0.2cm] (square1r) at (0.3, 1.5) {};
            \node[box, minimum size=0.2cm] (square2r) at (0.3, 1.78) {};
            \node[box, minimum size=0.2cm] (square1l) at (-0.3, 1.5) {};
            \node[box, minimum size=0.2cm] (square2l) at (-0.3, 1.78) {};
            \node at (\x, \y) {$A_2$};
            \node at (-\x, \y) {$A_1$};
            \draw[<-,thick,>=stealth] (square1r.south) to[out=-90, in=60, looseness=1.2] ($(center.north)+(0.1,-0.01)$);
            \draw[<-,thick,>=stealth] (square1l.south) to[out=-90, in=120, looseness=1.2] ($(center.north)+(-0.1,-0.01)$);
            % Fondamentali
            \draw[<-,thick,>=stealth] (bot.north) -- node[left] {$Q$} (center);
            \draw[<-,thick,>=stealth] (center) -- node[right] {$\tilde Q$} (bot2.north);
            % Fermi
            \draw[dashed] (bot) -- node[above] {$\Psi_{\!\scriptscriptstyle M}$} (bot2);
            % Fermi antisimmetrici
            \pgfmathsetmacro{\u}{1.5}
            \pgfmathsetmacro{\v}{1}
            \pgfmathsetmacro{\w}{0.28}
            \node[box2, minimum size=0.2cm] (ssquare1r) at (\u, -\v) {};
            \node[box2, minimum size=0.2cm] (ssquare2r) at (\u, -\v+\w) {};
            \node[box2, minimum size=0.2cm] (ssquare1l) at (\u+0.5, -\v) {};
            \node[box2, minimum size=0.2cm] (ssquare2l) at (\u+0.5, -\v+\w) {};
            \node at (\u+0.6, -\v+\w+0.5) {$\Psi_{\!\scriptscriptstyle B_2}$};
            \node at (\u-0.1, -\v+\w+0.5) {$\Psi_{\!\scriptscriptstyle B_1}$};
            \draw[dashed] (ssquare1r.south) to[out=-90, in=0, looseness=1] ($(bot2.east)+(0,0.05)$);
            \draw[dashed] (ssquare1l.south) to[out=-90, in=0, looseness=1.1] ($(bot2.east)+(0,-0.05)$);
            \end{tikzpicture}
            }
        \end{minipage}
        % STEP 2
        \begin{minipage}[b]{0.45\linewidth}
            \centering
            \makebox[\textwidth][c]{
            \begin{tikzpicture}[
                every node/.style={font=\footnotesize},
                box/.style={rectangle, draw, thick},
                box2/.style={rectangle, draw, dashed}
            ]
            \pgfmathsetmacro{\d}{2}
            \pgfmathsetmacro{\c}{0.75}
            \pgfmathsetmacro{\a}{2}
            \pgfmathsetmacro{\b}{2}
            % Nodi
            \node[fill=USPcol,circle,draw,thick] (uspl) at (-\a,0) {$2n$};
            \node[fill=USPcol,circle,draw,thick] (uspr) at (\a, 0) {$2n$};
            \node[fill=SUcol,circle,draw,thick] (center) at (0, 0) {$2n\!+\!1$};
            \node[box] (top) at (0, \d) {$1$};
            \node[box] (bot) at (0, -\d) {$3$};
            % Fondamentali
            \draw[<-, thick, >=stealth] (uspl.east) -- node[above] {$P_1$} (center.west);
            \draw[<-, thick, >=stealth] (uspr.west) -- node[above] {$P_2$} (center.east);
            \draw[->,thick,>=stealth] (bot.north) -- node[left] {$\tilde Q$} (center);
            % Mesoni
            \draw[thick] (top.east) to[out=0, in=100, looseness=1] node[above, overlay, xshift=2pt] {$R_2$} (uspr.north);
            \draw[thick] (top.west) to[out=180, in=80, looseness=1] node[below, overlay, xshift=2pt] {$R_1$} (uspl.north);
            % Fermi
            \draw[dashed] (center) -- node[right] {$\psi$} (top);
            \node (ph) at (-\a-1,0) {};
            \node (phname) at (-\a-1.2,0) {$\Psi_{\!\scriptscriptstyle M}$};
            \draw[dashed] ($(top.west)+(0,0.05)$) to[out=180, in=90, looseness=1] (ph.east);
            \draw[dashed] ($(bot.west)+(0,-0.05)$) to[out=180, in=-90, looseness=1] (ph.east);
            % Fermi antisimmetrici
            \pgfmathsetmacro{\u}{1.2}
            \pgfmathsetmacro{\v}{1.5}
            \pgfmathsetmacro{\w}{0.28}
            \node[box2, minimum size=0.2cm] (ssquare1r) at (\u, -\v) {};
            \node[box2, minimum size=0.2cm] (ssquare2r) at (\u, -\v+\w) {};
            \node[box2, minimum size=0.2cm] (ssquare1l) at (\u+0.5, -\v) {};
            \node[box2, minimum size=0.2cm] (ssquare2l) at (\u+0.5, -\v+\w) {};
            \node at (\u+0.6, -\v+\w+0.5) {$\Psi_{\!\scriptscriptstyle B_2}$};
            \node at (\u-0.1, -\v+\w+0.5) {$\Psi_{\!\scriptscriptstyle B_1}$};
            \draw[dashed] (ssquare1r.south) to[out=-90, in=0, looseness=1] ($(bot.east)+(0,0.05)$);
            \draw[dashed] (ssquare1l.south) to[out=-90, in=0, looseness=1.1] ($(bot.east)+(0,-0.15)$);
            \end{tikzpicture}
            }
        \end{minipage}
        \\[0.5cm]
        % STEP 3
        \begin{minipage}[b]{0.45\linewidth}
            \centering
            \makebox[\textwidth][c]{
            \begin{tikzpicture}[
                every node/.style={font=\footnotesize},
                box/.style={rectangle, draw, thick},
                box2/.style={rectangle, draw, dashed}
            ]
            \pgfmathsetmacro{\d}{1.75}
            \pgfmathsetmacro{\c}{0.75}
            \pgfmathsetmacro{\a}{2}
            \pgfmathsetmacro{\b}{2}
            % Nodi
            \node[fill=USPcol,circle,draw,thick] (uspl) at (-\a,0) {$2n$};
            \node[fill=USPcol,circle,draw,thick] (uspr) at (\a, 0) {$2n$};
            \node[fill=SUcol,circle,draw,thick] (center) at (0, 0) {$2n\!-\!1$};
            \node[box] (top) at (0, \d) {$1$};
            \node[box] (bot) at (0, -\d) {$3$};
            % Fondamentali
            \draw[->,thick, >=stealth] (center) -- node[right] {$q$} (top);
            \draw[->, thick, >=stealth] (uspl.east) -- node[above] {$\tilde P_1$} (center.west);
            \draw[->, thick, >=stealth] (uspr.west) -- node[above] {$\tilde P_2$} (center.east);
            % Mesoni
            \draw[thick] (bot.east) to[out=0, in=-100, looseness=1] node[below, overlay, xshift=2pt] {$N_2$} (uspr.south);
            \draw[thick] (bot.west) to[out=180, in=-80, looseness=1] node[above, overlay, xshift=4pt] {$N_1$} (uspl.south);
            % Fermi
            \draw[dashed] (bot.north) -- node[left] {$\lambda$} (center);
            \node (phname) at (-\a-1.2,0) {$\Psi_{\!\scriptscriptstyle M}$};
            \draw[dashed] ($(top.west)+(0,0.05)$) to[out=180, in=90, looseness=1] (ph.east);
            \draw[dashed] ($(bot.west)+(0,-0.05)$) to[out=180, in=-90, looseness=1] (ph.east);
            % Fermi antisimmetrici
            \pgfmathsetmacro{\u}{-0.3}
            \pgfmathsetmacro{\v}{3.2}
            \pgfmathsetmacro{\w}{0.28}
            \pgfmathsetmacro{\cacca}{0.6}
            \node[box2, minimum size=0.2cm] (square1r) at (\u, -\v) {};
            \node[box2, minimum size=0.2cm] (square2r) at (\u, -\v+\w) {};
            \node[box2, minimum size=0.2cm] (square1l) at (-\u, -\v) {};
            \node[box2, minimum size=0.2cm] (square2l) at (-\u, -\v+\w) {};
            \node at (\u-\cacca, -\v+\w-0.2) {$\Psi_{\!\scriptscriptstyle B_1}$};
            \node at (\u+0.6+\cacca, -\v+\w-0.2) {$\Psi_{\!\scriptscriptstyle B_2}$};
            \draw[dashed] (square2r.north) to[out=90, in=-120, looseness=1] ($(bot.south)+(-0.1,0)$);
            \draw[dashed] (square2l.north) to[out=90, in=-60, looseness=1] ($(bot.south)+(0.1,0)$);
            \end{tikzpicture}
            }
        \end{minipage}
        % STEP 4
        \begin{minipage}[b]{0.45\linewidth}
            \centering
            \makebox[\textwidth][c]{
            \begin{tikzpicture}[
                every node/.style={font=\footnotesize},
                box/.style={rectangle, draw, thick},
                box2/.style={rectangle, draw, dashed}
            ]
            \pgfmathsetmacro{\c}{1.75}
            \pgfmathsetmacro{\b}{0.75}
            \pgfmathsetmacro{\x}{0.8}
            \pgfmathsetmacro{\y}{1.6}
            \pgfmathsetmacro{\u}{2}
            \pgfmathsetmacro{\v}{1}
            \pgfmathsetmacro{\w}{0.28}
            % Nodi
            \node[box] (bot) at (-\b, -\c) {$1$};
            \node[box] (bot2) at (\b, -\c) {$3$};
            \node[fill=SUcol,circle,draw,thick] (center) at (0, 0) {$2n\!-\!1$};
            % Antisimmetriche
            \node[box, minimum size=0.2cm] (square1r) at (0.3, 1.5) {};
            \node[box, minimum size=0.2cm] (square2r) at (0.3, 1.78) {};
            \node[box, minimum size=0.2cm] (square1l) at (-0.3, 1.5) {};
            \node[box, minimum size=0.2cm] (square2l) at (-0.3, 1.78) {};
            \node at (\x, \y) {$\tilde A_2$};
            \node at (-\x, \y) {$\tilde A_1$};
            \draw[->,thick,>=stealth] (square1r.south) to[out=-90, in=60, looseness=1.2] ($(center.north)+(0.1,-0.01)$);
            \draw[->,thick,>=stealth] (square1l.south) to[out=-90, in=120, looseness=1.2] ($(center.north)+(-0.1,-0.01)$);
            % Fondamenali
            \draw[<-,thick,>=stealth] (center) -- node[right] {$\tilde q$} (bot2.north);
            \draw[<-,thick,>=stealth] (bot.north) -- node[left] {$q$} (center);
            % Fermi
            \draw[dashed] (bot) -- node[above] {$\Psi_{\!\scriptscriptstyle M}$} (bot2);
            \end{tikzpicture}
        }
        \end{minipage}
    \caption{Deconfinement steps of $\SU(2n+1)$ with two antisymmetrics, one fundamental and three antifundamentals.}
    \label{fig:SUdispari13}
\end{figure}
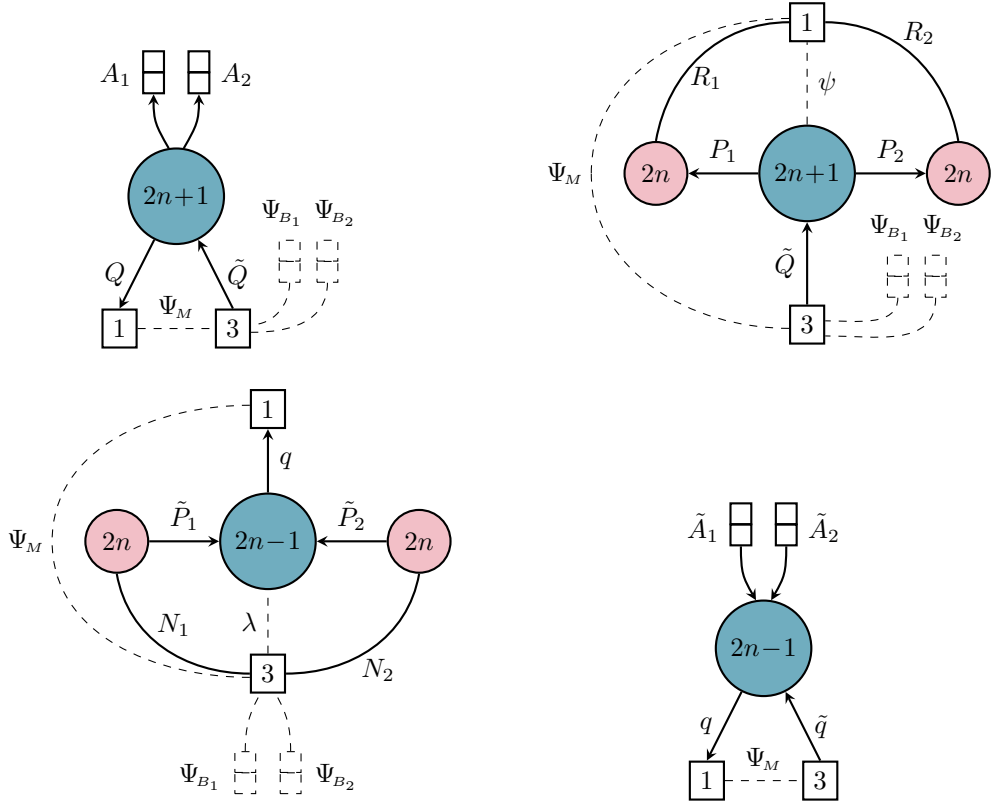
%%%%%%%%%%%%%%%%%%%%%%%%%%%%%%%%%%%%%%%%%%%%%%%%%%%%%%%%%%%%%%%%%%%%%%%

In the following we show the  matching of the elliptic genera 
along the same lines of the discussion performed in the other examples.
In this case we found more convenient\footnote{There are other possible patterns and flippers that can be used to prove the duality through tensor deconfinement, but we found that the choice made here allows for a simpler derivation.} to flip the meson $M$ and the baryons $B_{1,2}$, 
$A_{1,2}^{n} Q$ and $A_{1,2}^n A_{2,1} \tilde Q_{2,1}$ in the dual LG and we will show that the expected $J$-term in (\ref{JPSI7}) 
can be reconstructed by our procedure  using the $J$-term 
expected from the duality of subsection \ref{subsec:3.2}.

At the level of the elliptic genus the identity
associated to this duality is
\begin{eqnarray}
\label{toprovesec3.8}
&&    I^{(1,3,\cdot,2,\cdot)}_{\SU(2n+1)}( u;\vec v;\cdot;\vec t;\cdot)=    \frac{\prod_{j=0}^{2n-4}  
\theta(q/(t_1^{j+3}t_2^{2n-j-1} u v_1 v_2 v_3))}
{       
\prod_a  \theta (u v_a) 
\prod_{a<b} (\theta (t_{1,2} v_a v_b) )
\prod_{j=0}^{n-3}   \theta (t_1^{n-1-j} t_2^{j+2} u v_a v_b)
}
 \nonumber \\
 && 
 \times \frac{1}{
 \prod_{j=0}^{n-4}   \theta (t_1^{n-j-1} t_2^{j+3}  v_1 v_2 v_3)
 \prod_{j=0}^n \theta(t_1^{n-j} t_2^j u)
 \prod_{j=0}^{n-1} \prod_a \theta(t_1^{n-j} t_2^{j+1}  v_a)
 } \,.
\end{eqnarray}

In order to prove this relation we deconfine the two antisymmetric as in Figure \ref{fig:SUdispari13}, obtaining two $\USp(2n)$ gauge nodes, with two bifundamentals $P_{1,2}$.

Actually the procedure adopted here is slighlty different from the cases studied above and it deserves some further comment.
In the various cases studied so far the two symplectic gauge nodes could be reconfined simultaneously to give the original 
model. Here such reconfinement must be done in two steps. Indeed there is a field, corresponding to the Fermi $\psi$ in the second quiver of Figure \ref{fig:SUdispari13} that gives a non-vanishing $J$-term and which requires, after reconfining one of the two $\USp(2n)$ gauge nodes, to integrate out some massive combinations of chirals and Fermi.

The $J$-term for the fields $\psi$ is 
\begin{equation}
J_\psi=P_1 R_1 + P_2 R_2\,.
\end{equation}
In addition in this phase there are other $J$-terms, originating from the flippers for $M$ and $B_{1,2}$ discussed above

\begin{equation}
    J_{\Psi_M} = \tilde Q (P_1 R_1-P_2 R_2), \quad J_{\Psi_{B_{1,2}}}=P_{1,2}^2 \tilde Q^2.
\end{equation}

The $\SU(2n+1)$ gauge node has now $4n$ fundamentals, three antifundamentals and one Fermi. 
It can be dualized according to the rules explained in subsection \ref{SUnGen}, but in this case with $y=1$.
The dual phase corresponds to the  third quiver in Figure \ref{fig:SUdispari13}.
In this case the new $J$-term generated by the duality is  
$J_{\lambda} = N_2 \tilde P_2 + N_1 \tilde P_1$.
In addition the other $J$-terms in this phase are 
\begin{equation}
J_{\Psi_{B_{1}}}= N_1^2,\quad
J_{\Psi_{B_{2}}}= N_2^2, \quad
J_{\Psi_M} = q (N_1 \tilde P_1 -N_2 \tilde P_2)
\end{equation}
where the last term is obtained after integrating out the massive fields.

The last step consists of dualizing the two $\USp(2n)$ groups, each one with $2n+2$ chiral fundamentals, to an LG.
The final quiver is the fourth one  in  Figure \ref{fig:SUdispari13}.
In this case there are two $J$-terms associated to two Fermi singlets $\Psi_{1,2}^{(0)}$. They read $J_{\Psi_{1,2}^{(0)}} = \tilde A_{1,2}^{n-2} \tilde q^3$. There is a further $J$ term for the field $\Psi_M$ that reads $J_{\Psi_M} = q \tilde q$.

Again we have, up to an overall charge conjugation, a model already described in \ref{subsec:3.2}. We can the use the duality of such model with an LG in order to prove the  validity of (\ref{toprovesec3.8}) and to reconstruct (up to flippers) the expected $J$-term (\ref{JPSI7}).
The details of the derivation are straightforward and we leave them to the interested reader.

\subsection{$\SU(2n)$ with four antifundamentals}

In this case the dual LG model has four chiral gauge invariant operators corresponding to chiral fields interacting through a $J$-term with a Fermi multiplet.
The field content of the gauge theory and of the dual LG model is represented in the table below.
\begin{equation} 
    \begin{array}{c|c|c|c|c|c|c|}
    & \SU(2n) & \SU(2) & \SU(4) & \U(1)_A & \U(1)_{\tilde Q} & \U(1)_{R_0} \\
    \hline
    A & \begin{array}{c}
\square \vspace{-2.85mm} \\
\square 
\end{array}  & \square & \cdot & 1 & 0 & 0 \\
    \tilde Q & \square & \cdot & \square & 0 & 1 & 0 \\       
    \hline
    B & \cdot & \square & \begin{array}{c}
\square \vspace{-2.85mm} \\
\square 
\end{array}  & 1 & 2 & 0 \\
    T_{n} & \cdot & \otimes_{\mathrm{sym}}^{n} \square& \cdot & n & 0 & 0 \\
    T_{n-1} & \cdot & \otimes_{\mathrm{sym}}^{n-3} \square& \begin{array}{c}
\square \vspace{-2.85mm} \\
\square 
\end{array}  & n+1 & 2 & 0 \\
    T_{n-2} & \cdot & \otimes_{\mathrm{sym}}^{n-6} \square& \cdot & n+2 & 4 & 0 \\
    \Psi & \cdot & \otimes_{\mathrm{sym}}^{2n-6} \square& \cdot & -2n-2 & -4 & 2 \\
    \end{array}
\end{equation}

The chirals $T_j$ and $B$ correspond to the following gauge invariant combinations of the charged fields
\begin{equation}
T_{n} = A^n \,,\qquad
T_{n-1} = A^{n-1} (A \tilde Q)^2\,,\quad
T_{n-2} = A^{n-2} (A \tilde Q)^4\,, \quad
B = A \tilde Q^2\,,
\end{equation}
while the $J$-term compatible with the global symmetry is
\begin{equation}
\label{JPSI8}
J_{\Psi} = T_n T_{n-2} + T_{n-1}^2 + B^2 T_n^2 + B T_n T_{n-1} \,.
\end{equation}
We computed and matched the 't Hooft anomalies in the two dual theories. Explicitly we have
\begin{comment}
\begin{align}
    &\kappa_{\SU(2)_A^2} = \tfrac{n(2n+1)}{2}\,,\quad
    &&\kappa_{AA} = -\kappa_{AR_0} = 2n(2n+1)\,,\nonumber \\
    &\kappa_{\SU(2)_Q^2} =\kappa_{\SU(2)_{\tilde Q^2}} = \tfrac{2n+1}{2}\,,\quad
    &&\kappa_{QQ} = -\kappa_{QR_0} =\kappa_{\tilde Q\tilde Q} = -\kappa_{\tilde QR_0} = 4n+2\,,\\  
    &\kappa_{R_0 R_0}  = 6n+4\,,\quad
    &&\kappa_{AQ} =\kappa_{A\tilde Q} = 0\,.\nonumber
\end{align}
\end{comment}
\begin{align}
    &\kappa_{\SU(2)^2} = \tfrac{n(2n-1)}{2}\,,\quad
    &&\kappa_{AA} = -\kappa_{AR_0} = 2n(2n-1)\,,\nonumber \\
    &\kappa_{\SU(4)^2} = n\,,\quad
    &&\kappa_{\tilde Q\tilde Q} = -\kappa_{\tilde QR_0} = 8n\,,\\
    &\kappa_{R_0 R_0}  = 6n+1\,,\quad
    &&\kappa_{A\tilde Q} = 0\,.\nonumber
\end{align}

%%%%%%%%%%%%%%%%%%%%%%%%%%%%%%%%%%%%%%%%%%%%%%%%%%%%%%%%%%%%%%%%%%%%%%%
% FIGURE: SU(2n) w 2A 4aF deconfinement
\begin{figure}[h!]
    \centering
        % STEP 1
        \begin{minipage}{0.45\textwidth}
            \centering
            \makebox[\textwidth][c]{
            \begin{tikzpicture}[
                every node/.style={font=\footnotesize},
                box/.style={rectangle, draw, thick},
                box2/.style={rectangle, draw, dashed}
            ]
            \pgfmathsetmacro{\x}{0.8}
            \pgfmathsetmacro{\y}{1.6}
            % Nodi
            \node[box] (bot) at (0, -1.75) {$4$};
            \node[fill=SUcol,circle,draw,thick] (center) at (0, 0) {$2n$};
            % Antisimmetriche
            \node[box, minimum size=0.2cm] (square1r) at (0.3, 1.5) {};
            \node[box, minimum size=0.2cm] (square2r) at (0.3, 1.78) {};
            \node[box, minimum size=0.2cm] (square1l) at (-0.3, 1.5) {};
            \node[box, minimum size=0.2cm] (square2l) at (-0.3, 1.78) {};
            \node at (\x, \y) {$A_2$};
            \node at (-\x, \y) {$A_1$};
            \draw[<-,thick,>=stealth] (square1r.south) to[out=-90, in=60, looseness=1.2] ($(center.north)+(0.1,-0.01)$);
            \draw[<-,thick,>=stealth] (square1l.south) to[out=-90, in=120, looseness=1.2] ($(center.north)+(-0.1,-0.01)$);
            % Fondamentali
            \draw[->,thick,>=stealth] (bot.north) -- node[right] {$\tilde Q$} (center);
            % Fermi antisimmetrici
            \pgfmathsetmacro{\u}{1}
            \pgfmathsetmacro{\v}{1}
            \pgfmathsetmacro{\w}{0.28}
            \node[box2, minimum size=0.2cm] (ssquare1r) at (\u, -\v) {};
            \node[box2, minimum size=0.2cm] (ssquare2r) at (\u, -\v+\w) {};
            \node[box2, minimum size=0.2cm] (ssquare1l) at (\u+0.5, -\v) {};
            \node[box2, minimum size=0.2cm] (ssquare2l) at (\u+0.5, -\v+\w) {};
            \node at (\u+0.6, -\v+\w+0.5) {$\Psi_{\!\scriptscriptstyle B_2}$};
            \node at (\u-0.1, -\v+\w+0.5) {$\Psi_{\!\scriptscriptstyle B_1}$};
            \draw[dashed] (ssquare1r.south) to[out=-90, in=0, looseness=1] ($(bot.east)+(0,0.05)$);
            \draw[dashed] (ssquare1l.south) to[out=-90, in=0, looseness=1.1] ($(bot.east)+(0,-0.05)$);
            \end{tikzpicture}
            }
        \end{minipage}
        % STEP 2
        \begin{minipage}{0.45\textwidth}
            \centering
            \makebox[\textwidth][c]{
            \begin{tikzpicture}[
                every node/.style={font=\footnotesize},
                box/.style={rectangle, draw, thick},
                box2/.style={rectangle, draw, dashed}
            ]
            \pgfmathsetmacro{\a}{2}
            \pgfmathsetmacro{\b}{2}
            \pgfmathsetmacro{\aiuto}{1}
            % Nodi
            \node[fill=USPcol,circle,draw,thick] (uspl) at (-\a,\aiuto) {$2n\!-\!2$};
            \node[fill=USPcol,circle,draw,thick] (uspr) at (\a, \aiuto) {$2n\!-\!2$};
            \node[box] (bot) at (0, -\a) {$4$};
            \node[fill=SUcol,circle,draw,thick] (center) at (0, 0) {$2n$};
            % Fondamentali
            \draw[<-, thick, >=stealth] (uspl.south east) -- node[above] {$P_1$} (center.west);
            \draw[<-, thick, >=stealth] (uspr.south west) -- node[above] {$P_2$} (center.east);
            \draw[->, thick, >=stealth] (bot.north) -- (center);
            \node at (-0.65,-\a) {$\tilde Q$};
            % Fermi antisimmetrici
            \pgfmathsetmacro{\u}{1.5}
            \pgfmathsetmacro{\v}{1}
            \pgfmathsetmacro{\w}{0.28}
            \node[box2, minimum size=0.2cm] (ssquare1r) at (\u, -\v) {};
            \node[box2, minimum size=0.2cm] (ssquare2r) at (\u, -\v+\w) {};
            \node[box2, minimum size=0.2cm] (ssquare1l) at (\u+0.5, -\v) {};
            \node[box2, minimum size=0.2cm] (ssquare2l) at (\u+0.5, -\v+\w) {};
            \node at (\u+0.6, -\v+\w+0.5) {$\Psi_{\!\scriptscriptstyle B_2}$};
            \node at (\u-0.1, -\v+\w+0.5) {$\Psi_{\!\scriptscriptstyle B_1}$};
            \draw[dashed] (ssquare1r.south) to[out=-90, in=0, looseness=1] ($(bot.east)+(0,0.05)$);
            \draw[dashed] (ssquare1l.south) to[out=-90, in=0, looseness=1.1] ($(bot.east)+(0,-0.05)$);
            \end{tikzpicture}
            }
        \end{minipage}
        \\[0.5cm]
        % STEP 3
        \begin{minipage}{0.45\textwidth}
            \centering
            \makebox[\textwidth][c]{
            \begin{tikzpicture}[
                every node/.style={font=\footnotesize},
                box/.style={rectangle, draw, thick},
                box2/.style={rectangle, draw, dashed}
            ]
            \pgfmathsetmacro{\a}{2}
            \pgfmathsetmacro{\b}{2}
            \pgfmathsetmacro{\aiuto}{1}
            % Nodi
            \node[fill=USPcol,circle,draw,thick] (uspl) at (-\a,\aiuto) {$2n\!-\!2$};
            \node[fill=USPcol,circle,draw,thick] (uspr) at (\a, \aiuto) {$2n\!-\!2$};
            \node[box] (bot) at (0, -\a) {$4$};
            \node[fill=SUcol,circle,draw,thick] (center) at (0, 0) {$2n\!-\!4$};
            % Fondamentali
            \draw[->, thick, >=stealth] (uspl.south east) -- node[above] {$\tilde P_1$} (center.west);
            \draw[->, thick, >=stealth] (uspr.south west) -- node[above] {$\tilde P_2$} (center.east);
            % Mesoni
            \draw[thick] (bot.east) to[out=0, in=-100, looseness=1] node[above, overlay, xshift=-5pt] {$N_2$} (uspr.south);
            \draw[thick] (bot.west) to[out=180, in=-80, looseness=1] node[above, overlay, xshift=5pt] {$N_1$} (uspl.south);
            % Fermi
            \draw[dashed] (bot.north) -- node[left] {$\lambda$} (center);
            % Fermi antisimmetrici
            \pgfmathsetmacro{\u}{2.5}
            \pgfmathsetmacro{\v}{1}
            \pgfmathsetmacro{\w}{0.28}
            \node[box2, minimum size=0.2cm] (ssquare1r) at (\u, -\v) {};
            \node[box2, minimum size=0.2cm] (ssquare2r) at (\u, -\v+\w) {};
            \node[box2, minimum size=0.2cm] (ssquare1l) at (\u+0.5, -\v) {};
            \node[box2, minimum size=0.2cm] (ssquare2l) at (\u+0.5, -\v+\w) {};
            \node at (\u+0.6, -\v+\w+0.5) {$\Psi_{\!\scriptscriptstyle B_2}$};
            \node at (\u-0.1, -\v+\w+0.5) {$\Psi_{\!\scriptscriptstyle B_1}$};
            \draw[dashed] (ssquare1r.south) to[out=-90, in=0, looseness=1] ($(bot.east)+(0,-0.05)$);
            \draw[dashed] (ssquare1l.south) to[out=-90, in=0, looseness=1.1] ($(bot.east)+(0,-0.15)$);
            \end{tikzpicture}
            }
        \end{minipage}
        % STEP 4
        \begin{minipage}{0.45\textwidth}
            \centering
            \makebox[\textwidth][c]{
            \begin{tikzpicture}[
                every node/.style={font=\footnotesize},
                box/.style={rectangle, draw, thick},
            ]
            \pgfmathsetmacro{\a}{1.75}
            \pgfmathsetmacro{\x}{0.8}
            \pgfmathsetmacro{\y}{1.6}
            \pgfmathsetmacro{\u}{1.5}
            \pgfmathsetmacro{\v}{1}
            \pgfmathsetmacro{\w}{0.28}
            % Nodi        
            \node[box] (bot) at (0, -\a) {$4$};
            \node[fill=SUcol,circle,draw,thick] (center) at (0, 0) {$2n\!-\!1$};
            % Antisimmetriche
            \node[box, minimum size=0.2cm] (square1r) at (0.3, 1.5) {};
            \node[box, minimum size=0.2cm] (square2r) at (0.3, 1.78) {};
            \node[box, minimum size=0.2cm] (square1l) at (-0.3, 1.5) {};
            \node[box, minimum size=0.2cm] (square2l) at (-0.3, 1.78) {};
            \node at (\x, \y) {$\tilde A_2$};
            \node at (-\x, \y) {$\tilde A_1$};
            \draw[->, thick, >=stealth] (square1r.south) to[out=-90, in=45, looseness=1.2] ($(center.north)+(0.1,-0.01)$);
            \draw[->, thick, >=stealth] (square1l.south) to[out=-90, in=135, looseness=1.2] ($(center.north)+(-0.1,-0.01)$);
            % Fondamentali
            \draw[->, thick, >=stealth] (bot.north) -- node[right] {$\tilde q$} (center);
            \end{tikzpicture}
            }
        \end{minipage}
    \caption{Deconfinement steps of $\SU(2n)$ with two antisymmetrics and four antifundamentals.}
    \label{fig:SUpari04}
\end{figure}
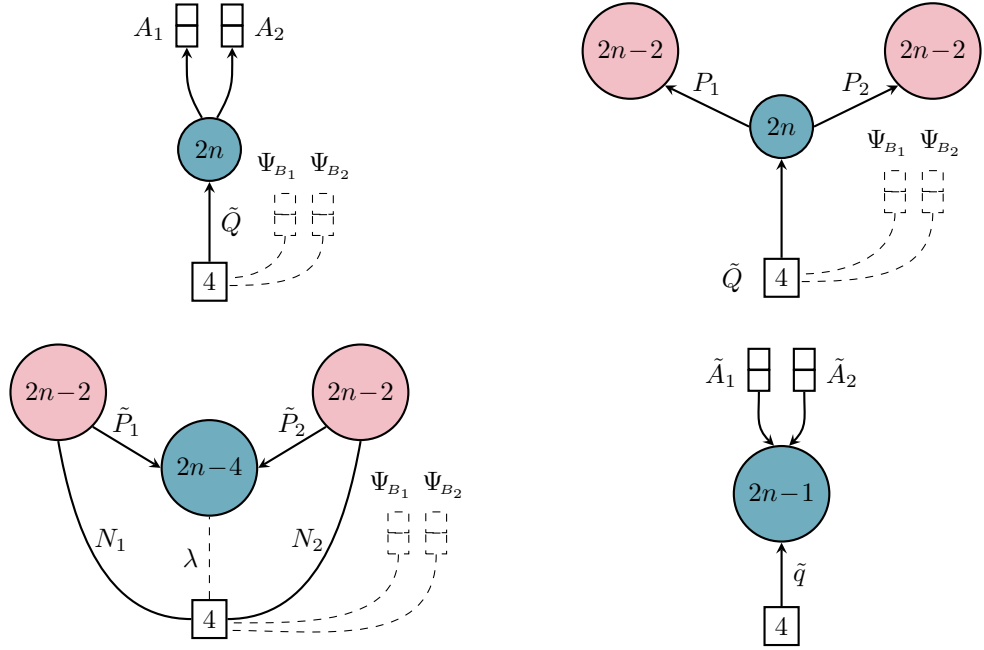
%%%%%%%%%%%%%%%%%%%%%%%%%%%%%%%%%%%%%%%%%%%%%%%%%%%%%%%%%%%%%%%%%%%%%%%

In the following we show the  matching of the elliptic genera 
along the same lines of the discussion performed in the other examples.
In this case we will flip the  baryons $B$ and
$\Pf A_{1,2}$ in the dual LG and we show that the expected $J$-term in (\ref{JPSI8}) 
can be reconstructed by our procedure  using the $J$-term 
expected from the duality of subsection \ref{subsec:3.2}.

At the level of the elliptic genus the identity
associated to this duality is
\begin{eqnarray}
\label{toprovesec3.9}
I^{(\cdot,4,\cdot,2,\cdot)}_{\SU(2n)}( \cdot ;\vec v;\cdot;\vec t;\cdot)&& =   
\frac{
\prod_{j=0}^{2n-6}  
\theta(q/(t_1^{j+4}t_2^{2n-j-2}  \prod_{a=1}^{4}v_a ))}
{       
\prod_{a<b} (\theta (t_{1,2} v_a v_b) 
\prod_{j=0}^{n-3}   \theta (t_1^{n-1-j} t_2^{j+2} v_a v_b)
)
}
 \nonumber \\
 && 
 \times \frac{1}{
 \prod_{j=0}^{n-6}   \theta (t_1^{n-j-2} t_2^{j+4} \prod_{a=1}^{4}v_a)
 \prod_{j=0}^n \theta(t_1^{n-j} t_2^j)
 }\,.
\end{eqnarray}

In order to prove this relation we deconfine the two antisymmetric as in Figure \ref{fig:SUpari04}, obtaining two $\USp(2n-2)$ gauge nodes, with two bifundamentals $P_{1,2}$.
The original antisymmetric chirals  $A_{1,2}$ correspond to the combinations $P_{1,2}^2$.
The $J$-term at this stage is the one flipping the operator $B$, corresponding to
\begin{equation}
J_{\Psi_{B_{1,2}}} = \tilde Q^2 P_{1,2}^2\,.
\end{equation}
The $\SU(2n)$ gauge theory has $4n-4$ fundamentals and four antifundamentals. It can be dualized according to the rules explained in subsection \ref{SUnGen}, and it gives origin to the third quiver in Figure \ref{fig:SUpari04}.

In this case the new $J$-term generated by the duality is  
$J_{\lambda} = N_2 \tilde P_2 + N_1 \tilde P_1$.
In addition the  $J$-terms for $\Psi_{B_1}$ and $\Psi_{B_2}$
become 
$J_{\Psi_{B_1}} = N_{1}^2$ and $J_{\Psi_{B_2}} = N_{2}^2$ (where the flavor indices are left implicit in these relations).

The last step consists of dualizing the two $\USp(2n-2)$ groups, each one with $2n$ chiral fundamentals, to an LG.
The final quiver is the fourth one  in  Figure \ref{fig:SUpari13}.
In this case there are two $J$-terms associated to two Fermi singlets $\Psi_{1,2}^{(0)}$. They read $J_{\Psi_{1,2}^{(0)}} = \tilde A_{1,2}^{n-4} \tilde q^4$. 

Again we have, up to an overall charge conjugation, a model already described in \ref{subsec:3.2}. We can the use the duality of such model with an LG in order to prove the  validity of (\ref{toprovesec3.9}) and to reconstruct (up to flippers) the expected $J$-term (\ref{JPSI8}).
The details of the derivation are straightforward and we leave them to the interested reader.

\subsection{$\SU(2n+1)$ with four antifundamentals}

Here we conclude this section with the last case of our classification of 2d 
$\mathcal{N}=(0,2)$ gauge/LG dualities for $\SU(N)$ gauge theories with two antisymmetric chirals.
In this case the dual LG model has three chiral gauge invariant operators corresponding to chiral fields interacting through a $J$-term with a Fermi multiplet.
The field content of the gauge theory and of the dual LG model is represented in the table below.
\begin{equation} 
    \begin{array}{c|c|c|c|c|c|c|}
    & \SU(2n+1) & \SU(2) & \SU(4) & \U(1)_A & \U(1)_Q & \U(1)_{R_0} \\
    \hline
    A & \begin{array}{c}
\square \vspace{-2.85mm} \\
\square 
\end{array}  & \square & \cdot & 1 & 0 & 0 \\
    \tilde Q & \square & \cdot & \square & 0 & 1 & 0 \\       
    \hline
    B & \cdot & \square & \begin{array}{c}
\square \vspace{-2.85mm} \\
\square 
\end{array}  & 1 & 2 & 0 \\
    P_{n+1} & \cdot & \otimes_{\mathrm{sym}}^{n-1} \square& \square & n+1 & 1 & 0 \\
    P_{n+2} & \cdot & \otimes_{\mathrm{sym}}^{n-4} \square& \overline{\square} & n+2 & 3 & 0 \\
    \Psi & \cdot & \otimes_{\mathrm{sym}}^{2n-5} \square& \cdot & -2n-3 & -4 & 2 \\
    \end{array}
\end{equation}

The chirals $P_j$ and $B$  correspond to the following gauge invariant combinations of the charged fields
\begin{equation}
P_{n+1} = A^n (A \tilde Q)\,,\qquad
P_{n+2} = A^{n-1} (A \tilde Q)^3\,,\quad
B = \tilde A Q^2\,,
\end{equation}
while the most general $J$-term compatible with the global symmetry is
\begin{equation}
\label{JPSI9}
J_{\Psi} = P_{n+1} P_{n+2} + B P_{n+1}^2\,.
\end{equation}
We have computed the 't Hooft anomalies in the two phases and they match. Explicitly we have
\begin{align}
    &\kappa_{\SU(2)^2} = \tfrac{n(2n+1)}{2}\,,\quad
    &&\kappa_{AA} = -\kappa_{AR_0} = 2n(2n+1)\,,\nonumber \\
    &\kappa_{\SU(4)^2} = n\,,\quad
    &&\kappa_{\tilde Q\tilde Q} = -\kappa_{\tilde QR_0} = 8n\,,\\
    &\kappa_{R_0 R_0}  = 6n+4\,,\quad
    &&\kappa_{A\tilde Q} = 0\,.\nonumber
\end{align}

%%%%%%%%%%%%%%%%%%%%%%%%%%%%%%%%%%%%%%%%%%%%%%%%%%%%%%%%%%%%%%%%%%%%%%%
% FIGURE: SU(2n+1) w 2A 4aF deconfinement
\begin{figure}[h!]
    \centering
        % STEP 1
        \begin{minipage}{0.45\textwidth}
            \centering
            \makebox[\textwidth][c]{
            \begin{tikzpicture}[
                every node/.style={font=\footnotesize},
                box/.style={rectangle, draw, thick},
                box2/.style={rectangle, draw, dashed}
            ]
            \pgfmathsetmacro{\x}{0.8}
            \pgfmathsetmacro{\y}{1.6}
            \pgfmathsetmacro{\a}{1.75}
            \pgfmathsetmacro{\s}{0}
            \pgfmathsetmacro{\t}{0.6}
            % Nodi
            \node[box] (bot) at (0, -1.75) {$4$};
            \node[fill=SUcol,circle,draw,thick] (center) at (0, 0) {$2n\!+\!1$};
            % Antisimmetriche
            \node[box, minimum size=0.2cm] (square1r) at (0.3, 1.5) {};
            \node[box, minimum size=0.2cm] (square2r) at (0.3, 1.78) {};
            \node[box, minimum size=0.2cm] (square1l) at (-0.3, 1.5) {};
            \node[box, minimum size=0.2cm] (square2l) at (-0.3, 1.78) {};
            \node at (\x, \y) {$A_2$};
            \node at (-\x, \y) {$A_1$};
            \draw[<-,thick,>=stealth] (square1r.south) to[out=-90, in=60, looseness=1.2] ($(center.north)+(0.1,-0.01)$);
            \draw[<-,thick,>=stealth] (square1l.south) to[out=-90, in=120, looseness=1.2] ($(center.north)+(-0.1,-0.01)$);
            % Fondamentali
            \draw[->,thick,>=stealth] (bot.north) -- node[right] {$\tilde Q$} (center);
            % Fermi antisimmetrici
            \pgfmathsetmacro{\u}{1.2}
            \pgfmathsetmacro{\v}{1}
            \pgfmathsetmacro{\w}{0.28}
            \node[box2, minimum size=0.2cm] (ssquare1r) at (\u, -\v) {};
            \node[box2, minimum size=0.2cm] (ssquare2r) at (\u, -\v+\w) {};
            \node[box2, minimum size=0.2cm] (ssquare1l) at (\u+0.5, -\v) {};
            \node[box2, minimum size=0.2cm] (ssquare2l) at (\u+0.5, -\v+\w) {};
            \node at (\u+0.6, -\v+\w+0.5) {$\Psi_{\!\scriptscriptstyle B_2}$};
            \node at (\u-0.1, -\v+\w+0.5) {$\Psi_{\!\scriptscriptstyle B_1}$};
            \draw[dashed] (ssquare1r.south) to[out=-90, in=0, looseness=1] ($(bot.east)+(0,0.05)$);
            \draw[dashed] (ssquare1l.south) to[out=-90, in=0, looseness=1.1] ($(bot.east)+(0,-0.05)$);
            \end{tikzpicture}
            }
        \end{minipage}
        % STEP 2
        \begin{minipage}{0.45\textwidth}
            \centering
            \makebox[\textwidth][c]{
            \begin{tikzpicture}[
                every node/.style={font=\footnotesize},
                box/.style={rectangle, draw, thick},
                box2/.style={rectangle, draw, dashed}
            ]
            \pgfmathsetmacro{\a}{2}
            \pgfmathsetmacro{\b}{3}
            % Nodi
            \node[fill=USPcol,circle,draw,thick] (uspl) at (-\a,0) {$2n\!-\!2$};
            \node[fill=USPcol,circle,draw,thick] (uspr) at (\a, 0) {$2n\!-\!2$};
            \node[box] (botl) at (-\a, -\b) {$1$};
            \node[box] (botr) at (\a, -\b) {$1$};
            \node[box] (bot) at (0, -\a) {$4$};
            \node[fill=SUcol,circle,draw,thick] (center) at (0, 0) {$2n\!+\!1$};
            % Fondamentali
            \draw[<-, thick, >=stealth] (uspl.east) -- node[above] {$P_1$} (center.west);
            \draw[<-, thick, >=stealth] (uspr.west) -- node[above] {$P_2$} (center.east);
            \draw[->, thick, >=stealth] (bot.north) -- (center);
            \node at (0.65,-\a) {$\tilde Q$};
            \draw[->,thick,>=stealth] (botl.north east) -- node[left,yshift=5pt]{$\tilde R_1$}(center.south west);
            \draw[->,thick,>=stealth] (botr.north west) -- node[right,yshift=5pt]{$\tilde R_2$}(center.south east);
            % Fermi
            \draw[dashed] (uspl) -- node[left] {$\psi_1$} (botl);
            \draw[dashed] (uspr) -- node[right] {$\psi_2$} (botr);
            % Fermi antisimmetrici
            \pgfmathsetmacro{\u}{-0.3}
            \pgfmathsetmacro{\v}{3.2}
            \pgfmathsetmacro{\w}{0.28}
            \pgfmathsetmacro{\cacca}{0.6}
            \node[box2, minimum size=0.2cm] (square1r) at (\u, -\v) {};
            \node[box2, minimum size=0.2cm] (square2r) at (\u, -\v+\w) {};
            \node[box2, minimum size=0.2cm] (square1l) at (-\u, -\v) {};
            \node[box2, minimum size=0.2cm] (square2l) at (-\u, -\v+\w) {};
            \node at (\u-\cacca, -\v+\w-0.2) {$\Psi_{\!\scriptscriptstyle B_1}$};
            \node at (\u+0.6+\cacca, -\v+\w-0.2) {$\Psi_{\!\scriptscriptstyle B_2}$};
            \draw[dashed] (square2r.north) to[out=90, in=-120, looseness=1] ($(bot.south)+(-0.1,0)$);
            \draw[dashed] (square2l.north) to[out=90, in=-60, looseness=1] ($(bot.south)+(0.1,0)$);
            \end{tikzpicture}
            }
        \end{minipage}
        \\[0.5cm]
        % STEP 3
        \begin{minipage}{0.45\textwidth}
            \centering
            \makebox[\textwidth][c]{
            \begin{tikzpicture}[
                every node/.style={font=\footnotesize},
                box/.style={rectangle, draw, thick},
                box2/.style={rectangle, draw, dashed}
            ]
            \pgfmathsetmacro{\a}{2}
            \pgfmathsetmacro{\b}{2}
            % Nodi 
            \node[fill=USPcol,circle,draw,thick] (uspl) at (-\a,0) {$2n\!-\!2$};
            \node[fill=USPcol,circle,draw,thick] (uspr) at (\a, 0) {$2n\!-\!2$};
            \node[box] (botl) at (-\a, \b) {$1$};
            \node[box] (botr) at (\a, \b) {$1$};
            \node[box] (bot) at (0, -\b) {$4$};
            \node[fill=SUcol,circle,draw,thick] (center) at (0, 0) {$2n\!-\!5$};
            % Fondamentali
            \draw[->, thick, >=stealth] (uspl.east) -- node[below] {$\tilde P_1$} (center.west);
            \draw[->, thick, >=stealth] (uspr.west) -- node[below] {$\tilde P_2$} (center.east);
            \draw[thick] (uspl) -- node[left] {$L_1$} (botl);
            \draw[thick] (uspr) -- node[right] {$L_2$} (botr);
            % Mesoni
            \draw[thick] (bot.east) to[out=0, in=-100, looseness=1] node[below, overlay, xshift=2pt] {$N_2$} (uspr.south);
            \draw[thick] (bot.west) to[out=180, in=-80, looseness=1] node[below, overlay, xshift=-2pt] {$N_1$} (uspl.south);
            % Fermi
            \draw[dashed] (bot.north) -- node[left] {$\lambda$} (center);
            \draw[-,dashed,>=stealth] (botl.south east) -- node[right]  {$\eta_2$}(center.north west);
            \draw[-,dashed,>=stealth] (botr.south west) -- node[left]{$\eta_1$}(center.north east);
            % Fermi antisimmetrici
            \pgfmathsetmacro{\u}{-0.3}
            \pgfmathsetmacro{\v}{3.2}
            \pgfmathsetmacro{\w}{0.28}
            \pgfmathsetmacro{\cacca}{0.6}
            \node[box2, minimum size=0.2cm] (square1r) at (\u, -\v) {};
            \node[box2, minimum size=0.2cm] (square2r) at (\u, -\v+\w) {};
            \node[box2, minimum size=0.2cm] (square1l) at (-\u, -\v) {};
            \node[box2, minimum size=0.2cm] (square2l) at (-\u, -\v+\w) {};
            \node at (\u-\cacca, -\v+\w-0.2) {$\Psi_{\!\scriptscriptstyle B_1}$};
            \node at (\u+0.6+\cacca, -\v+\w-0.2) {$\Psi_{\!\scriptscriptstyle B_2}$};
            \draw[dashed] (square2r.north) to[out=90, in=-120, looseness=1] ($(bot.south)+(-0.1,0)$);
            \draw[dashed] (square2l.north) to[out=90, in=-60, looseness=1] ($(bot.south)+(0.1,0)$);
            \end{tikzpicture}
            }
        \end{minipage}
        % STEP 4
        \begin{minipage}{0.45\textwidth}
            \centering
            \makebox[\textwidth][c]{
            \begin{tikzpicture}[
                every node/.style={font=\footnotesize},
                box/.style={rectangle, draw, thick},
            ]
            \pgfmathsetmacro{\a}{1.75}
            \pgfmathsetmacro{\x}{0.8}
            \pgfmathsetmacro{\y}{1.6}
            \pgfmathsetmacro{\u}{1.5}
            \pgfmathsetmacro{\v}{1}
            \pgfmathsetmacro{\w}{0.28}
            \pgfmathsetmacro{\s}{0}
            \pgfmathsetmacro{\t}{0.6}
            % Nodi        
           % \node[box] (botl) at (-\a+\s, -\a) {$1$};
            %\node[box] (botl2) at (-\a+\s, -\a+1) {$1$};
           % \node at (-\a+\s-\t, -\a+1) {$C_1^a$};
            %\node at (-\a+\s-\t, -\a) {$C_2^a$};
            \node[box] (bot) at (0, -\a) {$4$};
            \node[fill=SUcol,circle,draw,thick] (center) at (0, 0) {$2n\!-\!1$};
            % Antisimmetriche
            \node[box, minimum size=0.2cm] (square1r) at (0.3, 1.5) {};
            \node[box, minimum size=0.2cm] (square2r) at (0.3, 1.78) {};
            \node[box, minimum size=0.2cm] (square1l) at (-0.3, 1.5) {};
            \node[box, minimum size=0.2cm] (square2l) at (-0.3, 1.78) {};
            \node at (\x, \y) {$\tilde A_2$};
            \node at (-\x, \y) {$\tilde A_1$};
            \draw[->, thick, >=stealth] (square1r.south) to[out=-90, in=45, looseness=1.2] ($(center.north)+(0.1,-0.01)$);
            \draw[->, thick, >=stealth] (square1l.south) to[out=-90, in=135, looseness=1.2] ($(center.north)+(-0.1,-0.01)$);
            % Fondamentali
            \draw[->, thick, >=stealth] (bot.north) -- node[right] {$\tilde q$} (center);
            % Mesoni
           % \draw[->, thick, >=stealth] ($(bot.west)+(0,-0.05)$) -- ($(botl.east)+(0,-0.05)$);
            %\draw[->,thick,>=stealth] ($(bot.west)+(0,0.05)$) to[out=180, in=-45, looseness=1] (botl2.south east);
            \end{tikzpicture}
            %\caption{}
            }
        \end{minipage}
    \caption{Deconfinement steps of $\SU(2n+1)$ with two antisymmetrics and four antifundamentals. %{\color{red} Cancelliamo i $C_i^a$.}
    }
    \label{fig:SUdisp04}
\end{figure}
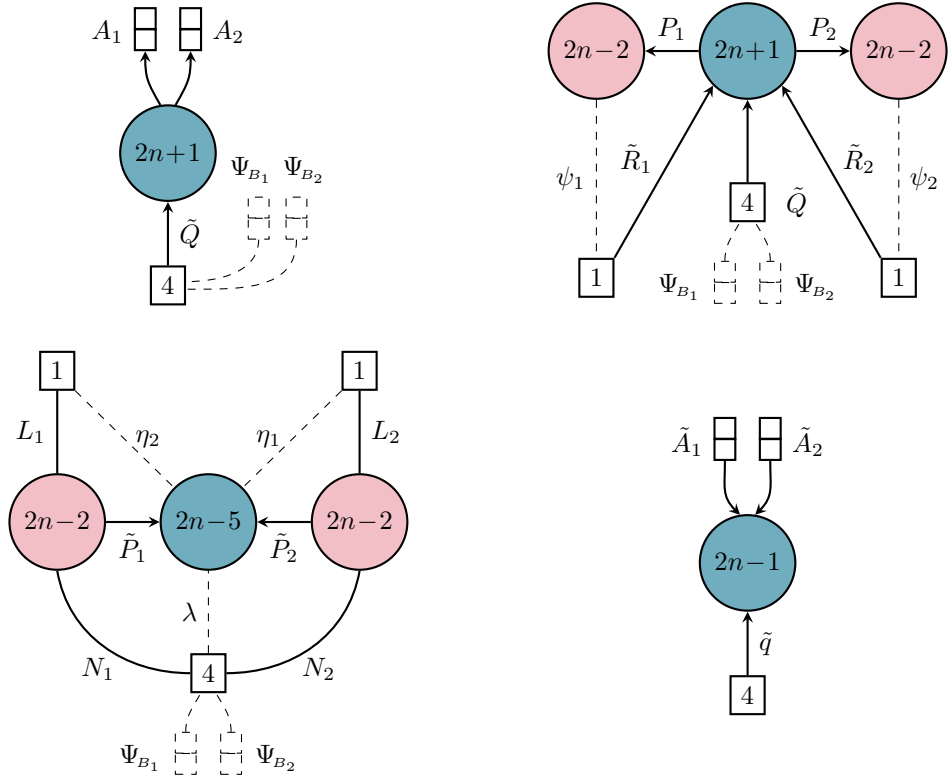
%%%%%%%%%%%%%%%%%%%%%%%%%%%%%%%%%%%%%%%%%%%%%%%%%%%%%%%%%%%%%%%%%%%%%%%

In the following we show the  matching of the elliptic genera 
along the same lines of the discussion performed in the other examples.

Actually in this case we need to adopt the same strategy used in the cases analyzed in subsections \ref{subsec:3.1} and  \ref{subsec:3.2} in order to trade the two antisymmetric with two symplectic gauge groups. Concretely, we flip them by using the  confining duality reviewed in subsection \ref{USpallaSacchi}.

In this case we also flip the  baryon $B$  in the dual LG and we show that the expected $J$-term in (\ref{JPSI9}) 
can be reconstructed by our procedure  using the $J$-term 
expected from the duality of subsection \ref{subsec:3.1}.

At the level of the elliptic genus the identity
associated to this duality is
\begin{eqnarray}
\label{toprovesec3.10}
&&I^{(\cdot,4,\cdot,2,\cdot)}_{\SU(2n+1)}( \cdot ;\vec v;\cdot;\vec t;\cdot)=   
\frac{
\prod_{j=0}^{2n-5}  
\theta(q/(t_1^{j+4}t_2^{2n-j-1}  \prod_{a=1}^{4}v_a))}
{       
\prod_{a<b} \theta (t_{1,2} v_a v_b) 
}
 \nonumber \\
 && 
 \times \frac{1}{
 \prod_{a<b<c} 
 \prod_{j=0}^{n-4}   \theta (t_1^{n-j-1} t_2^{j+3} v_a v_b v_c)
 \prod_a 
 \prod_{j=0}^{n-1} \theta(t_1^{n-j} t_2^{j+1} v_a)
 } \,.
\end{eqnarray}

In order to prove this relation we deconfine the two antisymmetrics as in Figure \ref{fig:SUdisp04}, obtaining two $\USp(2n-2)$ gauge nodes, with two bifundamentals $P_{1,2}$ and two fundamental Fermi $\psi_{1,2}$.
The original antisymmetric chirals  $A_{1,2}$ correspond to the combinations $P_{1,2}^2$. There are also two $J$-terms $J_{\psi_{1,2}} = P_{1,2} \tilde R_{1,2}$ and in addition there are  $J$-terms  flipping the operators $B_{1,2}$, corresponding to
\begin{equation}
J_{\Psi_{B_{1,2}}} = \tilde Q^2 P_{1,2}^2\,.
\end{equation}
As discussed in subsection \ref{subsec:3.1} in this phase we must impose a further constraint on the charges in order to prevent the generation of an axial symmetry for each $\USp(2n-2)$ gauge symmetry. We refer the reader to the discussion there for further comments on this problem.

The $\SU(2n+1)$ gauge theory has $4n-4$ fundamentals and four antifundamentals. It can be dualized according to the rules explained in subsection \ref{SUnGen}, and it gives origin to the third quiver in Figure \ref{fig:SUdisp04}.

In this case the new $J$-terms generated by the duality are
$J_{\lambda} = N_2 \tilde P_2 + N_1 \tilde P_1$ and $J_{\eta_{1,2}} = L_{1,2} \tilde P_{1,2}$.
In addition the  $J$-terms for $\Psi_{B_1}$ and  $\Psi_{B_2}$
become 
$J_{\Psi_{B_1}} = N_{1}^2$ and $J_{\Psi_{B_2}} = N_{2}^2$  respectively (again omitting the antisymmetric flavor indices).

The last step consists of dualizing the two $\USp(2n-2)$ groups, each one with $2n$ chiral fundamentals, to an LG.
The final quiver is the fourth one  in  Figure \ref{fig:SUdispari13}.
In this case there are two $J$-terms associated to two Fermi singlets $\Psi_{1,2}^{(0)}$. They read $J_{\Psi_{1,2}^{(0)}} = C_{1,2} \tilde A_{1,2}^{n-4} \tilde q^3$. 

Again we have, up to an overall charge conjugation, a model already described in \ref{subsec:3.2}. We can the use the duality of such model with an LG in order to prove the  validity of (\ref{toprovesec3.10}) and to reconstruct (up to flippers) the expected $J$-term (\ref{JPSI9}).
The details of the derivation are straightforward and we leave them to the interested reader.

%%%%%%%%%%%%%%%%%%%%%%%%%%%%%%%%%%%%
%%%%%%%%%%%%%%%%%%%%%%%%%%%%%%%%%%%%%
%%%%%%%%%%%%%%%%%%%%%%%%%%%%%%%%%%%%%
\section{A 4d parent for $\SU(2n)$ with $(A,\tilde A)$ and $4$ $ \square$}

In this section we discuss a 4d  duality that, once topologically twisted on a $S^2$ using the prescription of \cite{Gadde:2015wta}, gives origin to a 2d gauge/LG duality.
The 2d model corresponds to a 2d duality studied in \cite{Amariti:2024usp} using tensor deconfinement. The claim made in \cite{Amariti:2024usp} was that the model did not have a 4d origin in terms of any known 4d s-confining duality in the classification of \cite{Csaki:1996zb}.
Such a classification is complete for models with vanishing superpotential. 

Despite this fact, here we obtain a parent 4d duality between two gauge theories
in which the dual phase reduces in 2d to the expected LG model.

%%%%%%%%%%%%%%%%%%%%%%%%%%%%%%%%%%%%%%%%%%%%%%%%%%%%%%%%%%%%%%%%%%%%%%%
% FIGURE: Duality between SU(n+1) w 1 as F and LG
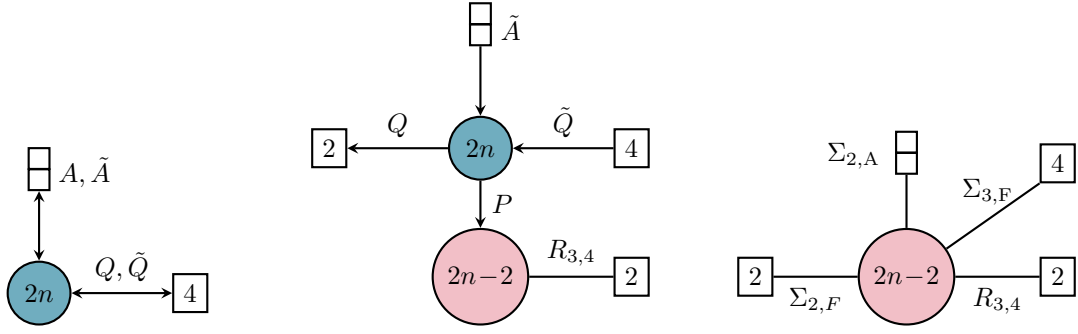
\begin{figure}[h!]
    \centering
    \begin{minipage}{\textwidth}
        \centering
        % STEP 1
        \makebox[0.32\textwidth][c]{
        \begin{tikzpicture}[
            every node/.style={font=\footnotesize},
            box/.style={rectangle, draw, thick},
            box2/.style={rectangle, draw, dashed}
        ]
        \pgfmathsetmacro{\x}{0.6}
        \pgfmathsetmacro{\y}{1.6}
        \pgfmathsetmacro{\a}{2}
        % Nodi
        \node[box] (r) at (\a,0) {$4$};
        \node[fill=SUcol,circle,draw,thick] (center) at (0, 0) {$2n$};
        % Fondamentali
        \draw[<->,thick,>=stealth] (r) -- node[above] {$Q, \tilde Q$} (center);
        % Antisimmetriche
        \node[box, minimum size=0.2cm] (square1r) at (0, 1.5) {};
        \node[box, minimum size=0.2cm] (square2r) at (0, 1.78) {};
        \node at (\x, \y) {$A, \tilde A$};
        \draw[<->,thick,>=stealth] (square1r.south) to[out=-90, in=90, looseness=0] (center.north);
        \end{tikzpicture}
        }
        \hfill
        % STEP 2
        \hspace{-30pt}
        \makebox[0.32\textwidth][c]{
        \begin{tikzpicture}[
            every node/.style={font=\footnotesize},
            box/.style={rectangle, draw, thick},
            box2/.style={rectangle, draw, dashed}
        ]
        \pgfmathsetmacro{\x}{0.4}
        \pgfmathsetmacro{\y}{1.6}
        \pgfmathsetmacro{\a}{2}
        \pgfmathsetmacro{\b}{1.7}
        % Nodi
        \node[box] (r) at (\a,0) {$4$};
        \node[box] (l) at (-\a,0) {$2$};
        \node[box] (botr) at (\a,-\b) {$2$};
        \node[fill=SUcol,circle,draw,thick] (center) at (0, 0) {$2n$};
        \node[fill=USPcol,circle,draw,thick] (bot) at (0, -\b) {$2n\!-\!2$};
        % Fondamentali
        \draw[<-,thick,>=stealth] (center) -- node[above] {$\tilde Q$} (r);
        \draw[->,thick,>=stealth] (center) -- node[above] {$Q$} (l);
        \draw[->,thick,>=stealth] (center) -- node[right] {$P$} (bot);
        \draw[-,thick,>=stealth] (bot) -- node[above] {$R_{3,4}$} (botr);
        % Antisimmetriche
        \node[box, minimum size=0.2cm] (square1r) at (0, 1.5) {};
        \node[box, minimum size=0.2cm] (square2r) at (0, 1.78) {};
        \node at (\x, \y) {$\tilde A$};
        \draw[->,thick,>=stealth] (square1r.south) to[out=-90, in=90, looseness=0] (center.north);
        \end{tikzpicture}
        }
        \hfill
        % STEP 3
        \hspace{-10pt}
        \makebox[0.32\textwidth][c]{
        \begin{tikzpicture}[
            every node/.style={font=\footnotesize},
            box/.style={rectangle, draw, thick},
            box2/.style={rectangle, draw, dashed}
        ]
        \pgfmathsetmacro{\x}{-0.7}
        \pgfmathsetmacro{\y}{1.6}
        \pgfmathsetmacro{\a}{2}
        \pgfmathsetmacro{\b}{1.5}
        % Nodi
        \node[box] (r) at (\a,0) {$2$};
         \node[box] (l) at (-\a,0) {$2$};
        \node[box] (ne) at (\a,\b) {$4$};
        \node[fill=USPcol,circle,draw,thick] (center) at (0, 0) {$2n\!-\!2$};
        % Fondamentali
        \draw[-,thick,>=stealth] (r) -- node[below] {$R_{3,4}$} (center);
        \draw[-,thick,>=stealth] (l) -- node[below] {$\Sigma_{2,F}$} (center);
        \draw[-,thick,>=stealth] (ne.south west) -- node[above, xshift=-2pt] {$\Sigma_{3,\mathrm{F}}$} (center);
        % Antisimmetriche
        \node[box, minimum size=0.2cm] (square1r) at (0, 1.5) {};
        \node[box, minimum size=0.2cm] (square2r) at (0, 1.78) {};
        \node at (\x, \y) {$\Sigma_{2,\mathrm{A}}$};
        \draw[-,thick,>=stealth] (square1r.south) to[out=-90, in=90, looseness=0] (center.north);
        \end{tikzpicture}
        }
    \end{minipage}
    \caption{In this Figure we provide a quiver description of the 4d confining gauge theory studied in this section. The original model corresponds to $\SU(2n)$ with 
    four fundamental flavors and one antisymmetric flavor. The theory has a non-trivial superpotential given in formula (\ref{spot4d1}). The second quiver corresponds to the model where we have deconfined the $\SU(2n)$ antisymmetric tensor. The third model is obtained after using the s-confining duality of \cite{Csaki:1996zb} on the $\SU(2n)$ gauge node.}
    \label{Fig4d}
\end{figure}
%%%%%%%%%%%%%%%%%%%%%%%%%%%%%%%%%%%%%%%%%%%%%%%%%%%%%%%%%%%%%%%%%%%%%%%

Here, in order to prove the 4d  duality we will use tensor deconfinement. Furthermore, the presence of the superpotential will trigger non-trivial RG flows in the dual phases. 
The 4d original model corresponds to $\SU(2n)$ SQCD with four fundamental flavors and one antisymmetric flavor.
We turn on the following superpotential 
\begin{equation}
\label{spot4d1}
W = A^{n-1} Q_3 Q_4 +\sum_{i=1}^{n-1} \beta_i \text{Tr} (\!A \tilde A)^{i}
+ s_A \Pf A + s_{\tilde A} \Pf \tilde A + M_{a} Q_{a} \tilde Q+
\gamma \tilde A^{n-2} \tilde Q^4 \,.
\end{equation}
with $a=1,2$.
We then deconfine the antisymmetric $A$ obtaining a new $\USp(2n-2)$ gauge group with an $\SU/\USp$ bifundamental $P$. In addition we consider two $\USp(2n-2)$ fundamentals $R_{3,4}$ as in Figure \ref{Fig4d}.
The original fields $A$ and $Q_{3,4}$ correspond in this phase to the combinations $P^2$ and $P R_{3,4}$ respectively.
The superpotential for this phase is
\begin{equation}
\label{spotII}
W =  \sum_{i=1}^{n-1} \beta_i \text{Tr} (P^2 \tilde A)^{i} + s_{\tilde A} \Pf \tilde A 
+ M_a Q_a \tilde Q+\gamma \tilde A^{n-2} \tilde Q^4\,.
\end{equation}
To proceed we notice that the $\SU(2n)$ gauge node has one conjugate antisymmetric, $2n$ fundamentals and $4$ antifundamentals. The model is then s-confining in terms of the following $\SU(2n)$ gauge invariant combinations
\begin{equation}
\begin{array}{lllll}
\Sigma_1 =\tilde A^{n-1} \tilde Q^2, \quad \quad &
\Sigma_{2,A} =\tilde A P^2, \quad\quad & 
\Sigma_{2,F} =\tilde A P Q,\quad \quad&
\Sigma_{2,s} =\tilde A Q^2 \\
\Sigma_{3,F} = \tilde Q P, \quad&
\Sigma_{3,s} = \tilde Q Q, \quad&
\Sigma_4 = Q^2 P^{2n-2},\quad\quad&
\Sigma_5 = \text{Pf } \tilde A,\quad\quad&
\Sigma_6 =  \tilde A^{n-2} \tilde Q^4
\end{array}
\end{equation} 
where $\Sigma_{2,A} $ is in the (tracless) antisymmetric representation 
of the leftover $\USp(2n-2)$ gauge group, while $\Sigma_{3,F} $ is in the fundamental of such $\USp(2n-2)$. 
The fields $\Sigma_{5,6}$ and $\Sigma_{3,s}$ are set to zero in the chiral ring by the superpotential (\ref{spotII}) and we can ignore them in the following.
The other $\Sigma$-fields  are $\USp(2n-2)$
singlets.
The superpotential for this phase, after integrating out the massive fields, is
\begin{equation}
\label{finalW}
W = \Sigma_1 \Sigma_{3,F}^2 (\Sigma_{2,s} \Sigma_{2,A}^{n-2}+\Sigma_{2,A}^{n-3} \Sigma_{2,F}^2)+\sum_{i=2}^{n-1} \beta_i \Tr (\Sigma_{2,A})^{i}+\Sigma_4 \Sigma_1^2 \, .
\end{equation}
In this way we have obtained a duality between $\SU(2n)$ with an antisymmetric flavor, four fundamental flavors and superpotential 
(\ref{spot4d1}), and an $\USp(2n-2)$ gauge theory with eight fundamental, one antisymmetric and superpotential (\ref{finalW}).
This last theory is not confining but it has 72 many (self \!\footnote{In the sense that the dual theories have an $\USp(2n-2)$ gauge group with an antisymmetric and eight fundamentals but a different structure of singlets and superpotential.}--) dual phases (with a different structure of flippers. Indeed the global symmetry is known to enhance to $E_7 \times \U(1)$ \cite{Razamat:2017hda}).

This duality gives origin to the 2d gauge/LG duality discussed in \cite{Amariti:2024usp} once we compactify the models using the prescription of \cite{Gadde:2015wta} by fixing the $R$ charges of the fields $R_{\tilde Q,\tilde A,A}=0$ and $R_{Q}=1$. The former become chirals in 2d $\mathcal{N}=(0,2)$ 
while the latter do not give rise to any 2d field.
Then, the singlets $\beta_i$, $s_A$, $\tilde s_{\tilde A}$  and $\gamma$ have $R$ charge $R=2$ and become Fermi fields in 2d, that we denote as $\Psi_{\beta_i}$, $\Psi_{s_A}$,$\Psi_{s_{\tilde A}}$ and $\Psi_{\gamma}$ respectively. 
The singlets $M_a$, on the other hand, have $R$ charge $1$ and they do not give rise to  any 2d field.

The 4d superpotential (\ref{spot4d1})  of the electric theory gives origin to the 2d $J$-terms
\begin{equation}
\label{J2d1}
J_{\psi_{\beta_i}} =  \text{Tr} (A \tilde A)^{i}, \quad 
J_{\psi_{s_A}}= \Pf  A, \quad 
J_{\psi_{s_{\tilde A}}}= \Pf \tilde A, \quad 
J_{\psi_\gamma} =  \tilde A^{n-2} \tilde Q^4,\,
\end{equation}
with $i=2,\dots,n-1$, where we kept the same names for the 2d $\mathcal{N}=(0,2)$  chirals of the corresponding 4d $\mathcal{N}=1$ chiral multiplets.

By following the duality map, in the dual $\USp(2n-2)$ gauge theory the antisymmetric $\Sigma_{2,A}$, the fundamentals $\Sigma_{3,F}$ and the singlet $\Sigma_1$ have $R$ charge $R=0$ and they survive as 2d $\mathcal{N}=(0,2)$ chirals while the fields
$\Sigma_{2,F}$ and $R_{3,4}$ have $R$ charge $R=1$ and do not survive in 2d.

On the other hand, the singlets $\Sigma_{2,s}$  and $\Sigma_4$ have $R$ charge $R=2$ and they give origin to the Fermi fields 
$\psi_{\Sigma_{2,s}}$  and $\psi_{\Sigma_4}$ respectively.
The 4d superpotential (\ref{finalW})  of the dual theory gives origin to the 2d $J$-terms
\begin{equation}
\label{J2d2}
J_{\psi_{\beta_i}} =  \Tr \left(\Sigma_{2,A}\right)^{i},\quad
J_{\psi_{\Sigma_{2,s}}} =  \Sigma_1 \Sigma_{3,F}^2 \Sigma_{2,A}^{n-2},\quad
J_{\psi_{\Sigma_{4}}}= \Sigma_1^2\,, 
\end{equation}
with $i=2,\dots,n-2$,
where again we kept the same names for the 2d $\mathcal{N}=(0,2)$  chirals of the corresponding 4d $\mathcal{N}=1$ chiral multiplets.

The $J$-terms in (\ref{J2d1}) and  (\ref{J2d2})
are equivalent to the 2d superpotentials discussed in formulas (4.72) and 
(4.76) of \cite{Amariti:2024usp}, where the dual phase was further dualized to an LG.

Summarizing, in this section we have obtained a parent 4d duality between two gauge theories that, once compactified to 2d, gives origin to the 2d gauge/LG duality studied in \cite{Amariti:2024usp}. 
The 4d duality in this case does not correspond to an s-confining  theory, even if it is still possible that there exist other 4d s-confining models that can be reduced to the same 2d dualities. 
We leave such possibility to future investigations.

\section{Dualities involving antisymmetric Fermi fields}

In this section we focus on $\SU(n)$ and $\USp(2n)$ gauge theories with (anti)-fundamental chirals and with a Fermi field in the antisymmetric representation.
The situation is different from the cases treated above and in \cite{Sacchi:2020pet,Jiang:2024ifv,Amariti:2024usp}, because in these cases the technique of tensor deconfinement does not apply. 
This is due to the fact that, to the best of our knowledge, the fundamental gauge/LG dualities proposed in the literature never include antisymmetric Fermi fields charged under a non-abelian flavor symmetry. 
This constitutes an obstruction to proving dualities using the deconfining technique.
Despite this fact, here we observe that some dualities with charged antisymmetric Fermi fields can be obtained by using the prescription of \cite{Gadde:2015wta} from 4d parent dualities proposed in the literature starting from the matching of the superconformal index.
We further check the 2d dualities obtained by matching the 't Hooft anomalies.

\subsection{A gauge/LG duality with an antisymmetric Fermi}

The first 4d duality under investigation has been proposed in \cite{
SPIRIDONOV200691,Spiridonov:2009za}
and further investigated in \cite{Nazzal:2021tiu}, where it was referred to as SWV duality. 
A proof of the dualities in terms of other fundamental dualities was then provided in \cite{Amariti:2023wts}.

The electric theory in this duality is an  $\SU(n+1)$  gauge theory with an antisymmetric $A$, $2n$ antifundamentals $\tilde Q$  and $n+3$ fundamentals $Q$ with 4d superpotential
\begin{equation}
W_{\mathrm{SWV}}^{\mathrm{ele}} =A \tilde Q^2 \,.
\end{equation}

The dual theory is a WZ model, described by the electric baryon $B =Q^{n+1}$ and meson $M=Q \tilde Q$ with superpotential 
\begin{equation}
W_{\mathrm{SWV}}^{\mathrm{mag}} =  B M^2\,.
\end{equation}

We are interested in a flipped version of this duality, where we break the global $\SU(n+3)$ symmetry by partially flipping the baryonic operator $B$.
On the electric side, we consider the split $Q \rightarrow (Q_1,Q_2)$,
corresponding to the decomposition  of  $\SU(n+3) \rightarrow \SU(n+1) \times \SU(2)$.
Such symmetry breaking pattern is enforced in the electric superpotential by adding a singlet $b$ and considering the flipped superpotential 
\begin{equation}
\label{baryonicflipper}
W_{\mathrm{SWV}}^{\mathrm{ele}}=A \tilde Q^2 + b Q_1^{n+1}\,.
\end{equation} 

On the magnetic side, under the symmetry breaking pattern above, the baryon $B$ is decomposed into three fields: an $\SU(n+1)$ antisymmetric $B_A$, an $\SU(n+1) \times \SU(2)$ bifundamental $B_V$ and a singlet $B_S$.
Similarly, the meson $M$ splits into $M_1$ and $M_2$ respectively in the fundamental representation of $\SU(n+1)$ and $\SU(2)$.
The dual superpotential becomes 
\begin{equation}
W_{\mathrm{SWV}}^{\mathrm{mag}} =  B_A M_1^2 + B_V M_1 M_2 +B_S M_2^2+ b B_S\,,
\end{equation}
and after integrating out the massive fields it reduces to 
\begin{equation}
W_{\mathrm{SWV}}^{\mathrm{mag}} =  B_A M_1^2 + B_V M_1 M_2 \,,
\end{equation}
as expected from the addition of the baryonic flipper in (\ref{baryonicflipper}).

The superconformal index still matches across the dual phases, as the addition of the flipper just moves the relative elliptic gamma function from one side of the duality to the other without modifying the balancing condition, or equivalently, the requirements from the anomaly freedom.

In the next step we fix the non-negative integer $R$ charges of the flipped version of the duality. We fix the $R$ charge of the antisymmetric to $R_A=2$ and, consistently with the anomaly freedom condition, we  fix the $R$ charges of the fields $Q_2$ to $R_{Q_2}=1$. The charges of $Q_1$ and $\tilde Q$ are then vanishing, consistently with the global symmetries. 
The singlet $b$ in (\ref{baryonicflipper}) on the other hand has $R_b=2$.

The 2d theory obtained using such assignation of charges is an $\SU(n+1)$ electric gauge theory with an antisymmetric Fermi $\Psi_A$, $2n$ antifundamental chirals $\tilde Q$ and $n+1$ fundamental chirals $Q$ with 2d $J$-terms 
\begin{equation}
J_{\Psi_A}= \tilde Q^2\,, \qquad J_{\Psi_B}= Q^{n+1}\,,
\end{equation}
where the Fermi $\psi_b$ descend from the 4d singlet $b$.

In the 4d dual theory the $R$-charges that can be read from the duality map are $R_{B_A}=2$, $R_{M_1}=0$ and $R_{M_2} = R_{B_V} = 1$. The field $B_A$ becomes a Fermi field, denoted as $\psi$, in the antisymmetric representation of  the $\SU(n+1)$ flavor symmetry. The meson $M_1$ becomes a chiral, denoted by $\phi_M$, in the antifundamental representation of  the $\SU(n+1)$ flavor symmetry (and in the fundamental of the other $\USp(2n)$ flavor symmetry). On the other hand, the fields $B_S$ and $M_2$ do not give rise to any 2d field.

The magnetic description is an LG model with $J$-term $J_\Psi= \Phi_M^2$ where we can further associate $\Phi_M$ to the combination $Q \tilde Q$.
This relation and the structure of the $J$-term fix the charges of the fields in the dual picture. This results into the table of global charges below.

\begin{equation}
\begin{array}{c|c|ccccc}
              &  \USp(2n)   &   \SU(n+1)   &   \U(1)_A   &  \U(1)_Q &  \U(1)_R \\
                \hline
\Psi_A          &    \cdot                   &    \cdot      &        1          &     0         &     1         \\
\tilde Q    &   \square               &     \cdot      &  -\frac{1}{2} &      0        &     0         \\
Q                 &    \cdot                  &  \square     &        0         &      1        &      0        \\
\Psi_B        &   \cdot                   &    \cdot      &         0        &       -n-1       &        1      \\
\hline
\Phi_M &    \square                     &     \square      &      1      &      -\frac{1}{2}       &        0      \\
\Psi &   \cdot                   &    \overline{\begin{array}{c}
\square \vspace{-2.85mm} \\
\square 
\end{array} }      &         -2        &      1       &        1      \\

\end{array}
\end{equation}

%%%%%%%%%%%%%%%%%%%%%%%%%%%%%%%%%%%%%%%%%%%%%%%%%%%%%%%%%%%%%%%%%%%%%%%
% FIGURE: Duality between SU(n+1) w 1 as F and LG
\begin{figure}[h!]
    \centering
        % STEP 1
        \begin{minipage}{0.45\textwidth}
            \centering
            \makebox[\textwidth][c]{
            \begin{tikzpicture}[
                every node/.style={font=\footnotesize},
                box/.style={rectangle, draw, thick},
                box2/.style={rectangle, draw, dashed}
            ]
            \pgfmathsetmacro{\x}{0.6}
            \pgfmathsetmacro{\y}{1.6}
            \pgfmathsetmacro{\a}{2}
            % Nodi
            \node[box] (l) at (-\a,0) {$n\!+\!1$};
            \node[box] (r) at (\a,0) {$2n$};
            \node[fill=SUcol,circle,draw,thick] (center) at (0, 0)
             {$n\!+\!1$};
            \node[circle, fill=black, inner sep=2pt] (dot) at (-\a, 1.5) {};
            % Fondamentali
            \draw[<-,thick,>=stealth] (l) -- node[above] {$Q$} (center);
            \draw[->,thick,>=stealth] (r) -- node[above] {$\tilde Q$} (center);
            % Fermi antisimmetrici
            \node[box2, minimum size=0.2cm] (square1r) at (0, 1.5) {};
            \node[box2, minimum size=0.2cm] (square2r) at (0, 1.78) {};
            \node at (\x, \y) {$\Psi_A$};
            \node at (-\a+0.5, \y) {$\Psi_B$};
            \draw[<-,dashed,>=stealth] (square1r.south) to[out=-90, in=90, looseness=0] (center.north);
            \draw[-,dashed,>=stealth] (l.north) to[out=90, in=-90, looseness=0] (dot.south);
            \end{tikzpicture}
            }
        \end{minipage}
        % STEP 2
        \begin{minipage}{0.45\textwidth}
            \centering
            \makebox[\textwidth][c]{
            \begin{tikzpicture}[
                every node/.style={font=\footnotesize},
                box/.style={rectangle, draw, thick},
                box2/.style={rectangle, draw, dashed}
            ]
            \pgfmathsetmacro{\x}{0.6}
            \pgfmathsetmacro{\y}{1.6}
            \pgfmathsetmacro{\a}{1.5}
            % Nodi
            \node[box] (l) at (-\a,0) {$n\!+\!1$};
            \node[box] (r) at (\a,0) {$2n$};
            % Fondamentali
            \draw[->,thick,>=stealth] (r) -- node[above] {$\phi_M$} (l);
            % Fermi antisimmetrici
            \node[box2, minimum size=0.2cm] (square1r) at (-\a, 1.5) {};
            \node[box2, minimum size=0.2cm] (square2r) at (-\a, 1.78) {};
            \node at (-\a+0.5, \y) {$\Psi$};
            \draw[<-,dashed,>=stealth] (square1r.south) to[out=-90, in=90, looseness=0] (l.north);
            \end{tikzpicture}
            %\caption{}
            }
        \end{minipage}
    \caption{In this figure we represent the quivers associated to the duality between $\SU(n+1)$ with an antisymmetric Fermi and the LG model. Dashed lines are for Fermi fields while continuous lines are for chiral multiplets. The dots refer to the singlets.}
    \label{Figconfining2}
\end{figure}
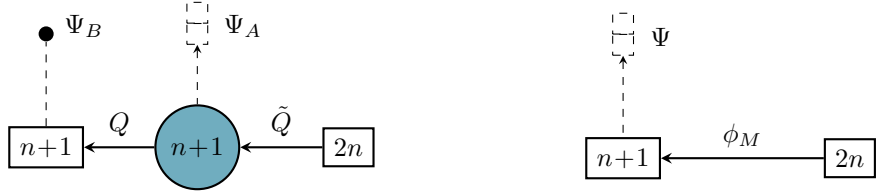
%%%%%%%%%%%%%%%%%%%%%%%%%%%%%%%%%%%%%%%%%%%%%%%%%%%%%%%%%%%%%%%%%%%%%%%

We now provide some checks of the duality proposed here obtained from the reduction of the parent 4d SWV duality. First of all, we have computed the 't Hooft anomalies and checked their matching. Explicitly, we have 
\begin{align}
&\kappa_{\SU(n+1)^2}=\frac{n+1}{2}\,,
\quad
&&\kappa_{\USp(2n)^2}=n+1\,,
\quad
&&\kappa_{QQ}=\kappa_{AA}=\kappa_{QA}=0\,,
\nonumber \\
&\kappa_{RR}=\frac{3n(n+1)}{2}\,,
\quad
&&\kappa_{AR}=\frac{n(n+1)}{2}\,,
\quad
&&\kappa_{RQ}=-n(n+1)\,.
\end{align}

The validity of the duality also implies the relation between the ellitpic genera
\begin{eqnarray}
I_{\text{ele}} &=&
\frac{(q;q)_\infty^{2n+2}}{(n+1)!} \oint_{\mathrm{JK}} \prod_{i=1}^{n+1} \frac{\dd z_i}{2 \pi \mi z_i} 
\frac{\prod_{1\leq i<j \leq n+1}\theta(q T z_i z_j)\theta((z_i/z_j)^{\pm 1})}{\prod_{i=1}^{n+1} (\prod_{m=1}^{n+1} \theta(z_i s_m) \prod_{k=1}^{n} \theta(t_k^{\pm 1}z_i^{-1}/\sqrt T))}
\nonumber \\
& \times &  \theta\left(q \prod_{\ell=1}^{n+1} s_\ell^{-1}\right) =\frac{\prod_{1\leq m < \ell \leq n+1} \theta(q T s_m^{-1} s_\ell^{-1})}{\prod_{m=1}^{n+1}\prod_{k=1}^{n} \theta(s_m t_k^{\pm 1}/\sqrt T)} = I_{\text{mag}}
\,,
\end{eqnarray}
with $\prod_{i=1}^{n+1} z_i=1$.
We explitly checked the validity of this identity for the cases $n=1,2$, where the duality reduces to a flipped version of the ones studied in subsection \ref{SUnGen}. Indeed, in the first case the antisymmetric Fermi disappears, while in the second case it is equivalent to an antifundamental Fermi field.
 
A further check can be obtained through a non-anomalous gauging of a subgroup of the $\USp(2n)$ flavor symmetry group.
For example, let us start by considering the case of $n=2m-1$ such that the flavor symmetry is $\USp(4m-2)$.
Then, we gauge an $\USp(2m-2)$ subgroup of this symmetry. The final $\USp(2m) \times \USp(2m-2)$ is then non-anomalous and the superpotential for this model becomes
\begin{equation}
W = \Psi_A \tilde Q_1^2 + \Psi_A \tilde Q_2^2 + \Psi_B Q^{2m}\,,
\end{equation} 
where $Q_1$ is the $\SU(2m) \times \USp(2m-2)$ bifundamentals while $Q_2$ represents the other $2m$ $\SU(2m)$ fundamentals whose flavor symmetry is not gauged.

The $\USp(2m-2)$ gauge theory has $2m$ fundamentals and it can be dualized into an LG model. The superpotential after this duality becomes 
\begin{equation}
W = \Psi_A  A + \Psi_A \tilde Q_2^2 +\psi_{\sigma} \text{Pf} A+ \Psi_B Q^{2m}\,.
\end{equation} 
The Fermi $\Psi_A$ and the chiral antisymmetric $A = Q_1^2$ are massive and can be integrated out, leaving us with 
\begin{equation}
W = \Psi_A  A + \Psi_A \tilde Q_2^2 +\psi_{\sigma} \text{Pf} A+ \Psi_B Q^{2m}\,.
\end{equation}
By integrating out the massive fields the superpotential for the remaining $\SU(2m)$ gauge theory becomes
\begin{equation}
W =   \psi_{\sigma} \tilde Q_2^{2m}+ \Psi_B Q^{2m}\,,
\end{equation}
that is dual to an LG with a meson $\Phi_M = Q \tilde Q_2$
with superpotential
\begin{equation}
\label{ateso}
W =   \hat \Psi \det \Phi_M\,.
\end{equation}
The same gauging can be performed in the dual phase in which the model becomes
\begin{equation}
 W = \Psi (\Phi_{M_1}^2+\Phi_{M_2}^2)\,,
 \end{equation}
 where $\Phi_{M_1}$ is charged under the $\USp(2m-2)$ gauge symmetry, while  $\Phi_{M_2}$ is charged under the  $\USp(2m)$
 flavor symmetry.
 The $\USp(2m-2)$ gauge group is dual to an LG model and we get
 \begin{equation}
 W = \Psi (\mathcal{A} +\Phi_{M_2}^2)+\Psi_0 \Pf \mathcal{A}\,,
 \end{equation}
 with $\mathcal{A} =\Phi_{M_1}^2$.
 By integrating out the massive fields, such superpotential becomes
  \begin{equation}
 W =\Psi_0 \Pf(\Phi_{M_2})^2 = \Psi_0 \det\Phi_{M_2}\,,
 \end{equation}
 which is equivalent to (\ref{ateso}).

\subsection{A gauge/gauge duality with an antisymmetric Fermi}

The second duality  in presence of charged antisymmetric Fermi fields that we consider relates two symplectic gauge theories with $\USp(2n)$ and $\USp(2n-2)$ gauge group respectively.
Again, the duality is obtained from a 4d parent, that was originally proposed in \cite{Spiridonov:2009za} from the matching of the associated superconformal indices obtained in \cite{van2009elliptic}.
The duality was then derived from a physical perspective in \cite{Amariti:2024sde}.

The 4d duality relates models characterized by symplectic gauge groups, an antisymmetric field together with fundamental flavors which interact with the antisymmetric, breaking the flavor symmetry with various possible patterns.
Here we focus on a specific pattern, suitable to give origin to a 2d $\mathcal{N}=(0,2)$ duality with a charged antisymmetric Fermi field.
In this case the flavor symmetry is $\SU(4) \times \USp(4n-2)$. There are four fundamentals $P$ and $4n-2$ fundamentals $Q$, which interact with the $\USp(2n)$ antisymmetric $A$ through a superpotential $W= A Q^2$.
The dual picture corresponds to an $\USp(2n-2)$ gauge theory with four fundamentals $p$, $4n-2$ fundamentals $q$ and an antisymmetric, denoted as $a$. The superpotential of this dual phase is 
\begin{equation}
W = N M^2 + M p q + a q^2,
\end{equation}
where the singlets $N$ and $M$ correspond in the electric phase to the gauge invariant combinations $P^2$ and $P Q$.

Again, we consider a flipped version of this duality by breaking the $\SU(4)$ flavor symmetry to $\SU(2)_1 \times \SU(2)_2$, denoting the two fundamentals as $P_1$ and $P_2$ and flipping the combination $P_2^2$ by adding a singlet $B$ in the electric phase. The electric theory has then superpotential 
\begin{equation}
W_{USp(2n)} = A Q^2 + B P_2^2.
\end{equation}
On the magnetic side the superpotential becomes
\begin{equation}
W_{USp(2n-2)} =  M_1 p_1 q+M_2 p_2 q + a q^2 + M_1^2 N_1 + M_1 M_2 L,
\end{equation}
where $M_{1} = Q P_1$, $M_{2} = Q P_2$, $L = P_1 P_2$ and $N_1 = P_1^2$.

We then compactify the theory on $S^2$ by fixing $R_{P_1} = R_Q=0$, $R_A=R_S=2$ and
 $R_{P_2}=1$. In this way we obtain a 2d $\mathcal{N}=(0,2)$  $\USp(2n)$ electric  gauge theory with an antisymmetric Fermi $\Psi_A$, $4n-2$ fundamental chirals $Q$ and $2$ fundamental chirals $P_1$ with $J$-terms $Y_{\Psi_A} = Q^2$ and $J_{\Psi_B} =  P_1^2$.
In the dual theory the $R$ charges, read from the duality map, are $R_{q}=R_{M_1}=0$, $R_a=R_{N_1} =R_{p_1}=2$ and  $R_{M_2}=R_L=R_{p_2}=1$.
The magnetic description is an $\USp(2n-2)$ gauge theory with an antisymmetric Fermi $\Psi_a$, $4n-2$ fundamental chirals $q$ and $4$ fundamentals Fermi multiplets $\Psi_{p_1}$, with $J$-terms $ J_{ \Psi_{p_1}}=q \Phi_{M_1}$, $J_{ \Psi_{a}}=q^2$ and $J_{\Psi_{N_1}}= \Phi_{M_1}^2$.

%%%%%%%%%%%%%%%%%%%%%%%%%%%%%%%%%%%%%%%%%%%%%%%%%%%%%%%%%%%%%%%%%%%%%%%
% FIGURE: Duality between SU(n+1) w 1 as F and LG
\begin{figure}[h!]
    \centering
        % STEP 1
        \begin{minipage}{0.45\textwidth}
            \centering
            \makebox[\textwidth][c]{
            \begin{tikzpicture}[
                every node/.style={font=\footnotesize},
                box/.style={rectangle, draw, thick},
                box2/.style={rectangle, draw, dashed}
            ]
            \pgfmathsetmacro{\x}{0.6}
            \pgfmathsetmacro{\y}{1.6}
            \pgfmathsetmacro{\a}{2}
            % Nodi
            \node[box] (l) at (-\a,0) {$2$};
            \node[box] (r) at (\a,0) {$4n\!-\!2$};
            \node[fill=USPcol,circle,draw,thick] (center) at (0, 0)
             {$2n$};
            \node[circle, fill=black, inner sep=2pt] (dot) at (-\a, 1.5) {};
            % Fondamentali
            \draw[-,thick,>=stealth] (l) -- node[above] {$P_{1}$} (center);
            \draw[-,thick,>=stealth] (r) -- node[above] {$Q$} (center);
            % Fermi antisimmetrici
            \node[box2, minimum size=0.2cm] (square1r) at (0, 1.5) {};
            \node[box2, minimum size=0.2cm] (square2r) at (0, 1.78) {};
            \node at (\x, \y) {$\Psi_A$};
            \node at (-\a+0.5, \y) {$\Psi_B$};
            \draw[-,dashed,>=stealth] (square1r.south) to[out=-90, in=90, looseness=0] (center.north);
            \draw[-,dashed,>=stealth] (l.north) to[out=90, in=-90, looseness=0] (dot.south);
            \end{tikzpicture}
            }
        \end{minipage}
        % STEP 2
        \begin{minipage}{0.45\textwidth}
            \centering
            \makebox[\textwidth][c]{
            \begin{tikzpicture}[
                every node/.style={font=\footnotesize},
                box/.style={rectangle, draw, thick},
                box2/.style={rectangle, draw, dashed}
            ]
            \pgfmathsetmacro{\x}{0.6}
            \pgfmathsetmacro{\y}{1.6}
            \pgfmathsetmacro{\a}{2}
            % Nodi
            \node[box] (l) at (-\a,0) {$2$};
            \node[box] (r) at (\a,0) {$4n\!-\!2$};
            \node[fill=USPcol,circle,draw,thick] (center) at (0, 0){$2n\!-\!2$};
            \node[circle, fill=black, inner sep=2pt] (dot) at (-\a, 1.5) {};
            % Fondamentali
            \draw[-,thick,>=stealth] (r) -- node[above] {$q$} (center);
            % Mesoni
            \draw[-,thick,>=stealth] (l.south) to[out=-60, in=-120, looseness=1] (r.south);
            \node at (0, -1) {$\Phi_{M_1}$};
            % Fermi
            \draw[-,dashed,>=stealth] (l) -- node[above] {$\Psi_{p_{1}}$} (center);
            % Fermi antisimmetrici
            \node[box2, minimum size=0.2cm] (square1r) at (0, 1.5) {};
            \node[box2, minimum size=0.2cm] (square2r) at (0, 1.78) {};
            \node at (\x, \y) {$\Psi_a$};
            \node at (-\a+0.5, \y) {$\Psi_{N_1}$};
            \draw[-,dashed,>=stealth] (square1r.south) to[out=-90, in=90, looseness=0] (center.north);
            \draw[-,dashed,>=stealth] (l.north) to[out=90, in=-90, looseness=0] (dot.south);
            \end{tikzpicture}
            }
        \end{minipage}
    \caption{In this figure we represent the quivers associated to the duality between $\USp(2n)$ and $\USp(2n-2)$ with an antisymmetric Fermi. Dashed lines are for Fermi fields while continuous lines are for chiral multiplets. The dots refer to the singlets.}
    \label{Figduality}
\end{figure}
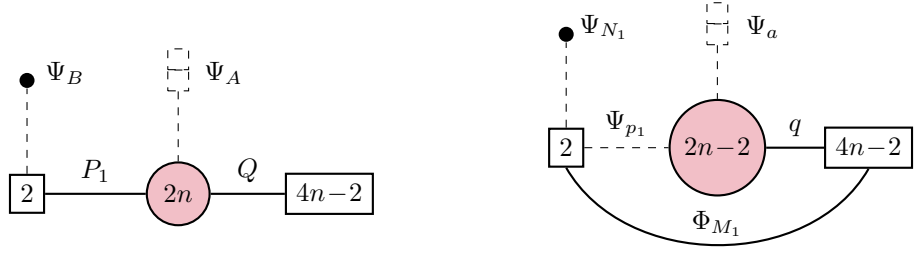
%%%%%%%%%%%%%%%%%%%%%%%%%%%%%%%%%%%%%%%%%%%%%%%%%%%%%%%%%%%%%%%%%%%%%%%
The duality map in 2d can be read once we associate the chiral $ \Phi_{M_1}$ to the electric mesonic combination $P_1 Q$. The global charges for this duality are then summarized in the table below.
\begin{equation}
\begin{array}{c|c|ccccc}
              &  \USp(2n)  & \USp(4n-2)    &   \SU(2)   &   \U(1)_A   &  \U(1)_P &  \U(1)_R \\
                \hline
\Psi_A           & \begin{array}{c}
\square \vspace{-2.85mm} \\
\square 
\end{array}   &    \cdot                   &    \cdot      &        1          &     0         &     1         \\
Q     &\square &   \square               &     \cdot      &  -\frac{1}{2} &      0        &     0         \\
P_1             &   \square      &    \cdot                  &  \square     &        0         &      1        &      0        \\
\Psi_B       &    \cdot     &   \cdot                   &    \cdot      &         0        &       -2       &        1      \\
\hline
   &\USp(2n-2)  & \USp(4n-2)    &   \SU(2)   &   \U(1)_A   &  \U(1)_P &  \U(1)_R \\
                \hline
 q              &    \square      &     \square    &  \cdot          &    -\frac{1}{2}  &    0       &       0     \\
 \Phi_{M_1}    &     \cdot      &       \square     &    \square      &   -\frac{1}{2}  &   1        &       0    \\
 \Psi_{p_1}    &     \square     &      \cdot       &    \square     &      1              &    1       &        1   \\
 \Psi_a    &   \begin{array}{c}
\square \vspace{-2.85mm} \\
\square 
\end{array}        &         \cdot             &     \cdot        &      1              &    0       &        1  \\
 \Psi_{N_1}    &     \cdot      &     \cdot             &     \cdot        &      1              &    -2      &        1 \\
\end{array}
\end{equation}
As a check we computed the 't Hooft anomalies for this duality and we have showed that they match. We have 
\begin{align}
&
\kappa_{PP} = \kappa_{AA} = \kappa_{AP} = 0\,,
\quad
&&\kappa_{RP} =2(1-2n)\,,
\quad \,\,
&&\kappa_{RA} =n(2n-1)\,,
\nonumber \\
&
\kappa_{\SU(2)^2} =n\,,
\quad\quad\quad\quad\quad\quad
&&\kappa_{\USp(4n-2)^2} =2n\,,
\quad
&&\kappa_{RR} =4n^2-1\,.
\end{align}
For completeness we also provide the expression for the elliptic genus in the electric and in the magnetic case. On the electric side we have
\begin{eqnarray}
I_{\text{ele}} = \frac{(q;q)_{\infty}^{2n} \theta(q T)^{n-1}\theta \left(\frac{q}{t_1 t_2}\right)}{n!  \, 2^n } \oint_{\mathrm{JK}} \prod_{i=1}^{n} \frac{\prod_{i<j}\theta(z_i^{\pm 1}z_j^{\pm 1}) \theta(q T z_i^{\pm 1}z_j^{\pm 1})\prod_{i=1}^{n} \theta(z_i^{\pm 2}) }{\prod_{i=1}^n\prod_{a=1}^{2}  \theta(z_{i}^{\pm 1} t_{a}) \prod_{\ell=1}^{2n-1}\theta(z_i^{\pm 1} T^{-\frac{1}{2}} s_\ell^{\pm 1} )}\,
\end{eqnarray}
while on the dual side we have
\begin{eqnarray}
&&
I_{\text{mag}} = \frac{(q;q)_{\infty}^{2n-2} \theta(q T)^{n-2} \theta \left(\frac{q T}{t_1 t_2}\right) }{(n-1)! \, 2^{n-1} \prod_{\ell=1}
^{2n-1} \theta(s_\ell^{\pm 1} t_{1,2}T^{-\frac{1}{2}}  )}
\nonumber \\
&& \oint_{\mathrm{JK}} \prod_{i=1}^{n-1} \frac{\prod_{i<j}\theta(z_i^{\pm 1}z_j^{\pm 1}) \theta(q T z_i^{\pm 1}z_j^{\pm 1})\prod_{i=1}^{n-1} \left( \theta(z_i^{\pm 2})\prod_{a=1}^{2} \theta\left(\frac{q T}{t_{a}} z_i^{\pm 1 }\right) \right)}{\prod_{i=1}^{n-1}  \prod_{\ell=1}^{2n-1}\theta(z_i^{\pm 1} T^{-\frac{1}{2}}  s_\ell^{\pm 1 })}\,.
\end{eqnarray}

\section{Conclusions}

In this paper we have studied 2d $\mathcal{N}=(0,2)$ dualities with antisymmetric matter, extending the program started in \cite{Amariti:2024usp}. 
The first class of dualities under investigation correspond to $\SU(N)$ gauge theories with 2 antisymmetric, $n_f$ fundamental and $n_a$ antifundamental chirals, with $n_f +n_a=4$.
Our main claim is that such models have a dual description in terms of LG models, with chiral and Fermi multiplets and non-trivial $J$-terms. Such gauge/LG dualities cannot in general be predicted from the prescription of \cite{Gadde:2015wta} by a 4d/2d reduction of s-confining dualities and a check of their validity requires, in principle, a pure 2d analysis.
We have checked such claims by verifying the 't Hooft anomaly matching and by studying the elliptic genus.
Our analysis has shown that the matching of the elliptic genera follows from other \emph{basic} identities that can be obtained from the dimensional reduction of 4d s-confining dualities.
By assuming the matching of the 4d index on $S^2 \times T^2$ defined in 
\cite{Closset:2013sxa,Benini:2015noa,Honda:2015yha}, the matching of the elliptic genera for such \emph{basic} dualities follows from the prescription of \cite{Gadde:2015wta}. 
The basic identity employed here have been reviewed in section \ref{basic}.
A crucial relation is the one discussed in subsection \ref{USpallaSacchi}. Despite the fact that the field content in this case admits an additional axial symmetry, the identity \ref{idsacchi} holds only if this extra symmetry is absent. This suggests that, on the field theory side, some non-perturbative effect must be invoked on the gauge side of the duality to lift such dangerous symmetry.
As originally discussed in \cite{Gadde:2015wta,Sacchi:2020pet}, this expectation is reminiscent of similar results obtained in the circle reduction of 4d dualities, keeping the effects of the KK monopole when considering the 3d effective dualities at finite size of the circle \cite{Aharony:2013dha}. 
Here we have used the \emph{basic} identity for this 2d duality under the assumption that such a non-perturbative effect lifts the axial symmetry from the spectrum. As we commented in the introduction one can 
think of such a 2d duality as arising from a 2d boundary one \cite{Dimofte:2017tpi} of the 3d effective theory on the circle. In such a case the matching of the half-index does not depend on the axial symmetry, which is expected to be lifted by the presence of the monopole superpotential in the bulk theory. We refer the reader to \cite{Okazaki:2023hiv} for examples of this kind. 
In this paper we have left to future investigation an appropriate discussion on such non-perturbative effect, and looked at the consequences of the duality of subsection \ref{USpallaSacchi}.
We have observed that, by deconfining the antisymmetric chirals using this duality and then dualizing the original $\SU(N)$ gauge node, there is no need to call for non-perturbative effects anymore, and it is then possible to re--confine the antisymmetric chirals using the ordinary gauge/LG for $\USp(2n)$  with $2n+2$ fundamental chirals.
By iterating the procedure, we have been able to find the expected identities for the proposed dualities with $n_f=4$ and $n_a=0$. 
Furthermore, we have shown that all the other dualities with $n_a>0$ arise as a  consequence of the ones with $n_a=0$. 

In the second part of the paper we studied a 4d duality between a $\SU(2n)$ theory with an antisymmetric and four fundamental flavors, and a $\USp(2n)$ one with an antisymmetric and eight fundamentals.
In absence of a superpotential the former has a nine dimensional Cartan for the non-anomalous non--R global symmetry group, compatible with the enhancement to $D_6 \times \U(1)^3$ in presence of an opportune set of flippers \cite{Razamat:2018gbu}. Analogously, the latter is compatible with an $E_7 \times \U(1)$ enhancement \cite{Razamat:2017hda}. 
Here we have broken the nine dimensional Cartan of the $\SU(2n)$ gauge theory by introducing a 4d superpotential term between an antisymmetric and two fundamentals. Such a breaking has allowed us to state a 4d duality using the tensor deconfinement technique. Despite the fact that such a duality can be interesting \emph{per se}, here we have studied its fate upon compactification on $S^2$ with non-negative integer R-charges through the prescription of \cite{Gadde:2015wta}. We have found that 
$\SU(2n)$ with an antisymmetric flavor and four fundamental chirals is dual symplectic gauge theory is related to the LG theory expected from the analysis of \cite{Amariti:2024usp}. In this sense the 4d duality obtained here is a parent of the 2d duality proposed in \cite{Amariti:2024usp}.
In the last part of the paper we obtained 2d dualities in presence of charged antisymmetric Fermi fields. Such dualities are especially interesting because it is not possible to deduce them from a pure 2d perspective by using the deconfining technique, or in other words, because we are not aware of any candidate \emph{basic} 2d duality with Fermi fields in the antisymmetric representation of the flavor symmetry.

We conclude with some general comments on the IR dynamics.
We expect that the gauge/LG dualities discussed in the paper give rise to interacting CFT in the IR.
The reason behind such expectation is that in presence of non-compact target spaces
the global symmetries associated to such non-compact directions get removed from the c-extremization procedure. In particular, the trial R-symmetry discussed in the various example is also the exact one. The central charge 
is then positive and there is no violation of the unitarity bounds.  
The presence of a non-compact target space implies also that we cannot turn off the fugacities in the matching of the elliptic genera. This signals that the dualities discussed here have to be interpreted as dualities in the mass deformed case, where we are only in presence of isolated vacua. Similar observations 
have been made in the literature for dualities with charged antisymmetric chiral fields in \cite{Sacchi:2020pet,Jiang:2024ifv,Amariti:2024usp}.

\section*{Acknowledgments}
The work of the authors has been supported in part by the Italian Ministero dell'Istruzione, Università e Ricerca (MIUR), in part by Istituto Nazionale di Fisica Nucleare (INFN) through the “Gauge Theories, Strings, Supergravity” (GSS) research project.

\appendix

\section{Conventions on the elliptic genus}
\label{convEG}

In this appendix we fix the conventions used in the paper for the elliptic genus.
We refer here to the elliptic genus in the NSNS sector, defined originally in \cite{Gadde:2013dda,Gadde:2013wq} (see \cite{Benini:2013nda,Benini:2013xpa} for the definition in the 
RR sector \cite{Gadde:2013dda,Gadde:2013wq}).
The index is given by
\begin{equation}
\label{EG}
I(\vec u;q) \equiv I(\vec u) \equiv  \Tr_{\text{NSNS}}(-1)^F q^{L_0}\prod_{a} u_a^{c_a},
\end{equation}
where $q=\ee^{2 \pi \mi \tau}$ and $\tau$ refers to the complex structure of $T^2$. 
Formula (\ref{EG}) can be interpreted as a Witten index refined by the flavor fugacities $u_a$.
Flat directions render the index divergent in presence of non-compact target space, therefore, it is crucial to keep the flavor fugacities turned on. This can be interpreted as a restriction of the analysis to the massive theory.

Denoting the gauge group as $G$, the elliptic genus can be computed as an integral over the Cartan torus of $G$, expressed in terms of the gauge fugacities $z$
\begin{equation}
I(u) = \frac{1}{|W|} \oint_{\mathrm{JK}} \prod_{i=1}^{\text{rk} \, G} \frac{\dd z_i}{2 \pi \mi z_i} I_{V}(\vec z) I_{\chi} (\vec z,\vec u) I_{\psi} (\vec z,\vec u),
\end{equation}
where JK refers to the fact that the integral is computed by using the Jeffrey-Kirwan prescription.

The vector, the chiral and the Fermi multiplets contribute respectively as
\begin{eqnarray}
I_{V}(\vec z) &=& (q;q)_{\infty}^{2 \text{rk}\,G}\prod_{\alpha_G} \theta \left(z^{\alpha_G}\right),\nonumber \\
I_\chi(\vec z,\vec u) &=& \prod_{\rho_G,\rho_F} \frac{1}{\theta \left(q^{\frac{R_\chi}{2}} z^{\rho_G} u^{\rho_F}\right)},  \\
I_{\psi}(\vec z,\vec u) &=&\prod_{\rho_G,\rho_F} \theta\left( q^{\frac{R_{\psi}+1}{2}} z^{\rho_G} u^{\rho_F}  \right),\nonumber
\end{eqnarray} with
 $\theta(x) = (x;q)_{\infty}  (q x^{-1} ;q)_{\infty}$ and $(x;q)_\infty = \prod_{j=0}^{\infty}(1-x q^j)$.

The index of an $\SU(N)$ gauge theory with $F$ fundamentals $Q$,
$\tilde F$ antifundamentals $\tilde Q$, $H$ fundamental (or antifundamental) Fermi $\Lambda$,
$K$ antisymmetrics $A$ and $\tilde K$ conjugate antisymmetrics $\tilde A$ can be expressed as  
\begin{align}
  I_{\SU(N)}^{(F,\tilde F;H;K;\tilde K)}(\vec m;\vec n; \vec h;\vec r,\vec s)&=\frac{(q;q)_{\infty}^{2(N-1)} }{N!}\nonumber\\
  &\times \oint_{\mathrm{JK}} \prod_{i=1}^{N} \frac{\dd z_i}{2 \pi \mi z_i} 
  \frac{\prod_{i < j } \theta \left((z_i/ z_j)^{\pm 1}\right)
  \prod_{i=1}^{N} \prod_{a=1}^{H}  \theta (q^{\frac{R_\Lambda+1}{2}} z_i h_a )}{\prod_{i=1}^{N} \left(\prod_{a=1}^{F}
  \theta \left(q^{R_Q} z_i m_a \right) \cdot \prod_{a=1}^{\tilde F} \theta (q^{R_{\tilde Q}} z_i^{-1} n_a )\right)}\nonumber\\[5pt]
  &\times \frac{\delta(1-\prod_{i=1}^{N} z_i)}{\prod_{i < j } \left(\prod_{a=1}^{K} \theta ( q^{R_A} r_a z_i z_j) \cdot \prod_{a=1}^{\tilde K} \theta \left( q^{R_{\tilde A}} s_a z_i^{-1} z_j^{-1}\right)\right)} \,.
\end{align}
Throughout the paper, we have conventionally omitted the fugacities by leaving a $\cdot$ in the argument, when the associated field is absent.

The index of an $\USp(2N)$ gauge theory with $F$ fundamentals $Q$,
$H$ fundamental Fermi multiplets  $\Lambda$ and one traceless
antisymmetric chiral $A$ is denoted as
\begin{eqnarray}
I_{\USp(2N)}^{(F;H;1)} (\vec m; \vec h;r)
 &=& 
 \frac{(q;q)_\infty^{2N}}{2^N N!  \, \theta \left(q^{r_A}r\right)^{N-1}}
 \oint_{\mathrm{JK}} \prod_{i=1}^{N} \frac{\dd z_i}{2 \pi \mi z_i} 
\frac{\prod_{i < j } \theta \left(z_i^{\pm 1} z_j^{\pm 1}\right)
 \prod_{i=1}^{N}  \theta (z_i^{\pm 2 })}
 {\prod_{i < j } \theta \left(q^{r_A} z_i^{\pm 1} z_j^{\pm 1} r\right)}
\nonumber \\
 &\times& 
 \prod_{i=1}^{N}  \frac{\prod_{a=1}^{H}  \theta \left(q^{\frac{R_\Lambda+1}{2}} z_i ^{\pm 1} h_a \right)}{
\prod_{a=1}^{F} \theta (q^{R_Q} z_i^{\pm 1} m_a )}\,.
\end{eqnarray}

\newpage

\bibliographystyle{JHEP}
\bibliography{ref.bib}

\providecommand{\href}[2]{#2}\begingroup\raggedright\begin{thebibliography}{10}

\bibitem{Gadde:2015wta}
A.~Gadde, S.~S. Razamat and B.~Willett, \emph{{On the reduction of 4d $
  \mathcal{N}=1 $ theories on $ {\mathbb{S}}^2 $}},
  \href{https://doi.org/10.1007/JHEP11(2015)163}{\emph{JHEP} {\bfseries 11}
  (2015) 163} [\href{https://arxiv.org/abs/1506.08795}{{\ttfamily
  1506.08795}}].

\bibitem{Benini:2015noa}
F.~Benini and A.~Zaffaroni, \emph{{A topologically twisted index for
  three-dimensional supersymmetric theories}},
  \href{https://doi.org/10.1007/JHEP07(2015)127}{\emph{JHEP} {\bfseries 07}
  (2015) 127} [\href{https://arxiv.org/abs/1504.03698}{{\ttfamily
  1504.03698}}].

\bibitem{Benini:2012cz}
F.~Benini and N.~Bobev, \emph{{Exact two-dimensional superconformal R-symmetry
  and c-extremization}},
  \href{https://doi.org/10.1103/PhysRevLett.110.061601}{\emph{Phys. Rev. Lett.}
  {\bfseries 110} (2013) 061601}
  [\href{https://arxiv.org/abs/1211.4030}{{\ttfamily 1211.4030}}].

\bibitem{Benini:2013cda}
F.~Benini and N.~Bobev, \emph{{Two-dimensional SCFTs from wrapped branes and
  c-extremization}}, \href{https://doi.org/10.1007/JHEP06(2013)005}{\emph{JHEP}
  {\bfseries 06} (2013) 005} [\href{https://arxiv.org/abs/1302.4451}{{\ttfamily
  1302.4451}}].

\bibitem{Seiberg:1994pq}
N.~Seiberg, \emph{{Electric - magnetic duality in supersymmetric nonAbelian
  gauge theories}},
  \href{https://doi.org/10.1016/0550-3213(94)00023-8}{\emph{Nucl. Phys. B}
  {\bfseries 435} (1995) 129}
  [\href{https://arxiv.org/abs/hep-th/9411149}{{\ttfamily hep-th/9411149}}].

\bibitem{Intriligator:1995ne}
K.~A. Intriligator and P.~Pouliot, \emph{{Exact superpotentials, quantum vacua
  and duality in supersymmetric SP(N(c)) gauge theories}},
  \href{https://doi.org/10.1016/0370-2693(95)00618-U}{\emph{Phys. Lett. B}
  {\bfseries 353} (1995) 471}
  [\href{https://arxiv.org/abs/hep-th/9505006}{{\ttfamily hep-th/9505006}}].

\bibitem{Sacchi:2020pet}
M.~Sacchi, \emph{{New 2d $ \mathcal{N} $ = (0, 2) dualities from four
  dimensions}}, \href{https://doi.org/10.1007/JHEP12(2020)009}{\emph{JHEP}
  {\bfseries 12} (2020) 009}
  [\href{https://arxiv.org/abs/2004.13672}{{\ttfamily 2004.13672}}].

\bibitem{Jiang:2024ifv}
J.~Jiang, S.~Nawata and J.~Zheng, \emph{{2d dualities from 4d}},
  \href{https://arxiv.org/abs/2407.17350}{{\ttfamily 2407.17350}}.

\bibitem{Amariti:2024usp}
A.~Amariti, P.~Glorioso, F.~Mantegazza, D.~Morgante and A.~Zanetti,
  \emph{{Dualities from dualities in 2d $\mathcal{N}=(0,2)$}},
  \href{https://arxiv.org/abs/2410.12453}{{\ttfamily 2410.12453}}.

\bibitem{Berkooz:1995km}
M.~Berkooz, \emph{{The Dual of supersymmetric SU(2k) with an antisymmetric
  tensor and composite dualities}},
  \href{https://doi.org/10.1016/0550-3213(95)00400-M}{\emph{Nucl. Phys. B}
  {\bfseries 452} (1995) 513}
  [\href{https://arxiv.org/abs/hep-th/9505067}{{\ttfamily hep-th/9505067}}].

\bibitem{Luty:1996cg}
M.~A. Luty, M.~Schmaltz and J.~Terning, \emph{{A Sequence of duals for Sp(2N)
  supersymmetric gauge theories with adjoint matter}},
  \href{https://doi.org/10.1103/PhysRevD.54.7815}{\emph{Phys. Rev. D}
  {\bfseries 54} (1996) 7815}
  [\href{https://arxiv.org/abs/hep-th/9603034}{{\ttfamily hep-th/9603034}}].

\bibitem{Pasquetti:2019uop}
S.~Pasquetti and M.~Sacchi, \emph{{From 3$d$ dualities to 2$d$ free field
  correlators and back}},
  \href{https://doi.org/10.1007/JHEP11(2019)081}{\emph{JHEP} {\bfseries 11}
  (2019) 081} [\href{https://arxiv.org/abs/1903.10817}{{\ttfamily
  1903.10817}}].

\bibitem{Benvenuti:2020wpc}
S.~Benvenuti, I.~Garozzo and G.~Lo~Monaco, \emph{{Monopoles and dualities in
  3d$ \mathcal{N} $ = 2 quivers}},
  \href{https://doi.org/10.1007/JHEP10(2021)191}{\emph{JHEP} {\bfseries 10}
  (2021) 191} [\href{https://arxiv.org/abs/2012.08556}{{\ttfamily
  2012.08556}}].

\bibitem{Etxebarria:2021lmq}
I.~G. Etxebarria, B.~Heidenreich, M.~Lotito and A.~K. Sorout,
  \emph{{Deconfining $ \mathcal{N} $ = 2 SCFTs or the art of brane bending}},
  \href{https://doi.org/10.1007/JHEP03(2022)140}{\emph{JHEP} {\bfseries 03}
  (2022) 140} [\href{https://arxiv.org/abs/2111.08022}{{\ttfamily
  2111.08022}}].

\bibitem{Benvenuti:2021nwt}
S.~Benvenuti and G.~Lo~Monaco, \emph{{A toolkit for ortho-symplectic
  dualities}}, \href{https://doi.org/10.1007/JHEP09(2023)002}{\emph{JHEP}
  {\bfseries 09} (2023) 002}
  [\href{https://arxiv.org/abs/2112.12154}{{\ttfamily 2112.12154}}].

\bibitem{Bottini:2022vpy}
L.~E. Bottini, C.~Hwang, S.~Pasquetti and M.~Sacchi, \emph{{Dualities from
  dualities: the sequential deconfinement technique}},
  \href{https://doi.org/10.1007/JHEP05(2022)069}{\emph{JHEP} {\bfseries 05}
  (2022) 069} [\href{https://arxiv.org/abs/2201.11090}{{\ttfamily
  2201.11090}}].

\bibitem{Bajeot:2022lah}
S.~Bajeot and S.~Benvenuti, \emph{{Sequential deconfinement and self-dualities
  in 4d$ \mathcal{N} $ = 1 gauge theories}},
  \href{https://doi.org/10.1007/JHEP10(2022)007}{\emph{JHEP} {\bfseries 10}
  (2022) 007} [\href{https://arxiv.org/abs/2206.11364}{{\ttfamily
  2206.11364}}].

\bibitem{Bajeot:2022wmu}
S.~Bajeot and S.~Benvenuti, \emph{{4d$ \mathcal{N} $ = 1 dualities from 5d
  dualities}}, \href{https://doi.org/10.1007/JHEP08(2024)197}{\emph{JHEP}
  {\bfseries 08} (2024) 197}
  [\href{https://arxiv.org/abs/2212.11217}{{\ttfamily 2212.11217}}].

\bibitem{Amariti:2022wae}
A.~Amariti and S.~Rota, \emph{{3d N=2 SO/USp adjoint SQCD: s-confinement and
  exact identities}},
  \href{https://doi.org/10.1016/j.nuclphysb.2022.116068}{\emph{Nucl. Phys. B}
  {\bfseries 987} (2023) 116068}
  [\href{https://arxiv.org/abs/2202.06885}{{\ttfamily 2202.06885}}].

\bibitem{Amariti:2023wts}
A.~Amariti, F.~Mantegazza and D.~Morgante, \emph{{Sporadic dualities from
  tensor deconfinement}},
  \href{https://doi.org/10.1007/JHEP05(2024)188}{\emph{JHEP} {\bfseries 05}
  (2024) 188} [\href{https://arxiv.org/abs/2307.14146}{{\ttfamily
  2307.14146}}].

\bibitem{Amariti:2024sde}
A.~Amariti and F.~Mantegazza, \emph{{A new 4d $ \mathcal{N} $ = 1 duality from
  the superconformal index}},
  \href{https://doi.org/10.1007/JHEP06(2024)206}{\emph{JHEP} {\bfseries 06}
  (2024) 206} [\href{https://arxiv.org/abs/2402.00609}{{\ttfamily
  2402.00609}}].

\bibitem{Amariti:2024gco}
A.~Amariti and F.~Mantegazza, \emph{{Confinement for 3d $\mathcal{N}=2$$SU(N)$
  with a Symmetric tensor}},
  \href{https://arxiv.org/abs/2405.11972}{{\ttfamily 2405.11972}}.

\bibitem{Benvenuti:2024glr}
S.~Benvenuti, R.~Comi, S.~Pasquetti and M.~Sacchi, \emph{{Deconfinements,
  Kutasov-Schwimmer dualities and $D_p[SU(N)]$ theories}},
  \href{https://arxiv.org/abs/2407.11134}{{\ttfamily 2407.11134}}.

\bibitem{Hwang:2024hhy}
C.~Hwang and S.~Kim, \emph{{S-confinement of 3d Argyres-Douglas theories and
  the Seiberg-like duality with an adjoint matter}},
  \href{https://arxiv.org/abs/2407.11129}{{\ttfamily 2407.11129}}.

\bibitem{Bajeot:2022kwt}
S.~Bajeot and S.~Benvenuti, \emph{{S-confinements from deconfinements}},
  \href{https://doi.org/10.1007/JHEP11(2022)071}{\emph{JHEP} {\bfseries 11}
  (2022) 071} [\href{https://arxiv.org/abs/2201.11049}{{\ttfamily
  2201.11049}}].

\bibitem{Csaki:1996zb}
C.~Csaki, M.~Schmaltz and W.~Skiba, \emph{{Confinement in N=1 SUSY gauge
  theories and model building tools}},
  \href{https://doi.org/10.1103/PhysRevD.55.7840}{\emph{Phys. Rev. D}
  {\bfseries 55} (1997) 7840}
  [\href{https://arxiv.org/abs/hep-th/9612207}{{\ttfamily hep-th/9612207}}].

\bibitem{Nii:2019ebv}
K.~Nii, \emph{{On s-confinement in 3d $\mathcal{N}=2$ gauge theories with
  anti-symmetric tensors}},  \href{https://arxiv.org/abs/1906.03908}{{\ttfamily
  1906.03908}}.

\bibitem{Toappear}
A.~Amariti, F.~Mantegazza and S.~Rota, \emph{Rank-two tensors and deconfinement
  in $su(n)$ gauge theories with 4 supercharges}, {\emph{To Appear} (2025) }.

\bibitem{Dimofte:2017tpi}
T.~Dimofte, D.~Gaiotto and N.~M. Paquette, \emph{{Dual boundary conditions in
  3d SCFT\textquoteright{}s}},
  \href{https://doi.org/10.1007/JHEP05(2018)060}{\emph{JHEP} {\bfseries 05}
  (2018) 060} [\href{https://arxiv.org/abs/1712.07654}{{\ttfamily
  1712.07654}}].

\bibitem{Spiridonov:2009za}
V.~P. Spiridonov and G.~S. Vartanov, \emph{{Elliptic Hypergeometry of
  Supersymmetric Dualities}},
  \href{https://doi.org/10.1007/s00220-011-1218-9}{\emph{Commun. Math. Phys.}
  {\bfseries 304} (2011) 797}
  [\href{https://arxiv.org/abs/0910.5944}{{\ttfamily 0910.5944}}].

\bibitem{Aharony:2013dha}
O.~Aharony, S.~S. Razamat, N.~Seiberg and B.~Willett, \emph{{3d dualities from
  4d dualities}}, \href{https://doi.org/10.1007/JHEP07(2013)149}{\emph{JHEP}
  {\bfseries 07} (2013) 149} [\href{https://arxiv.org/abs/1305.3924}{{\ttfamily
  1305.3924}}].

\bibitem{Dedushenko:2017osi}
M.~Dedushenko and S.~Gukov, \emph{{IR duality in 2D $N=(0,2)$ gauge theory with
  noncompact dynamics}},
  \href{https://doi.org/10.1103/PhysRevD.99.066005}{\emph{Phys. Rev. D}
  {\bfseries 99} (2019) 066005}
  [\href{https://arxiv.org/abs/1712.07659}{{\ttfamily 1712.07659}}].

\bibitem{Razamat:2017hda}
S.~S. Razamat and G.~Zafrir, \emph{{E$_{8}$ orbits of IR dualities}},
  \href{https://doi.org/10.1007/JHEP11(2017)115}{\emph{JHEP} {\bfseries 11}
  (2017) 115} [\href{https://arxiv.org/abs/1709.06106}{{\ttfamily
  1709.06106}}].

\bibitem{SPIRIDONOV200691}
V.~P. Spiridonov and S.~O. Warnaar, \emph{Inversions of integral operators and
  elliptic beta integrals on root systems},
  \href{https://doi.org/https://doi.org/10.1016/j.aim.2005.11.007}{\emph{Advances
  in Mathematics} {\bfseries 207} (2006) 91}.

\bibitem{Nazzal:2021tiu}
B.~Nazzal, A.~Nedelin and S.~S. Razamat, \emph{{Minimal $(D,D)$ conformal
  matter and generalizations of the van Diejen model}},
  \href{https://doi.org/10.21468/SciPostPhys.12.4.140}{\emph{SciPost Phys.}
  {\bfseries 12} (2022) 140}
  [\href{https://arxiv.org/abs/2106.08335}{{\ttfamily 2106.08335}}].

\bibitem{van2009elliptic}
F.~J. van~de Bult, \emph{An elliptic hypergeometric beta integral
  transformation}, {\emph{arXiv preprint arXiv:0912.3812} (2009) }.

\bibitem{Closset:2013sxa}
C.~Closset and I.~Shamir, \emph{{The $\mathcal{N}=1$ Chiral Multiplet on
  $T^2\times S^2$ and Supersymmetric Localization}},
  \href{https://doi.org/10.1007/JHEP03(2014)040}{\emph{JHEP} {\bfseries 03}
  (2014) 040} [\href{https://arxiv.org/abs/1311.2430}{{\ttfamily 1311.2430}}].

\bibitem{Honda:2015yha}
M.~Honda and Y.~Yoshida, \emph{{Supersymmetric index on $T^2 \times S^2$ and
  elliptic genus}},  \href{https://arxiv.org/abs/1504.04355}{{\ttfamily
  1504.04355}}.

\bibitem{Okazaki:2023hiv}
T.~Okazaki and D.~J. Smith, \emph{{Boundary confining dualities and
  Askey-Wilson type q-beta integrals}},
  \href{https://doi.org/10.1007/JHEP08(2023)048}{\emph{JHEP} {\bfseries 08}
  (2023) 048} [\href{https://arxiv.org/abs/2305.00247}{{\ttfamily
  2305.00247}}].

\bibitem{Razamat:2018gbu}
S.~S. Razamat, O.~Sela and G.~Zafrir, \emph{{Curious patterns of IR symmetry
  enhancement}}, \href{https://doi.org/10.1007/JHEP10(2018)163}{\emph{JHEP}
  {\bfseries 10} (2018) 163}
  [\href{https://arxiv.org/abs/1809.00541}{{\ttfamily 1809.00541}}].

\bibitem{Gadde:2013dda}
A.~Gadde and S.~Gukov, \emph{{2d Index and Surface operators}},
  \href{https://doi.org/10.1007/JHEP03(2014)080}{\emph{JHEP} {\bfseries 03}
  (2014) 080} [\href{https://arxiv.org/abs/1305.0266}{{\ttfamily 1305.0266}}].

\bibitem{Gadde:2013wq}
A.~Gadde, S.~Gukov and P.~Putrov, \emph{{Walls, Lines, and Spectral Dualities
  in 3d Gauge Theories}},
  \href{https://doi.org/10.1007/JHEP05(2014)047}{\emph{JHEP} {\bfseries 05}
  (2014) 047} [\href{https://arxiv.org/abs/1302.0015}{{\ttfamily 1302.0015}}].

\bibitem{Benini:2013nda}
F.~Benini, R.~Eager, K.~Hori and Y.~Tachikawa, \emph{{Elliptic genera of
  two-dimensional N=2 gauge theories with rank-one gauge groups}},
  \href{https://doi.org/10.1007/s11005-013-0673-y}{\emph{Lett. Math. Phys.}
  {\bfseries 104} (2014) 465}
  [\href{https://arxiv.org/abs/1305.0533}{{\ttfamily 1305.0533}}].

\bibitem{Benini:2013xpa}
F.~Benini, R.~Eager, K.~Hori and Y.~Tachikawa, \emph{{Elliptic Genera of 2d
  ${\mathcal{N}}$ = 2 Gauge Theories}},
  \href{https://doi.org/10.1007/s00220-014-2210-y}{\emph{Commun. Math. Phys.}
  {\bfseries 333} (2015) 1241}
  [\href{https://arxiv.org/abs/1308.4896}{{\ttfamily 1308.4896}}].

\end{thebibliography}\endgroup

\end{document}